\documentclass[12pt,preprint]{emulateapj}
\pdfoutput=1
\usepackage{graphicx}
\usepackage{epstopdf}

\shortauthors{Bertincourt et al.}
\shorttitle{A {\it Spitzer} Unbiased Ultradeep Spectroscopic Survey (SUUSS)}

\begin{document}

\title{A {\it Spitzer} Unbiased Ultradeep Spectroscopic Survey}
\author{B. Bertincourt\altaffilmark{1,2}, G. Helou\altaffilmark{2}, P. Appleton\altaffilmark{2,3}, P. Ogle\altaffilmark{2}, G. Lagache\altaffilmark{1}, T. Brooke\altaffilmark{2}, J-D. Smith\altaffilmark{4}, K. Sheth\altaffilmark{2}, D. Dale\altaffilmark{5}, M. Harwit\altaffilmark{6}, J-L. Puget\altaffilmark{1}, H. Roussel\altaffilmark{7}.}

\altaffiltext{1} {Institut d'Astrophysique Spatiale (IAS), B\^atiment
121, F-91405 Orsay (France); Universit\'e Paris-Sud 11 and CNRS (UMR
8617);
[benjamin.bertincourt, guilaine.lagache, jean-loup.puget]@ias.u-psud.fr.}
\altaffiltext{2}{{\it Spitzer} Science Center, California Institute of Technology, 1200 East California Boulevard, Pasadena, CA 91125.}
\altaffiltext{3}{NASA {\it Herschel} Science Center, California Institute of Technology, Pasadena, CA 91125.}
\altaffiltext{4}{Ritter Astrophysical Observatory, University of Toledo, Toledo, OH 43606, USA.}
\altaffiltext{5}{Department of Physics and Astronomy, University of Wyoming, Laramie, WY 82071, USA.}
\altaffiltext{6}{511 H Street, SW, Washington DC 20024-2725, United States; Cornell University, Ithaca, New York, United States.}
\altaffiltext{7}{Institut d'Astrophysique de Paris, 75014 Paris, France; UPMC (Universit\'e Paris 6).}

\begin{abstract}
We carried out an unbiased, spectroscopic survey using the low-resolution module of the infrared spectrograph (IRS) on board Spitzer targeting two 2.6 square arcminute regions in the GOODS-North field. IRS was used in spectral mapping mode with 5 hours of effective integration time per pixel. One region was covered between 14 and 21$\mu m$ and the other between 20 and 35$\mu m$. We extracted spectra for 45 sources. About 84\% of the sources have reported detections by GOODS at $24\mu m$, with a median $f_\nu(24\mu m)\sim 100\mu Jy$. All but one source are detected in all four IRAC bands, $3.6$ to 8 $\mu$m. We use a new cross-correlation technique to measure redshifts and estimate IRS spectral types; this was successful for $\sim 60\%$ of the spectra. Fourteen sources show significant PAH emission, four mostly SiO absorption, eight present mixed spectral signatures (low PAH and/or SiO) and two show a single line in emission. For the remaining 17, no spectral features were detected. Redshifts range from $z\sim 0.2$ to $z\sim 2.2$, with a median of 1. 
IR Luminosities are roughly estimated from 24$\mu m$ flux densities, and have median values of $2.2 \times 10^{11} L_{\odot}$ and $7.5 \times 10^{11} L_{\odot}$ at $z\sim 1$ and $z\sim 2$ respectively.
This sample has fewer AGN than previous faint samples observed with IRS, which we attribute to the fainter luminosities reached here.  
\end{abstract}


\section{Introduction}

The Spitzer Space Telescope \citep{2004AdSpR..34..600W} has multiplied by several orders of magnitude the volumes previously surveyed in the  infrared for extragalactic objects \citep{2008ARA&A..46..201S}.  Mid-infrared continuum surveys turned out to be especially fertile ground  because of the superb sensitivity and speed of the 24$\mu m$ channel in the MIPS instrument \citep{2004ApJS..154...25R}.  These mid-infrared surveys have established the dramatic evolution of galaxy populations over the last 10-12 billion years (e.g. \citeauthor{2004ApJS..154..170L} \citeyear{2004ApJS..154..170L}, \citeyear{2005ApJ...632..169L} ; \citeauthor{2006ApJ...637..727C} \citeyear{2006ApJ...637..727C}; \citeauthor{2007ApJ...668...45P} \citeyear{2007ApJ...668...45P}) previously suggested by ISO surveys (e.g. \citeauthor{1999A&A...351L..37E} \citeyear{1999A&A...351L..37E}; \citealt{2000MNRAS.316..749O}; \citeauthor{2001A&A...372..364D} \citeyear{2001A&A...372..364D}; \citealt{2004MNRAS.351.1290R}), and yielded  rich samples of galaxies for followup.  Critical to the interpretation of the data is the availability of multiple-band detections, which are used to constrain the redshifts, luminosities and sometimes the powering mechanism of sources.  These constraints come about because the mid-infrared spectra of galaxies are often strongly structured (\citeauthor{2007ApJ...656..770S} \citeyear{2007ApJ...656..770S}; \citeauthor{2007ApJ...656..148A} \citeyear{2007ApJ...656..148A}), and sensitive to the presence of an AGN, to optical depth, and to heating and dust geometry.  By the same token, any single-band survey will have redshift-dependent  biases resulting in a sampling of the population that varies with redshift. The obvious way to avoid this sampling bias is to survey in the dispersed light, a technique pioneered from the ground using objective-prism imaging in the visible.

In this paper we report on a spectrally unbiased extragalactic census in the mid-infrared using the IRS instrument \citep{2004ApJS..154...18H} on Spitzer, and covering two spectral windows, 14 to 21$\mu m$ and 20 to 38$\mu m$. Besides the specific results reported here, this survey illustrates the advantages of spectral mapping in the infrared. The placement of the windows allows us to attach an observed spectrum to the corresponding 24$\mu m$ source, and could therefore potentially address the bolometric correction for each source.  The data  also allow us to use spectral shapes, features and emission lines to characterize all  detected sources, rather than only those sources  selected for follow-up by the inevitably rough criteria of filtering on colors within the mid-infrared or in combination with other bands (e.g. \citeauthor{2004ApJS..154...75Y} \citeyear{2004ApJS..154...75Y}; \citeauthor{2006ApJ...651..101W} \citeyear{2006ApJ...651..101W}; \citeauthor{2007ApJ...671..323H} \citeyear{2007ApJ...671..323H}; \citeauthor{2008ApJ...677..957F} \citeyear{2008ApJ...677..957F}). This unbiased examination of the emission line properties also  yields a fair assessment of the distribution of line strengths and line-to-continuum ratios.  Such an assessment constrains the frequency of line-dominated sources, and might even yield examples of sources radiating largely line emission, and therefore very rarely picked up in broad-band surveys. When our survey was proposed, a few examples of line-dominated sources in the mid-infrared were known, including NGC~1569 \citep{2003ApJ...588..199L}.  Today the most striking case known is probably the intergalactic shock in Stephan's Quintet, which emits more than 20\% of its total infrared emission in the main three molecular hydrogen pure rotational transitions \citep{2006ApJ...639L..51A}. There is no reason to rule out scaled up versions of these systems, which might be detectable with Spitzer.

Source confusion is often  a limiting factor for infrared surveys, since the need to cool telescopes limits the size of the primary. The confusion limit appears in a variety of forms and definitions (\citeauthor{1990LIACo..29..117H} \citeyear{1990LIACo..29..117H}; \citeauthor{2001ApJ...549..745R} \citeyear{2001ApJ...549..745R}; \citeauthor{2003MNRAS.338..555L} \citeyear{2003MNRAS.338..555L}; \citeauthor{2004ApJS..154...93D} \citeyear{2004ApJS..154...93D}), but is ultimately dictated by the source density and its dependence on flux.  A spectrally dispersed survey offers the spectral dimension as a discriminator among closely spaced sources, since the strongly featured shapes of mid-infrared spectra would cause sources to peak almost always at different wavelengths. A method based on PSF fitting at different wavelengths allows to estimate accurately the confused sources positions at their peak wavelength as well as their contribution in flux to the neighboring sources. This spectral separation, solely relying on the knowledge of the PSF of the instrument, can be used as a prior to extract distinct spectra and thus disentangle sources, even if their broad-band images overlap substantially.

This paper describes the survey design, observations and data reduction in Sect. 2, describes the extraction of redshifts and spectral types from the data in Sect. 3, presents results on the distribution of redshifts and relation to broadband surveys in Sect. 4, and discusses those results in Sect. 5.  

Throughout this paper, we assume a $\Lambda$-CDM cosmology with $H_{0} = 70$ km s$^{-1}$ Mpc$^{-1}$, $\Omega_{M} = 0.3$ and $\Omega_{\Lambda} = 0.7$.

\section{Observations and Data Reduction}

\begin{figure*}
\plotone{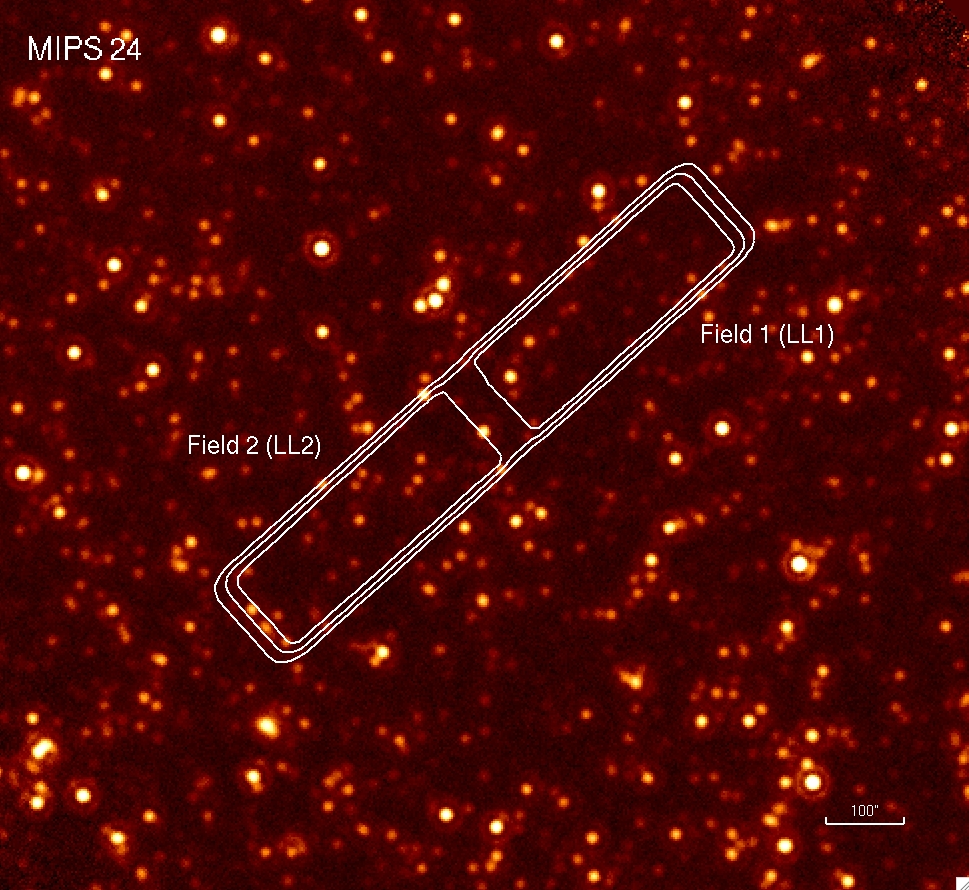}
\caption{Depth-of-coverage of the {\it Spitzer} Unbiased Ultradeep Spectroscopic Survey overplotted on MIPS 24$\mu m$ observation ({\it background}) of the GOODS-North field \citep{2003mglh.conf..324D}. We show depth contours in {\it white} at values of 48, 95 and 143 (number of times a given position is covered by the IRS slit), respectively from the innermost to the outermost contour.}
\label{AORdescription}
\end{figure*}

The goal was to carry out a  blind survey rather than targeting specific sources. However, the general location of the surveyed area on the sky could be optimized to minimize sky brightness, and improve access to ancillary data.  We chose to place the survey in the Hubble Deep Field North (HDF-N) covered in the infrared by Spitzer as part of the GOODS survey \citep{2003mglh.conf..324D}. The availability of deep imaging with Spitzer allows us to relate the IRS data to continuum imaging immediately, and to avoid bright sources that would generate artifacts and reduce the effective size of the survey area.
Additional ancillary data such as redshifts, ground-based photometry or high-resolution imaging all facilitate interpretation.  We also made an effort to place the survey area in a particularly low-density region of GOODS-N judging by the 24$\mu$m continuum map (Fig \ref{AORdescription}).

The GOODS 24$\mu$m data are 84\% complete around 80$\mu$Jy, and go as deep as 10$\mu$Jy, the weakest extractions reported by the GOODS team. 

\subsection{Observations}

We observed a region in the Hubble Deep Field North (Fig \ref{AORdescription}) with the Long-Low module of the Infrared Spectrograph (IRS) onboard Spitzer. Data are collected simultaneously from the two slits (LL1 and LL2) of the low resolution module of the IRS, which cover different spectral ranges. We chose to maximize the depth of coverage with each slit, and therefore obtained two spatially and spectrally disjoint sets of data.
The survey was executed using 8 AORs (Astronomical Observation Requests) for a total of 46.5 hours of observation resulting in an integration time at full depth of 5 hours per sky position. Each AOR consists of 65 slit exposures following a 13 by 5 raster map with a step of one pixel (5.1") between each exposure. Moreover, the 8 AORs are dithered by 1/3 of a pixel in order to improve the Point Spread Function (PSF) coverage at short wavelength. Those observations yield two adjacent areas, covering 14 to 21$\mu$m (LL2) and 20 to 38$\mu$m (LL1) respectively. Each field is covered to near-full depth over an area of $165^{\prime \prime} \times 56^{\prime \prime}$ (innermost contour in Fig \ref{AORdescription}), corresponding to a number of spatial resolution elements of $33 \times 11$ for both LL1 and LL2.
This is the most effective use of telescope time and instrument for an exploratory survey, rather than attempting to cover the same area with both sub-modules to get complete wavelength coverage. 

\subsection{Data filtering and Cube construction}

Processing was pushed farther than the usual IRS pipeline products to obtain the best sensitivity. We applied additional reduction steps starting with the \emph{Basic Calibrated Data} (BCD) level processed through the version S15 of the pipeline.  BCD are dispersed slit images that have been cleaned of radiation hit artifacts, rectified, and calibrated (see IRS data handbook: \url{http://ssc.spitzer.caltech.edu/irs/dh/dh32.pdf}).

First, we corrected  for an upward drift in the BCD, manifested as a monotonic increase  in the median signal with time for each set of consecutive AORs. This effect was assumed to be a detector effect, since the rise is too steep for any plausible phenomenon on the sky.  It was thus  removed as a function of time (robust polynomial fit). The robust fit was used to be resistant to a small number of outliers in the BCD offsets. The program used was {\it robust\_poly\_fit}, available as part of the IDL Astro Library.
This procedure also removes  any real time-invariant background, since we end up with a zero-median dataset. The data at this point contain solely sources plus rogue signals. The latter originate in rogue pixels,  whose dark current is abnormally high and varies with time and different sky brightness. 
Considering the impact of such artifacts on the faint sources we want to extract, a two-step removal was applied. 
The first step relies on sets of zero-median BCDs taken close in time within a AOR, which we call an ensemble. After choosing the length of the zero-median ensemble, rogue pixels were subtracted by removing pixel-by-pixel a running robust trimmed linear fit to the ensemble from each BCD. The program used was {\it robust\_linefit}, available as part of the IDL Astro Library. A robust linear fit rather than a robust mean was required to take into account some wavelength dependence in the drift described above.  However this residual drift was small and the fits were close to robust means. This method only removes rogue signals thus ensuring that no correlated-noise is added to the data.
We used IRSCLEAN \footnote{software provided by the SSC, \\http://ssc.spitzer.caltech.edu/archanaly/contributed/irsclean/\\IRSCLEAN\_MASK.html} after the ensemble subtraction to check how many additional bad pixels are found for ensembles of decreasing size. Changes in the number of bad-pixels detected were found to be less than 2\% when going below 1/4 of an AOR (32 frames) which was then chosen as the ensemble size. Additional rogue pixels were found and repaired at the single-BCD level using IRSCLEAN. IRSCLEAN fixes pixels by interpolating adjacent rows. One fixed iteration was done on each frame at a threshold of 4 $\sigma$.

\begin{figure}
\plotone{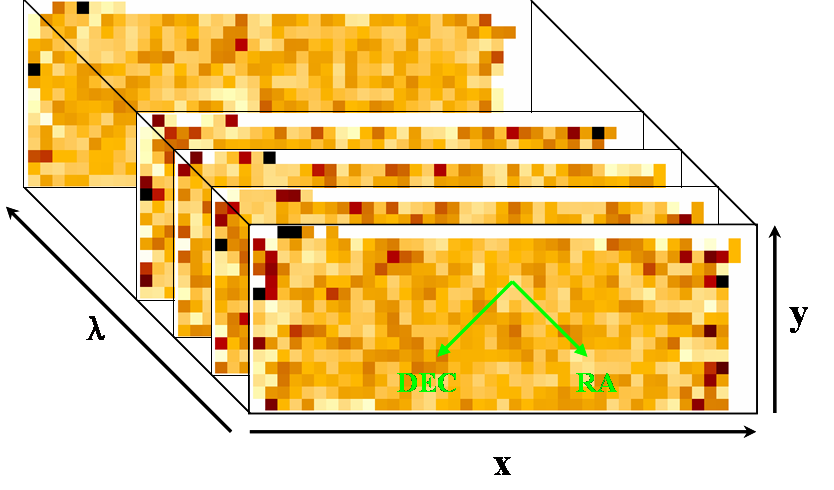}
\caption{Representation of a reconstructed datacube using the CUBISM software developped by the SINGS team (Smith et al. 2007).}
\label{SchemeCube}
\end{figure}

The second step removes the remaining rogue pixels in the map-making. We used \emph{CUBISM} developed  by \cite{2007PASP..119.1133S}, a custom tool created for the assembly and analysis of spectral cubes from IRS spectral maps. Zero-median BCD frames were recombined into two spectral cubes providing 2 spatial and 1 spectral dimensions for each of LL1 (Field 1) and LL2 (Field 2). CUBISM uses trimmed averages over the set of individual measurements going into the signal for each pixel in the cube. The measurements are first weighted by the overlap fraction between BCD pixel area and cube pixel area, then averaged with trimming.
We represent one of the cubes on a diagram in Fig \ref{SchemeCube}. The considerable redundancy in the spatial coverage and the dithering between AORs oversample the sky sufficiently  that  the data cubes could be built with  $(2.55'')^2$ pixels, oversampling by a factor of $\sim$2   the native spatial  pixels. The cubes were not oversampled in the spectral dimension, thus the sampling remains $0.092 \mu m/pixel$ in LL2 and ranges from $0.177$ to $0.187 \mu m/pixel$ in LL1. The resulting cube dimensions are 79 x 28 x 95 pixels covering  $201'' \times 71'' \times 17.3\mu m$ for Field 1 and 79 x 28 x 75 pixels covering $201'' \times 71'' \times 6.9\mu m$ for Field 2, corresponding to the outermost contours in Fig \ref{AORdescription}. 

\subsection{Characterization of the Noise}

\begin{figure}
\plotone{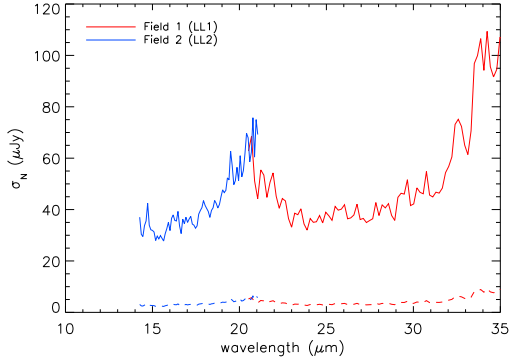}
\caption{Standard deviation of the noise in our dataset in each wavelength bin and for both fields (Field1 in {\it red} and Field 2 in {\it blue}). {\it Solid lines} illustrate the noise relative to each source extraction whereas the {\it dotted lines} correspond to the noise in one single pixel of $(2.55'')^2$.}
\label{Noise}
\end{figure}

In order to estimate the noise in the datacubes and extract its variation with wavelength, we integrated each cube over its whole spectral range creating 2D map-like planes (one per field). We applied an iterative sigma-clipping on those maps to mask pixels containing sources, leaving about a thousand pixels in each field, excluding the edges where the noise increases dramatically due to decreasing coverage in the map. These remaining pixels are assumed to be dominated by noise. Spectra at all of these pixels were then used to compute the 1$\sigma$-deviation of the noise at each wavelength. Note that due to the background substraction, these spectra have a zero mean. Integrated over one oversampled $(2.55'')^2$ pixel, this standard deviation ranges between $2.7\mu$Jy and $9.1\mu$Jy in Field 1 (LL1, between 20 and 35$\mu$m) and between $2.3\mu$Jy and $6.3\mu$Jy in Field 2 (LL2, between 14 and 21$\mu$m) (see {\it dashed lines} in Fig \ref{Noise}).
The samples of positions used to extract the noise were  then  split into several spatially distinct subsets to check for any spatial dependency but no significant variation was found.

All information at $\lambda > 35\mu m$ in Field 1 (LL1) was discarded due to very high red-end noise. In the following, LL1 spectral range refers to a 20 to 35$\mu m$ band.

\subsection{Source detection and extraction} 

\begin{figure*}
\plottwo{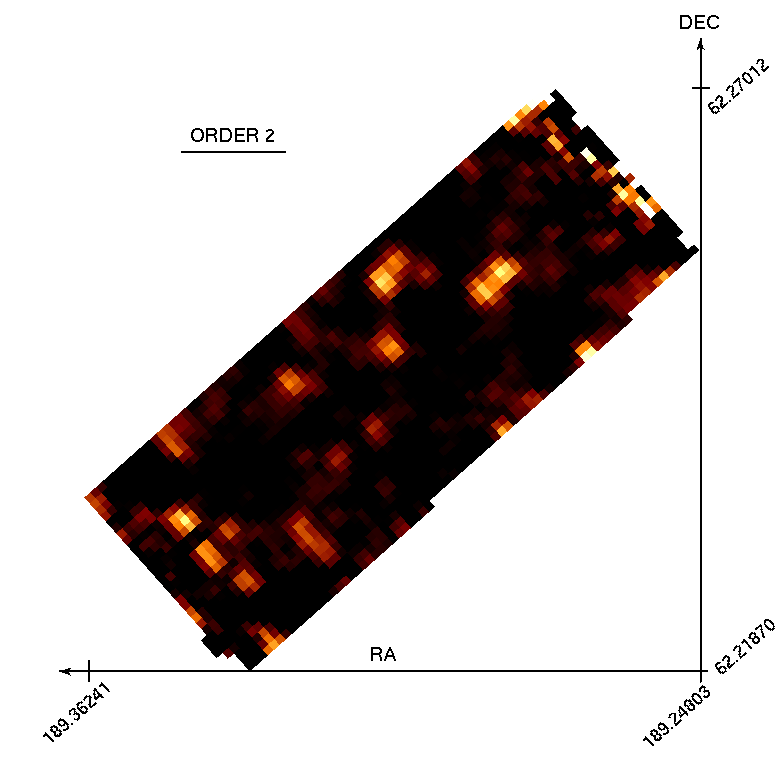}{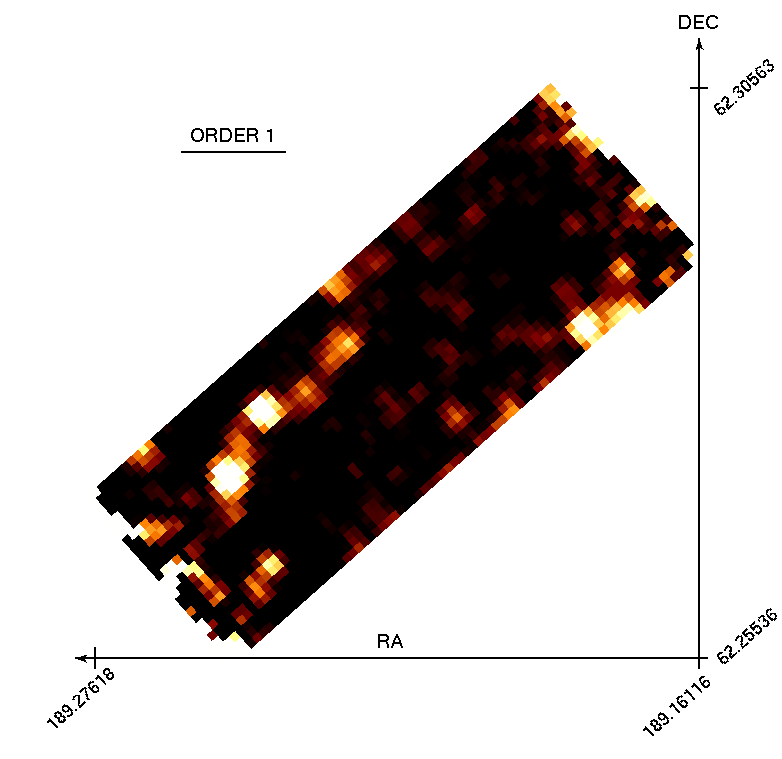}
\caption{On the {\it left}, spatial map (X-Y plane, cf Fig \ref{SchemeCube}) of the Field 2 cube integrated over 14-19$\mu$m.On the {\it right}, spatial map (X-Y plane) of the Field 1 cube integrated over 21-34$\mu$m. Both maps are projected onto the sky.}
\label{Maps}
\end{figure*}

\begin{figure*}
\plotone{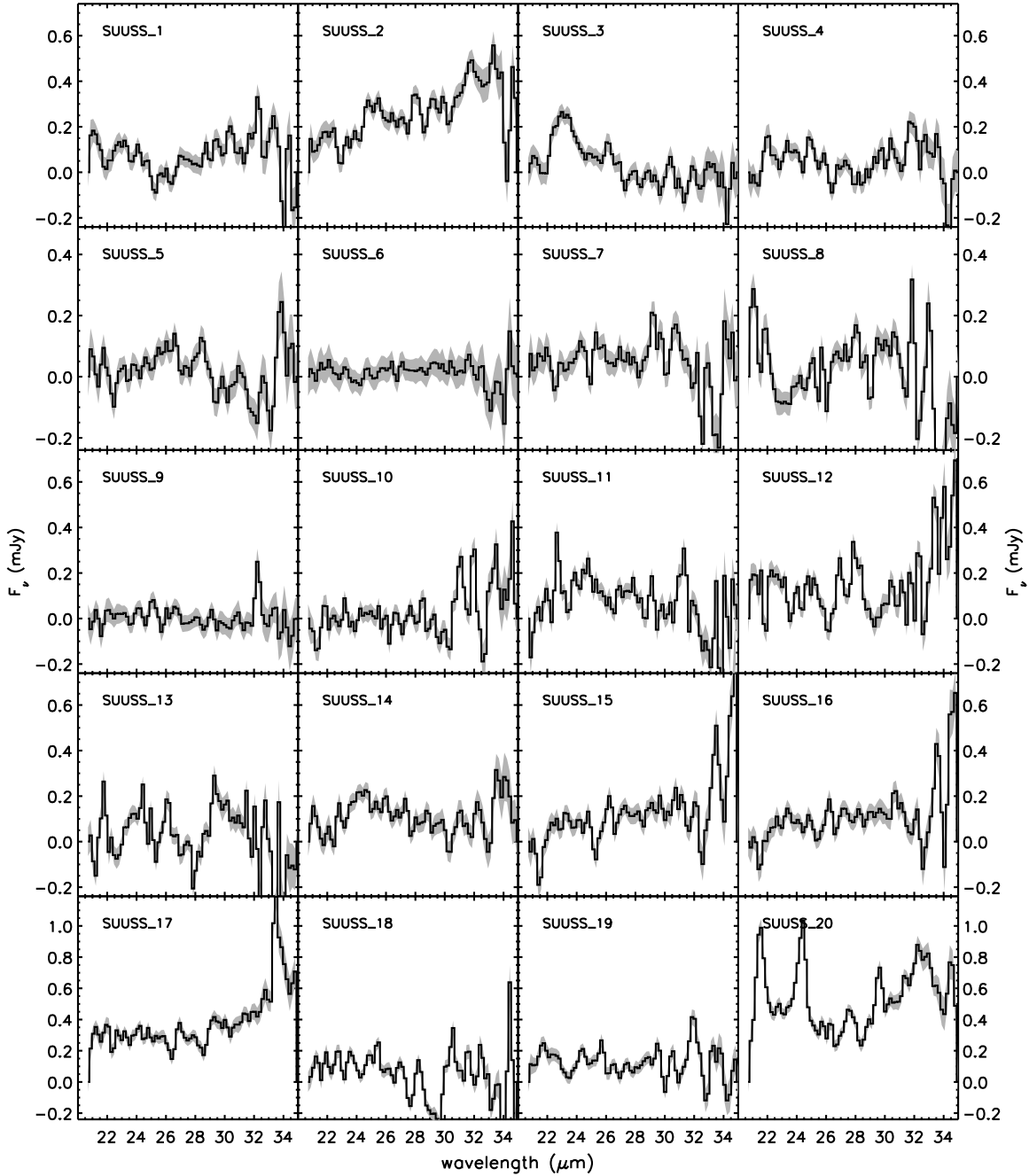}
\caption{Spectra of the 18 sources extracted from FIELD 1. In {\it shaded gray} we plot the $1\sigma$-deviation for each spectrum. Two additional source spectra were extracted from this field. They are discussed in more details in sect. 4.3.}
\label{MosaicLL1_1}
\end{figure*}

\begin{figure*}
\plotone{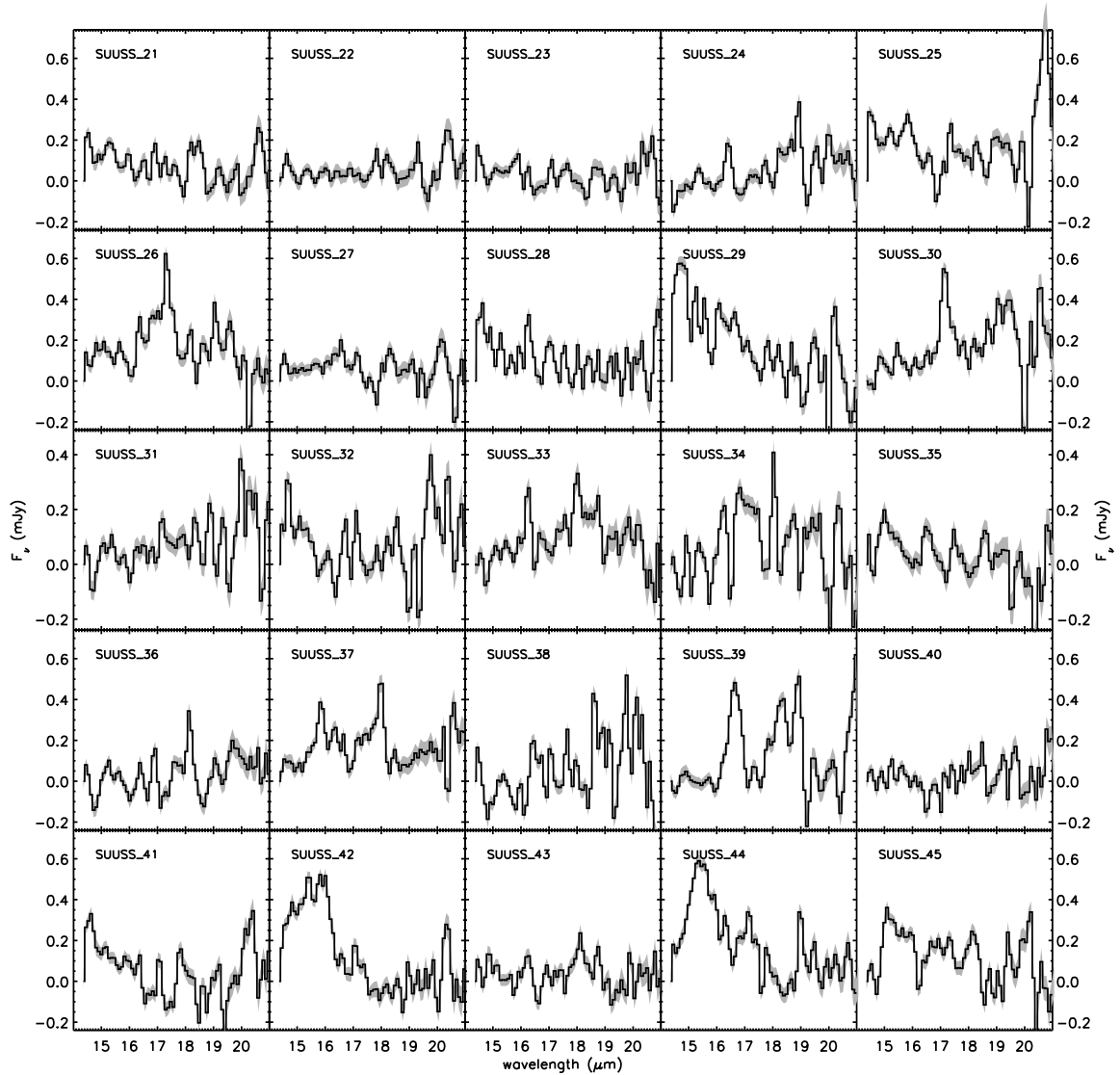}
\caption{Spectra of the 25 sources extracted from FIELD 2. In {\it shaded gray} we plot the $1\sigma$-deviation for each spectrum.}
\label{MosaicLL2_1}
\end{figure*}

\begin{figure*}
\plotone{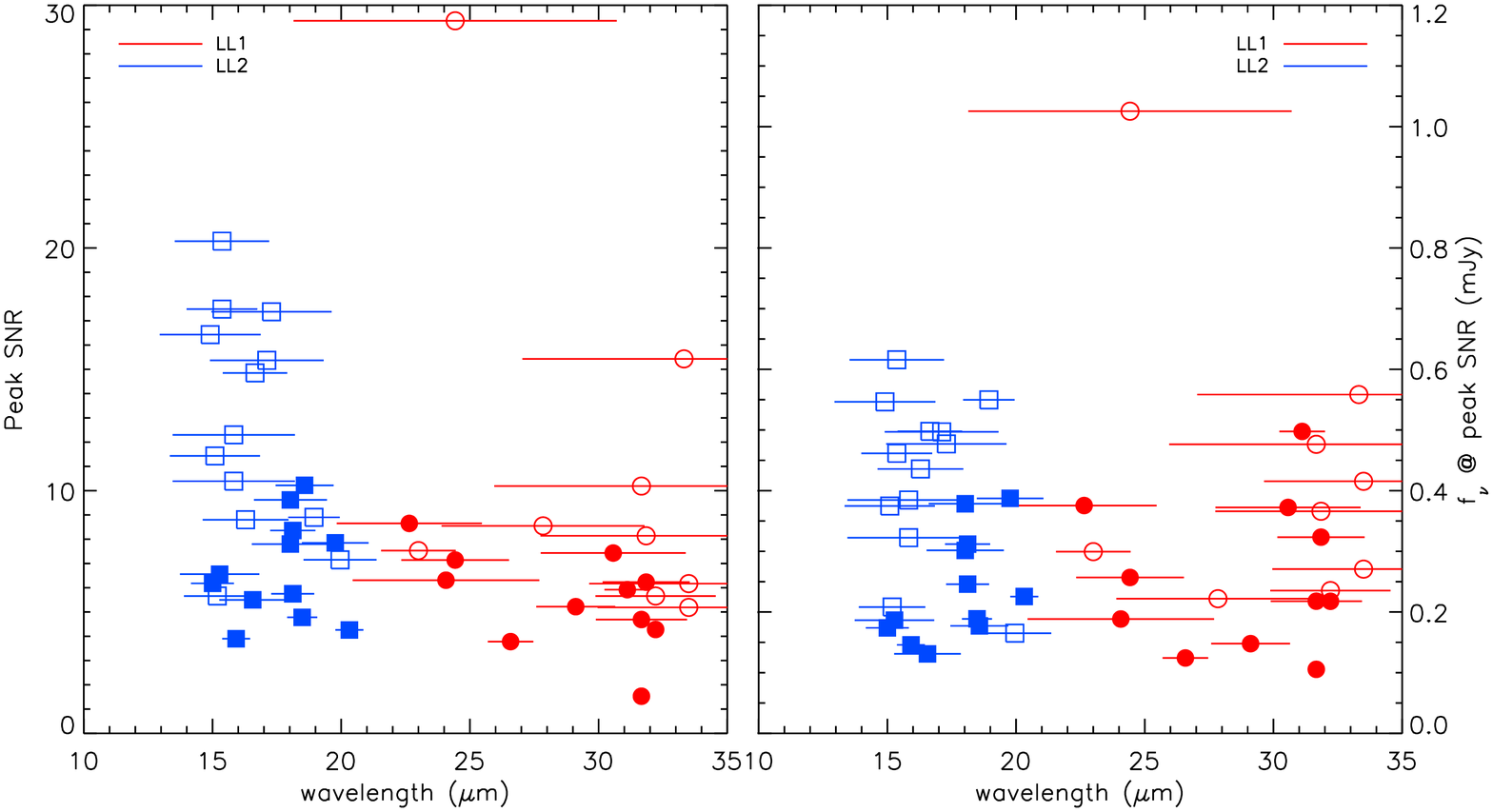}
\caption{On the {\it left}, maximum SNR achieved in each of our spectra versus the observed wavelength at which it peaks. On the {\it right}, flux density $f_\nu$ of the SUUSS sources where their spectral SNRs peak. In both plots, the bars in the wavelength direction relates to the contracted total spectral coverage over which each spectrum SNR is higher than 2. {\it Empty symbols} are sources for which MIPS $f_{\nu}(24\mu m) > 100 \mu Jy$, and {\it filled symbols} are sources for which MIPS $f_{\nu}(24\mu m) < 100 \mu Jy$ as well as sources for which we do not have a MIPS 24$\mu$m flux.}
\label{SpecSNR}
\end{figure*}

The three dimensional structure of the data was used to detect sources in both spectral and spatial dimensions. Sources were selected down to a low significance by scanning through X-$\lambda$ planes by eye and then noting their positions in X-Y planes (summed over a portion of the spectral range) as can be seen in Fig \ref{Maps}. 
The signal was then estimated in the spatial vicinity of each position and summed over the spectral range where the source signal is higher than $2\sigma$, using optimal extraction  \citep{2007ASPC..376..437N}, applying a matched filter to the data.  This estimated signal was then fitted by  a 2-dimensionnal Gaussian function in order to get the best positions and the best spectra for our sources. Finally, we estimate the integrated SNR for each optimally-extracted spectrum over the full spectral range of LL1 or LL2 and keep only sources with integrated SNR greater than 2.  We were able to extract 45 spectra, 20 in LL1 and 25 in LL2 (see Table \ref{SrcExtrTab} and Fig \ref{MosaicLL1_1} and \ref{MosaicLL2_1}).
We show in Fig \ref{SpecSNR} the Signal-to-Noise ratios at peak achieved in our spectra. We also show the contracted bandwidth in $\mu$m over which our spectra achieve a SNR greater than 2. It should be noted that in estimating signal to noise ratios, the noise values presented in the previous section 
need to be multiplied by a factor of 12 to account for the spatial integration implicit in optimal extraction (see {\it solid lines} in Fig \ref{Noise}).

\subsection{Comparison with 24$\mu$m \emph{GOODS} catalog}
\label{SectIRStoMIPS}

We correlated our sample with the MIPS 24$\mu$m catalog from the \emph{GOODS} survey \cite{2003mglh.conf..324D}. Source detection in the MIPS image uses  IRAC positions as priors, thus allowing to select sources to a lower level than expected in the presence of confusion. The final MIPS catalog is described in more detail by \cite{2007ASPC..380..375C}; it has a 1$\sigma$ depth of about 5$\mu$Jy and reaches a 84\% completeness limit of 80$\mu$Jy. We select the nearest MIPS source up to 3 arcseconds (corresponding to $\sim 2\sigma$ in IRS positioning accuracy) to be the counterpart of our IRS detections. We observe a mean difference of 1.7\" between IRS and MIPS counterpart coordinates. Results of the cross-identification (together with the IRAC catalog) are presented in Table \ref{SrcMultiwTab}. Considering both LL1 and LL2, 38 out of 45 sources (84\%) of our sample have a 24$\mu$m counterpart with a median $f_{\nu}(24\mu m)$ of $103 \mu Jy$. For LL1 only, the match rate is 90\% (18 out of 20 sources). All MIPS sources with $S_{24} > 80\mu Jy$ that fall into the region of near-full depth coverage of this survey are detected. The faintest MIPS counterpart to one of our extractions (SUUSS 18) has a 24$\mu$m flux density of 47 $\mu$Jy. This source is detected in LL1 with a maximum SNR of $\sim$7 at $30.5\mu m$.

\begin{figure}
\plotone{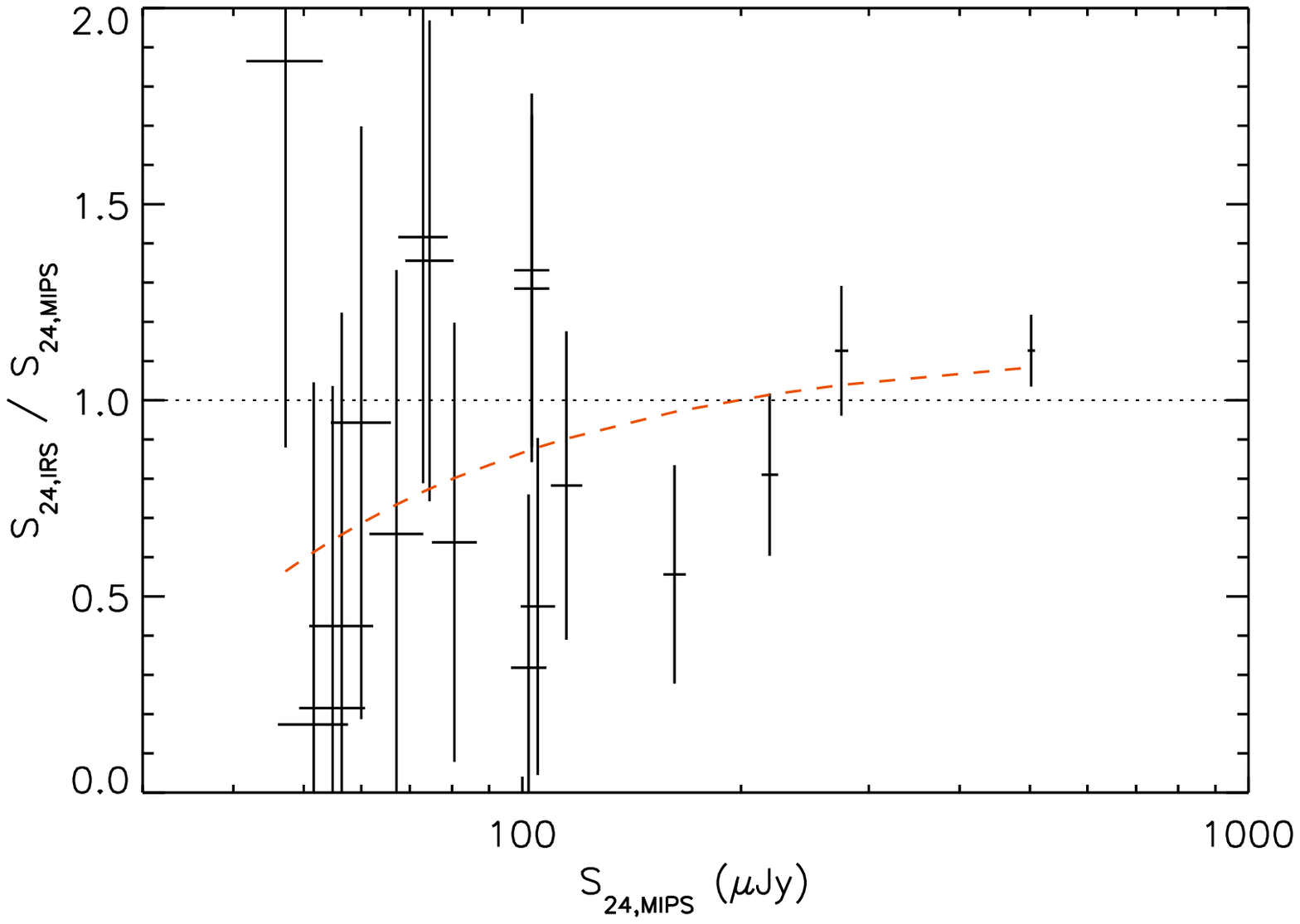}
\caption{Ratio of IRS derived 24$\mu m$ and MIPS 24$\mu m$ fluxes against MIPS $24\mu m$ flux for the SUUSS sources from Field 1 (which overlap MIPS $24\mu m$ band). The {\it dotted line} is the 1-to-1 reference. The {\it red dashed line} is the linear fit between $S_{24,IRS}$ and $S_{24,MIPS}$ presented in Sect \ref{SectIRStoMIPS}.}
\label{FluxCalib}
\end{figure}

The 24$\mu$m fluxes were used to check the flux calibration of our dataset on sources extracted from Field 1 where the spectral range overlaps the MIPS 24$\mu$m filter. 
We integrated the IRS LL1 spectra under the 24$\mu$m bandpass. Sources with SNR greater than 2 in this integration were compared to their GOODS 24$\mu$m flux counterpart in Fig \ref{FluxCalib}.   A linear regression on the data in Fig \ref{FluxCalib} yields   
\begin{eqnarray*}
S_{24, IRS} = 1.14 \left( \pm 0.09\right) \times S_{24,MIPS}  - 27.1 \left( \pm 16.2 \right)
\end{eqnarray*}
As one might expect, the main source of error in estimating the 24$\mu$m band fluxes using IRS spectra is the low SNR. We don't observe a clear bias or offset in the calibration.

\begin{figure*}
\plotone{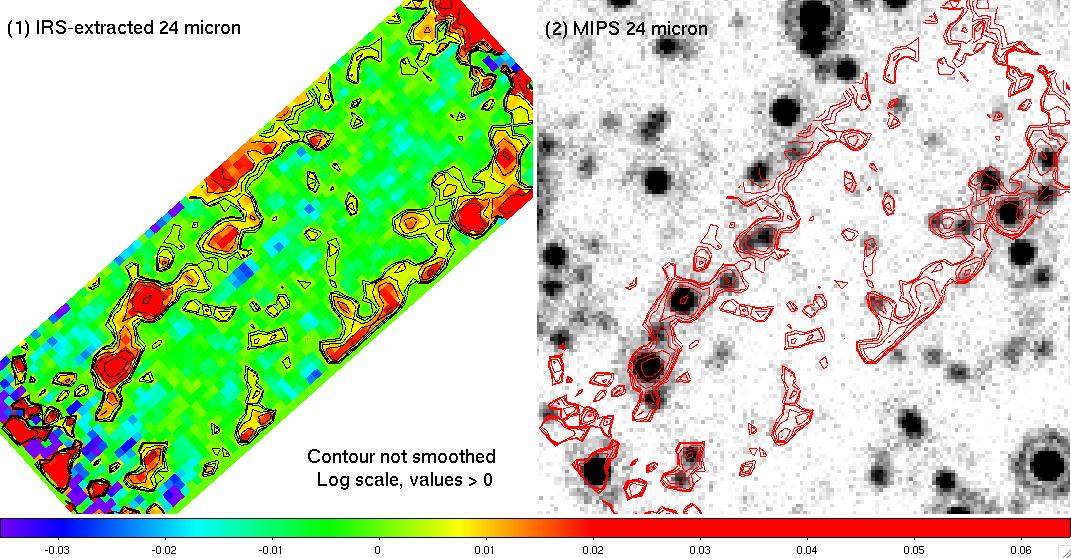}
\caption{On the \emph{left}, an IRS-computed 24$\mu$m map for Field 1. The  contour levels were computed in logarithmic scale in order to highlight faint sources. The first contours correspond to 0.3 $\sigma$. \emph{Right}, the contours extracted from the \emph{left} panel are overplotted on the \emph{GOODS-North} MIPS 24$\mu$m observations (Dickinson et al., 2003).}
\label{IRS24map}
\end{figure*}

In order to further compare our data to the photometric Mid-IR data provided by the \emph{GOODS} survey, we convolved the whole Field 1 cube by the MIPS 24$\mu$m filter. Contours were extracted of the resulting IRS 24$\mu$m map and overplotted on the released MIPS observation at the same wavelength (Fig \ref{IRS24map}). No bias or offset was observed in the comparison of source positions. We can detect a few examples of spatial confusion in the IRS map that are not present in the MIPS observation due to the better spatial resolution. These confused sources were however separated spectrally before extraction.

\section{Determining redshifts and spectral types from IRS spectra}

We have developed an original method to determine the redshift and spectral type of our sources using template spectra typical of various types of galaxies, and cross-correlation to fit the observed spectra to the templates. We describe here the method and present the results.

\begin{figure*}
\plotone{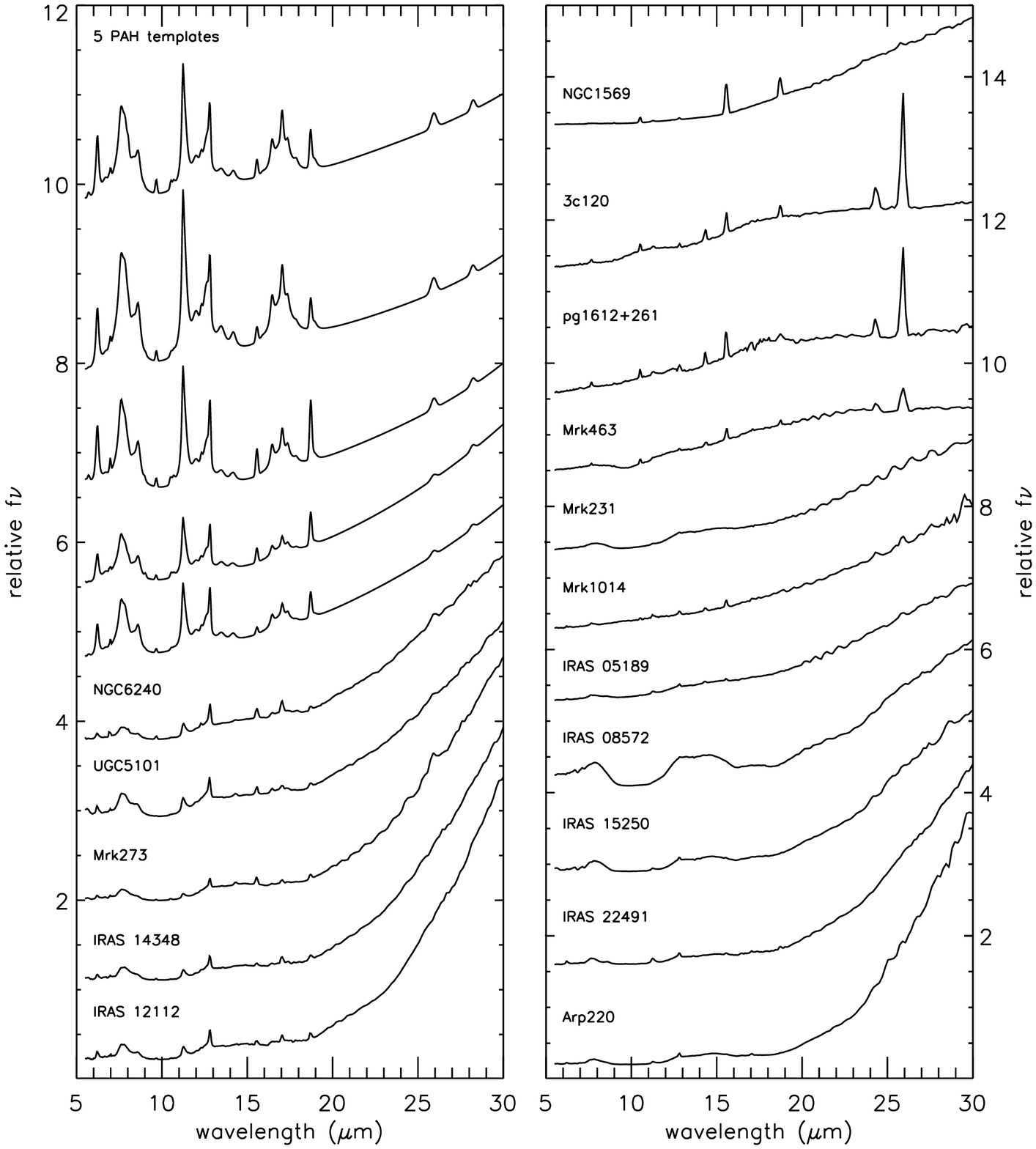}
\caption{Full IRS Low resolution spectra (SL and LL modules) of all the templates used in the Cross-correlation analysis. The 5 PAH templates are presented in Smith et al. 2007. IRAS sources (05189, 08572, 12112, 14348, 15250, 22491), Arp220, NGC6240, UGC5101 and Mrk 231, 273, 463 and 1014 are part of the IRAS Bright Sample. Their IRS Mid-IR spectra are discussed in Armus et al (2006). The 2 QSO spectra (3c120 and PG1612+261) are from P. Ogle (private communication). NGC1569 spectrum is presented in \cite{2006ApJ...639..157W}.}
\label{allTpls}
\end{figure*}

We selected a set of 21 template spectra: 5 templates dominated by Aromatic Feature emission presented in \cite{2007ApJ...656..770S}, 13 ULIRGs spectra from \cite{2007ApJ...656..148A}, two radio galaxy and quasar spectra (P. Ogle private communication) and the spectrum of the Wolf-Rayet galaxy NGC1569 \citep{2006ApJ...639..157W}.  For all those templates we have full IRS low resolution spectral coverage from 5 to 38 $\mu$m. All spectra were converted to restframe before being used. They cover a full range of source properties from star-forming  to AGN-dominated galaxies,  and include the various known mid-IR signatures: Aromatic Features, silicate absorption, both high and low ionization lines, non-thermal continuum emission, and steeply rising thermal dust emission  (see Fig \ref{allTpls} and Table \ref{TplCcorrTab}).\\

\subsection{Correlation analysis of individual sources}
The method is based on a 2 parameter (redshift and spectral template) cross-correlation to estimate IRS spectroscopic redshift and determine a best fit spectral type for our sources. For a given redshift z, we compute the Pearson product-moment correlation coefficient between data in the observed frame and each template spectrum, redshifted to z. The Pearson product is an estimate of the degree of linear relationship between two data vectors, usually noted $\rho_{S,T}$,
\begin{equation}
\rho_{S,T}(z) = \frac{Cov(S,T)}{\sigma_S . \sigma_T}
\label{eRho}
\end{equation}
where S and T are the data spectrum and the template spectrum, and $\sigma_S$ and $\sigma_T$ are the standard deviations of the source spectrum and the template spectrum in the wavelength range (LL1 or LL2) respectively.  Note that $\sigma_S$ is not the standard deviation of the noise presented in Sect. 2.3, but an estimate of the departure of each spectrum from a constant flux density (i.e. the usual definition of a standard deviation if we consider S as a random variable). Instead of a traditional cross-correlation with lag, we redshift the template by z,  compute the Pearson product, $\rho_{S,T}(z)$, for that value of z as in Eq. \ref{eRho} above, and refer to it as the cross-correlation function. We build cross-correlation functions for $0 < z < 3$ with $\Delta z = 0.01$. Such functions vary between 1 (perfect linear correlation) and -1 (perfect linear anti-correlation). A cross-correlation maximum indicates a redshift identification with high likelihood while the cross-correlation function's behaviour away from the peak reflects the shape of the template in comparison to the observed spectrum.

\subsection{Effect of Noise in the Spectra on Cross-correlation}
\label{CcorrNoise}
\begin{figure}
\plotone{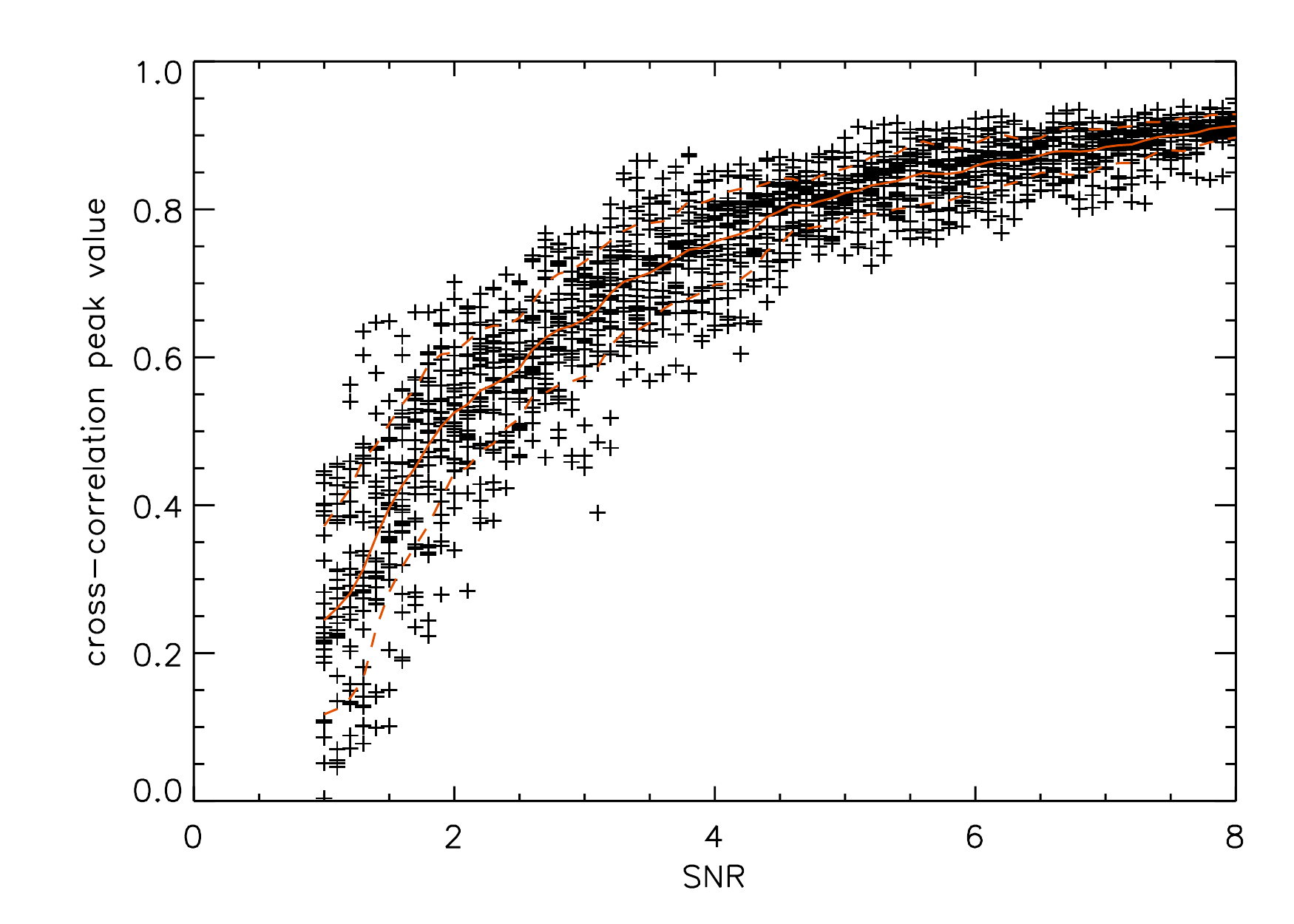}
\caption{Value of a peak in a cross-correlation function as a function of the SNR in the data. We used a PAH template from Smith et al. (2007) redshifted at $z=1$. The modified template (with noise added) was cross-correlated with the original PAH template and the value of the peak (at $z=1$ in the cross-correlation function) was tracked as a function of the noise. At each SNR, 30 realisations of the noise have been computed. The \emph{red curves} show the mean $\pm$ 1$\sigma$ of the distribution at each SNR value.}
\label{PeakToNoise}
\end{figure}

The noise in the data has an important effect on the cross-correlation. Assuming that the templates are virtually noiseless in comparison to our spectra the measured Pearson product-moment correlation coefficient is in fact
\begin{equation}
\tilde\rho_{S,T}(z) = \rho_{S,T}(z) \times \left(1 + \frac{\sigma_{N}^2}{\sigma_{S}^2}\right)^{-\frac{1}{2}}
\label{eRhoTilde}
\end{equation}
where $\sigma_{N}$ is the standard deviation of the noise in our dataset (see Fig \ref{Noise}). This equation shows that the noise will produce an overall decrease of the amplitude of the cross-correlation function. 
To quantify the effect of the noise on the cross-correlation function $\rho_{S,T}(z)$ we simulate data spectrum using the templates. The noise in our data has been characterized in Sect. 2.3 and its standard deviation, $\sigma_{N}$, extracted as a function of wavelength for both fields (LL1 and LL2) is shown in Fig \ref{Noise}. We define the SNR in our spectrum as 
\begin{equation}
SNR = \frac{< S >_{\Delta \lambda max}}{< \sigma_{N} >_{\Delta \lambda max}}
\label{eSNR}
\end{equation}
where $\Delta \lambda max$ is a small wavelength band around the maximum of the spectrum ($2\mu m$ and $1\mu m$ in LL1 and LL2 respectively).
We add noise in the template spectrum to reach a specific SNR then compute $\tilde\rho_{T(z),T}(z)$ (Template-with-noise-added vs Template) which is equivalent to an auto-correlation function. We then extract the real value of the peak of the auto-correlation and plot it against the SNR in the spectrum (see Fig \ref{PeakToNoise}). At each given SNR we make 30 realisations of the noise and average them to get the mean auto-correlation peak value. This relation, built on the template, is very helpful to estimate the SNR in our data from the value of the peak of the cross-correlation. As we can see from Eq. \ref{eRhoTilde}, $\sigma _S / \sigma_N$ controls the dynamic range of the cross-correlation function. That is equivalently a measure of the variability of the signal in the data spectrum.

\subsection{Application to this work: breaking the degeneracy}
When working with low SNR spectra over a reduced wavelength range, the cross-correlation functions often contain several peaks at different redshifts of similar amplitude. 
Since the noise reduces the amplitude of the peaks of the cross-correlation by a factor directly linked to the SNR in the data spectrum, the value of the cross-correlation alone is not enough to decide between several candidate redshifts and candidate templates. The additional information we need to decide on the best match is provided by the shape of the cross-correlation function away from the peak. 
A given template is a good match for a given data spectrum if, under the same observational constrains (SNR and spectral coverage in this case), its ``auto-correlation function'' provides a good fit to the cross-correlation function {\it over the whole range of redshifts}. 

\begin{figure*}
\plotone{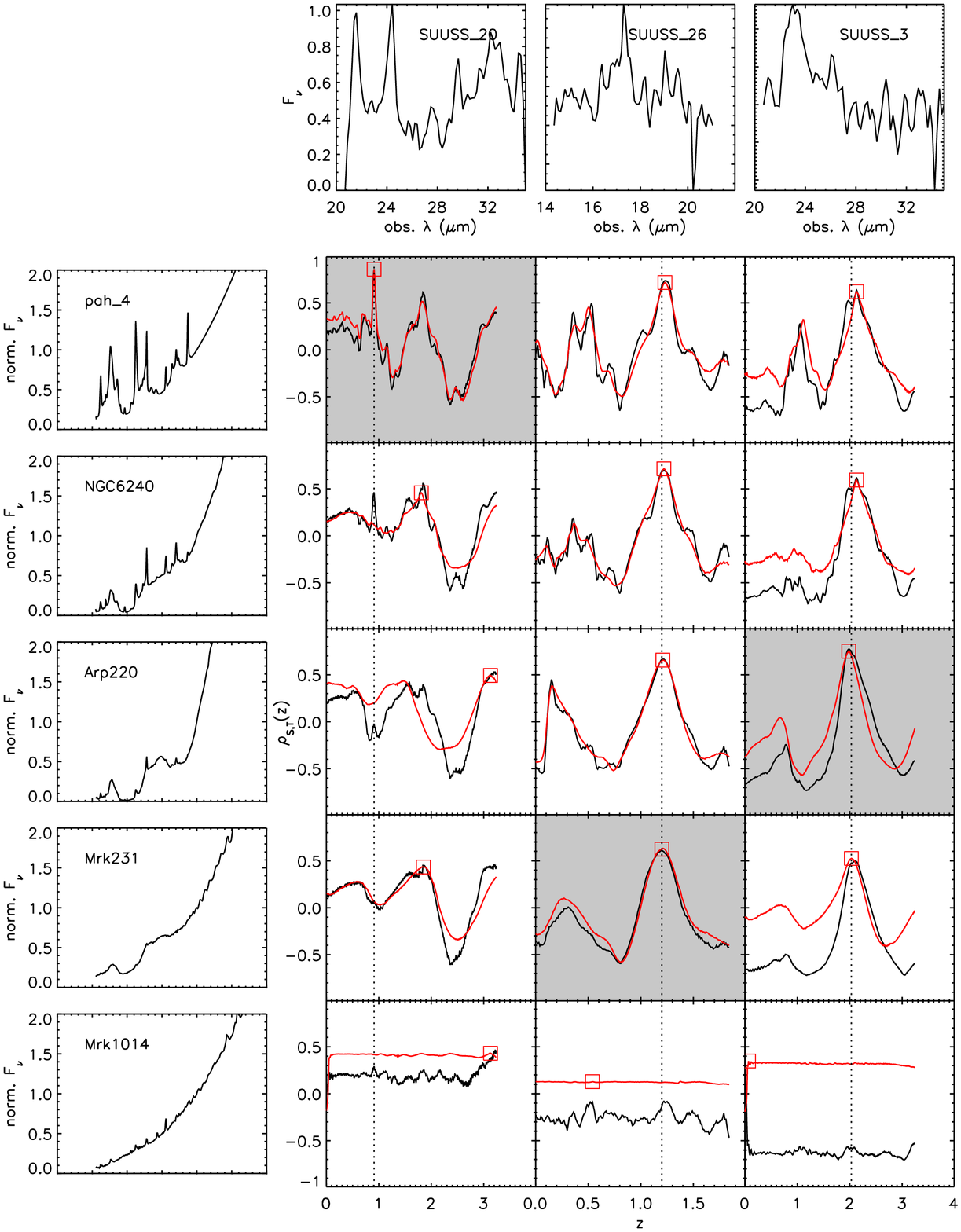}
\caption{We represent the cross-correlation between 3 different spectra and 5 different templates with very diverse mid-IR spectral signatures. In each case we show the cross-correlation function ({\it black}) and the auto-correlation function ({\it red}) for the best matching redshift ({\it red square}) with that specific template. The best candidate solution (redshift and template) for each source is shown as a {\it gray shaded} plot.}
\label{CCorrMethod}
\end{figure*}

Each maximum of the cross-correlation function yields a candidate solution consisting of a redshift-template combination. We use the template and redshift to simulate the observed spectrum and compute $\tilde\rho_{\tilde T(z_{max}),T}(z)$, that we call an ``auto-correlation function'' as above:
\begin{equation}
\rho_{\tilde T(z_{max}),T}(z) = \frac{Cov(\tilde T(z_{max}),T)}{\sigma_{\tilde T} . \sigma_T}
\label{eRhoTT}
\end{equation}
We use Fig \ref{PeakToNoise} to pick the amount of noise to add to the template to reach the SNR corresponding to the cross-correlation amplitude of the candidate solution. The comparison between the cross-correlation and the ``auto-correlation'' tells us whether the choice of z and template recover the shape of the correlation at lags other than zero, or equivalently redshifts other than the best-fit z.
We compute the chi-square between $\rho_{\tilde T(zmax),T}(z)$ and $\tilde \rho_{S,T}(z)$ for each $(z,T)$ candidate-solution. The minimum chi-square, finally, gives us the best matching result which we call the IRS redshift $z_{IRS}$. We illustrate in Fig \ref{CCorrMethod} the two-parameter nature of the fitting, which looks simultaneously for the best-fit redshift and best-fit template. This method was automatically applied to all extracted SUUSS spectra, without priors.

\section{Results}

\subsection{Redshift Distribution}

The previous method is successful in determining the redshift for 28 sources (62\% of the sample). For the remaining sources, the lack of redshift determination is mainly due to low SNR and/or the absence of any Mid-IR feature in our spectral window. Results are presented in Table \ref{SrcCcorrTab}. We note that we have a comparable rate of successful redshift determination in Field 2  at 14-21$\mu$m (17/25 sources or 68\%) and in Field 1 at 20-35$\mu$m (11/20 sources or 55\%). We find a median redshift of 1.05 slightly larger than the median redshift of 0.935 found for the 24$\mu$m flux-limited sample ($S_{24} > 80 \mu Jy$) by \cite{2007ApJ...660...97C}. Interestingly, mean redshifts for each field are quite different and are equal to 1.25 and 0.92 in Field 1 (LL1) and Field 2 (LL2) respectively. We identify more higher redshifts sources in Field 1 (4 / 20 sources with $z > 2$) than in Field 2 (1 / 25 sources with $z > 2$) which is probably due to the different rest-frame wavelength ranges observed by LL1 ($6.7$ to $11.3\mu m$ at $z = 2$) and LL2 ($4.7$ to $6.7\mu m$ at $z = 2$). The most distant galaxy is at $z\sim 2.2$. 

\begin{figure*}
\begin{center}
\plottwo{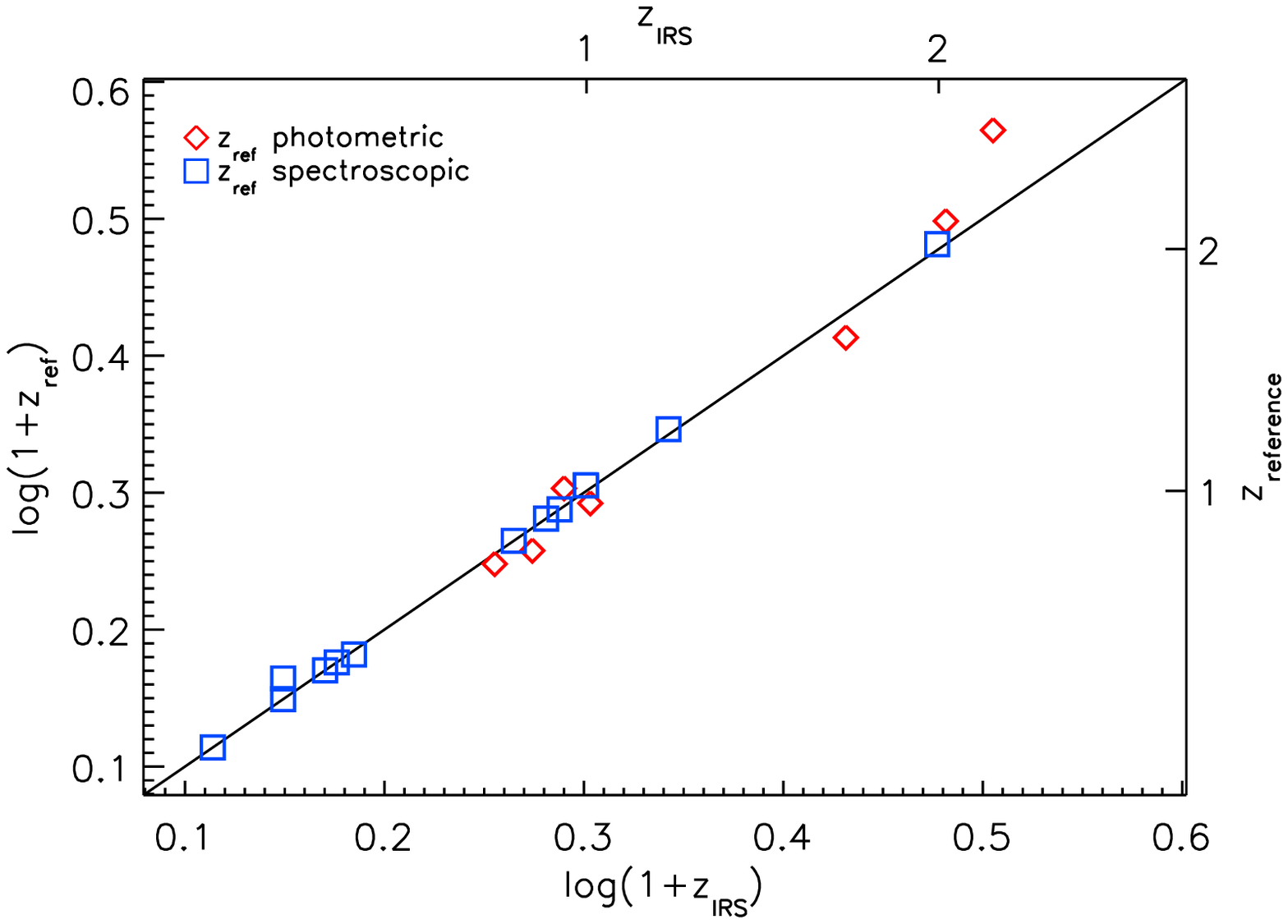}{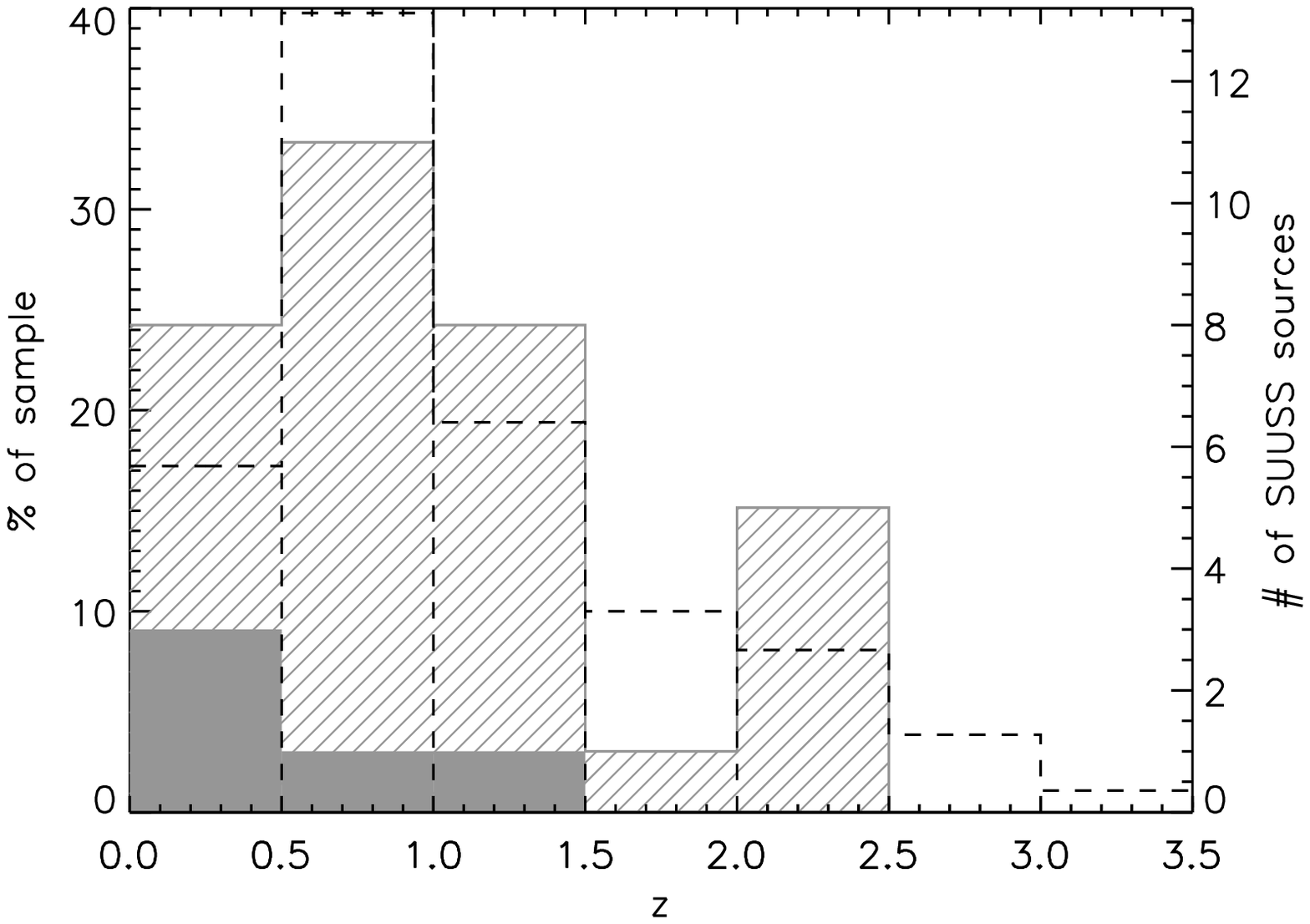}
\caption{\emph{Left}, IRS spectroscopic redshifts ($z_{IRS}$) from our study against existing spectroscopic (TKRS, Wirth et al., 2004) and photometric (Caputi et al., 2007) redshifts ($z_{ref}$). \emph{Right}, redshift distributions. The \emph{dashed-gray} histogram show the total redshift distribution for our sources ($z_{IRS}$ supplemented with $z_{ref}$ when $z_{irs}$ couldn't be obtained). The \emph{shaded} histogram shows the sources for which we had only $z_{ref}$. The \emph{dashed open} histogram is Caputi et al. (2006) sample distribution.}
\label{ZZHisto}
\end{center}
\end{figure*}

We compare the IRS spectroscopic redshift of our sample with the optical spectroscopic redshifts from \cite{2004AJ....127.3121W} (13 sources, see Table \ref{SrcCcorrTab} and Fig \ref{ZZHisto}). The mean of $\Delta z / (1+z)$ is equal to $4.3 \times 10^{-3}$ and its standard deviation is of $1.0\times 10^{-2}$. This shows that, even with very noisy IRS spectra, our method is able to determine mid-IR spectroscopic redshifts as accurate as 1\% of $(1+z)$. Where an optical spectroscopic redshift is not available, and the MIPS flux density is higher than $83\mu Jy$, we use photometric redshift from \cite{2007ApJ...660...97C}. The seven such sources in our sample yield a $<\Delta z / (1+z)>$ of $1.5\times 10^{-2}$ with a standard deviation of $6.6\times 10^{-2}$. This dispersion is higher than expected from the photometric redshift characterization in \cite{2007ApJ...660...97C}.

Of the 17 sources without an IRS redshift 5 have an optical spectroscopic redshift and 12 have no redshift at all.  As stated in Sect. 2.5, we do detect all sources with MIPS $S_{24} > 80 \mu Jy$ in our sample thus reaching comparable depth as the sample used in \cite{2007ApJ...660...97C} which was selected to have only sources with $S_{24} > 80 \mu Jy$. Figure \ref{ZZHisto} shows our redshift distribution and we compare it to that of \cite{2007ApJ...660...97C}.

\subsection{Spectral Type}

As presented above, the cross-correlation method provides spectroscopic redshifts along with an indication of the most similar  spectral type.  We have classified the sources in our sample by the  best-fit template found for each.  We divide the sources among four roughly defined populations: those distinguished by prominent aromatic emission features(14/45 sources), those characterized only by the presence of silicate absorption (4/45 sources), an intermediate population gathering mixed signatures from weak PAH and/or silicate absorption (8/45 sources),  and those for which we could not find a redshift, and thus a spectral type (17/45 sources) either because of very low SNR or because no features fall into the spectral band covered by this survey (14-21$\mu$m or 20-35$\mu$m). An absence of features in the spectrum can result from a high redshift ($z> 3.5$ in LL1 or $z>2$ in LL2) or the intrinsic properties of a continuum dominated source usually associated with a dominant AGN. Two sources for which we have been able to get a redshift are not part of any of the subsets described above. We will discuss these two in greater detail in the next section. It has been previously shown that H$_2$ lines can sometimes be detected even in low-resolution IRS spectroscopy (e.g. \citealp{2006ApJ...640..204A}, \citealp{2009arXiv0906.5271D}). We do not however detect any reliable H$_2$ emission in our spectra. We show the complete spectral type statistics of the SUUSS sample for each field in Table \ref{SrcStatTab}.

A majority of the identified matches (14/28 sources) in our sample present some PAH emission features. The spectral ranges available can not cover all the PAH bands at the same time, we refer to it as partial PAH emission.
In spectra with limited mid-IR spectral coverage (in this case 14-21$\mu m$ or 20-35$\mu m$), there is a degeneracy between a low SNR $7.7\mu m$ PAH feature on the one hand and silicate absorption that creates a false ``continuum bump'' around $8\mu m$ in the other. The easiest way to overcome this would be to ascertain the existence or absence of another PAH feature (e.g. 6.2 or 11.3$\mu m$) not affected by the presence of SiO absorption. This is however almost impossible here due to the redshifts of our sources that tends to throw $6.2$ or $11.3\mu m$ rest frame wavelength out of the observed spectral window.

\begin{figure*}
\plotone{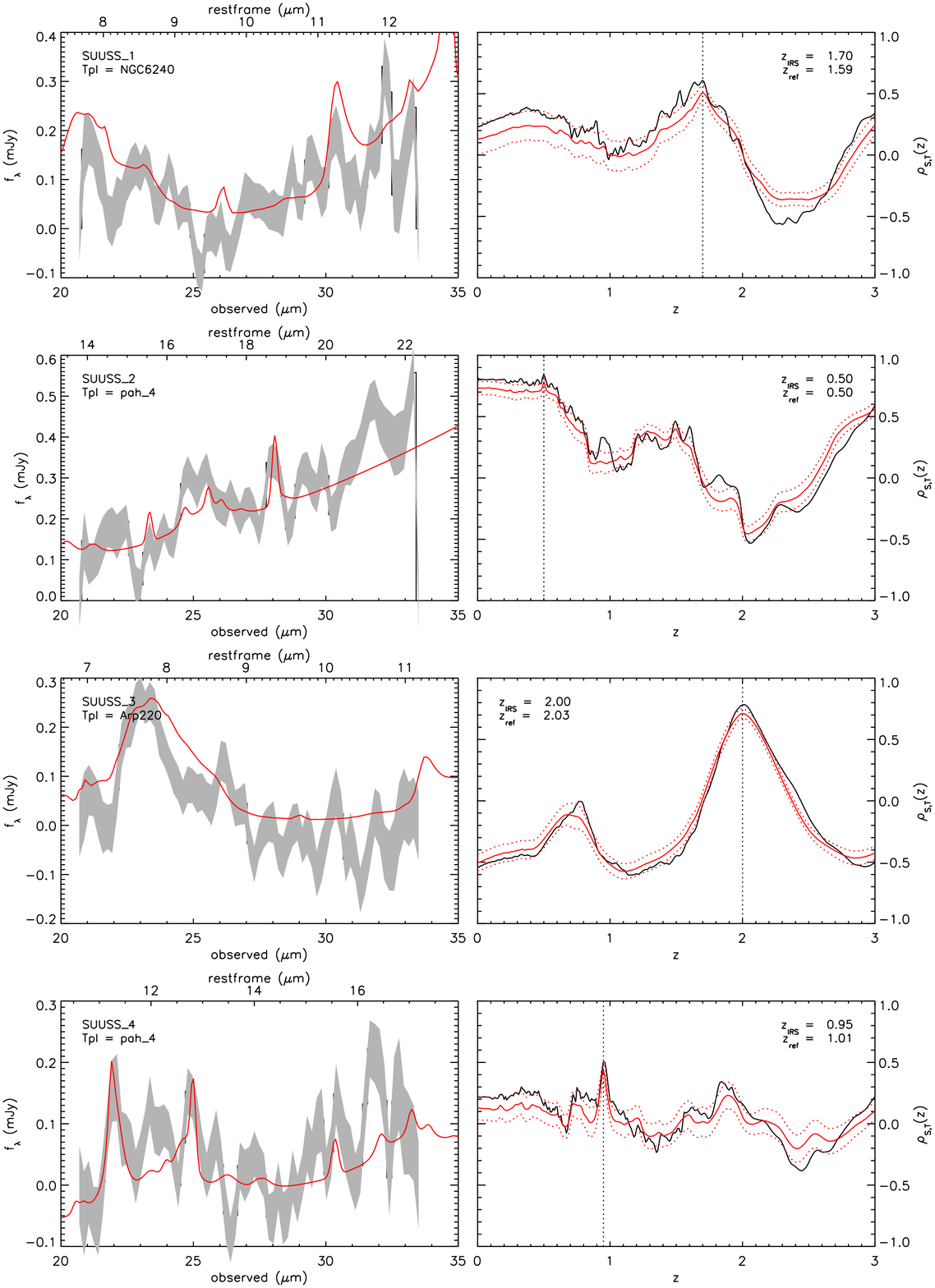}
\caption{On the \emph{left}, our sources spectra (\emph{black}) with best match template(\emph{red}) overplotted. The template is scaled to the mean flux of the spectrum. On the \emph{right}, the related cross-correlation function (in \emph{black}) and auto-correlation function (in \emph{red}). {\it Results for the first 4 sources are shown here as an example, plots for the whole sample can be found at the end of this document.}}
\label{ZspecMosaic0}
\end{figure*}

Figure \ref{ZspecMosaic0} shows spectra in restframe (using $z_{IRS}$ only) together with the best template. Templates have been normalized to the mean fluxes of our sources computed over the whole available spectral range. Provided a spectrum with high enough SNR (empirically $> 3$) and enough dynamic range (measured as $\frac{\sigma_S}{\sigma_N} > 1$, cf Eq. 2 and sect. \ref{CcorrNoise}), the effectiveness of the cross-correlation method will only be limited by the diversity of mid-IR SED shapes covered by our templates. These conditions lead to very good agreement as seen, for example, for sources SUUSS 20 or SUUSS 3 in Fig \ref{ZspecMosaic0}. Sharp cross-correlation peaks at the good redshift denote higher SED-feature ``frequencies'' (e.g. SUUSS 20, 30 or 25) such as $11.3\mu$m and $12.7\mu$m PAHs and/or atomic forbidden lines (e.g. [NeIII], [OIV]). On the other hand, wider cross-correlation maxima (usually ranging from high anti-correlation to high correlation) reveal a more diffuse identification based on the global shape of the spectrum (``lower signal frequencies'') that we can see with low SNR 7.7$\mu$m emission and SiO absorption (e.g. SUUSS 3, 41 or 42).

Analysing results from Table \ref{SrcCcorrTab} and Table \ref{SrcStatTab}, we see that PAH templates from \cite{2007ApJ...656..770S} provide most of the identifications (14 sources). Those templates show PAH emission (6.2, 7.7 and/or 8.6$\mu$m features, 12.8$\mu$m emission line, etc ), a ``higher-frequency'' signal (compared to continuum and silicate absorption) that provide sharp cross-correlation peaks and thus more accurate and reliable redshift identification. A few other templates also match PAH emission features such as NGC6240 (2 sources) or UGC5101 (1 source) with differences being mainly on the continuum and the $11.3 / 7.7$ ratio. Other identifications mostly rely on broad spectral behavior and thus preferentially match templates such as Arp 220, Mrk231, Mrk463 or IRAS 15250 (6 sources total). A few sources remain with in-between signal properties such as Mrk273 (2 sources). Overall, the cross-correlation analysis provides strong characterization of mid-infrared emission features in our sources, that we propose to use as a first order diagnostic to distinguish between starburst and AGN as power agent for these sources.
For the featureless spectra the cross-correlation method does not even yield a slope, and therefore generates limited information about the nature of the sources. We computed restframe composites for each group of sources with a redshift and similar spectral type. The increased SNR and spectral coverage are expected to allow us to determine additional global properties for those populations (cf Sect \ref{SectLir}). 

No spectral type information could be extracted for the 5 sources for which we have an optical spectroscopic redshift but no IRS redshift. At least 2 of those have a mid-IR spectrum ``polluted'' by a spatially close and brighter source also present in our sample. While we have been able to separate spectra of confused sources in the previously described redshift identifications, in these two cases the ''polluting'' source was too strong in comparison and contributions could not be spatially or spectrally separated. For one of them (SUUSS 19) there is a hint of PAH emission (11.3$\mu$m and 12.7$\mu$m) in agreement with the existing optical spectroscopic redshift of the counterpart of $z = 1.013$.

\subsection{Two special line-sources: [S\,\textsc{iv}] and [O\,\textsc{iv}] lines}
\label{SectLineSrc}

Among the detected sources, two (SUUSS 9 and 17) have only one prominent emission line each. In both cases, the redshift determination method had to be applied more carefully. 

\begin{figure}
\plotone{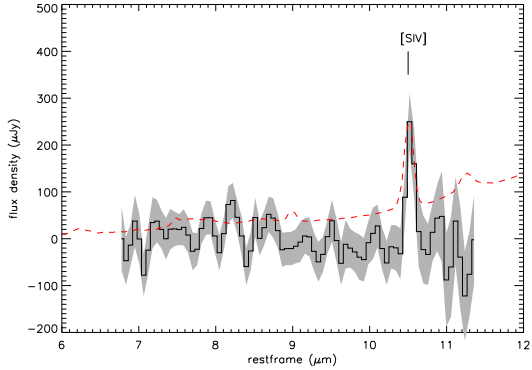}
\caption{Spectrum of the [SIV] ``line-source'' (SUUSS 9) in restframe. 1$\sigma$-deviation is shown as the {\it gray shading}. The cross-correlation gave us a redshift of $z=2.08$ for this source. In {\it dashed red} we show the spectrum of NGC1569 for which we obtained the best cross-correlation match. It is scaled to our source $10.5\mu m$ restframe flux density.}
\label{SIVsrc}
\end{figure}

We observe SUUSS 9 at position 189.210 +62.2869 with a spectrum showing only one emission line at 32.40$\mu m$ (see Fig \ref{SIVsrc}) with a total flux of  $\sim 8\times 10^{-19}\ \rm{W\ m^{-2}}$.  A spurious origin for this feature was ruled out by several means, one of which was to build 8 separate cubes (one from each AOR). The feature was found in 5 out of the 8 cubes, which is consistent with the lower SNR when using only one eighth of the data. The reliability of this source was further verified by inspection of its spatial profile. We integrated fluxes in a Hanning-window centered on $32.4\mu m$ and extracted the source PSF which matches that of brighter sources in our sample.
We address the problem of assigning a redshift based on a single line by relying on our knowledge of typical mid-IR spectra of known sources, and by using surveys at other wavelengths in the same part of the sky for additional evidence. There are three GOODS sources \citep{2007ASPC..380..375C} sufficiently close to be considered as potential counterparts, as reported in NED (NASA/IPAC Extragalactic Database, December 2008):  GOODS J123650.87+621712.8 is 3.4" away from our estimated position of SUSS9, and at a spectroscopic redshift of 2.133; GOODS J123650.22+621718.4 is 5.5" away from SUSS9, and at $z=0.51283$; GOODS J123649.44+621712.3 is 6.7" away from SUSS9, and at a photometric redshift of 0.04.

\cite{2007ApJ...656..770S} and \cite{2006ApJ...646..161D} have reported the detailed spectroscopic content of SINGS galaxies; similarly, \cite{2007ApJ...656..148A} reported on nearby ULIRGs.  We will examine which of the most prominent lines seen in those systems might be associated with this one line, in view of the mid-IR spectral  properties of those samples, and of the visible sources positionally associated with SUUSS9.
We should clarify however that mid-IR spectra are still yielding many surprises, as in the case of radio galaxies (Ogle et al 2009, {\it in prep}), so that plausibility arguments presented here cannot yet be quantified as to their likelihood.

Molecular hydrogen is known to dominate the mid-IR spectra of certain objects, as in the case of Stephan's Quintet intergalactic shock \citep{2006ApJ...639L..51A}, or the Radio Galaxy system 3C\,326 \citep{2007ApJ...668..699O}. One possibility is that this could be the H$_2$ S(1) line at $z=0.902$; one would then expect the S(2) line at 23.35$\mu$m and the [Ne\,{\sc ii}] line at 24.36$\mu$m, neither one of which is detected.  In Stephan's Quintet, S(1) is the most luminous line, and each of [Ne\,{\sc ii}] and S(2) carries one third or more of the luminosity in S(1), so their absence in SUUSS 9 is a significant argument against this possibility, as is the absence of any nearby visible sources at redshifts close to 0.9. Similarly, this could be the H$_2$ S(0) 28$\mu$m line at $z=0.148$, with the S(1) line expected at 19.55$\mu$m, outside the survey spectral coverage. The lack of visible sources in the redshift vicinity of 0.148 argues against this possibility.  Finally, if we assume this is the H$_2$ S(2) line at $z=1.639$, then the S(3) would be expected at 25.5$\mu$m, and it is not detected. The weak signal peaking around 16.97$\mu$m (restframe) is significantly removed from the expected wavelength. This non-detection is a strong argument given the typical weakness of the S(2) line, so we conclude this is an unlikely assignment.

The most prominent fine-structure lines that might be associated with the detection at 32.40$\mu m$ are:  [\rm{S\,\textsc{iv}}] at $z=2.083$, [\rm{Ne\,\textsc{ii}}] at $z=1.529$, [\rm{Ne\,\textsc{v}}] at $z=1.262$, [\rm{Ne\,\textsc{iii}}] at $z=1.083$, [\rm{S\,\textsc{iii}}] at $z=0.731$, [\rm{O\,\textsc{iv}}] at $z=0.251$, or [\rm{Fe\,\textsc{ii}}] at $z=0.247$. Among all these possibilities, we favor the [\rm{S\,\textsc{iv}}] assignment, for two main reasons, one being the closest of the visible sources has a very similar redshift at $z=2.133$, and the other being the existence of a similar mid-IR spectrum from the local Universe, namely NGC1569 \citep{2006ApJ...639..157W}.  The other potential assignments are less likely  because the observed spectrum does not contain expected lines or Aromatic features, and none correspond to any of the redshifts of nearby sources. For instance, [\rm{Ne\,\textsc{ii}}]  is quite unlikely to appear without Aromatic features, and we would have detected the 11.3$\mu$m band at 28.6$\mu$m for $z=1.529$. However, the radio galaxy 3C317 does display a high ratio of [\rm{Ne\,\textsc{ii}}] to Aromatic Features (Ogle et al 2009, {\it in prep}).  Similarly, [\rm{Ne\,\textsc{v}}] is generally much weaker than  [\rm{Ne\,\textsc{ii}}], and the latter would have been detected at 26.7$\mu$m for $z=1.083$. [\rm{Ne\,\textsc{iii}}] on the other hand can be much more luminous than [\rm{Ne\,\textsc{ii}}], thus appearing as the only line in a spectrum, but only in a small minority of cases.  Dale et al (2009) find that the [\rm{Ne\,\textsc{iii}}] flux exceeds twice the [\rm{Ne\,\textsc{ii}}] flux in less than 10\% of the cases for a variety of SINGS objects.
While it is possible that the line we detect is [\rm{Ne\,\textsc{iii}}], and that [\rm{Ne\,\textsc{ii}}] has escaped detection at 28.99$\mu$m for $z=1.262$, this redshift does not correspond to any objects in the field. The same argument applies to [\rm{S\,\textsc{iii}}], since [\rm{Ne\,\textsc{ii}}] would have been detected at 22.19$\mu$m for $z=0.731$, and it is rare for [\rm{S\,\textsc{iii}}] to exceed twice the flux of [\rm{Ne\,\textsc{ii}}]. The lines of [\rm{O\,\textsc{iv}}] or [\rm{Fe\,\textsc{ii}}] near $z=0.25$ are again unlikely candidates, since the [\rm{S\,\textsc{iii}}] line is expected to be at least comparable in flux and detectable at 23.4$\mu$m.  The exception would be that [\rm{O\,\textsc{iv}}] might exceed [\rm{S\,\textsc{iii}}] in AGN-dominated sources.  However, the lack of a  detectable continuum and the lack of a similar redshift in this vicinity are both arguments against this identification.  

The close association of SUUSS9 with GOODS J123650.87+621712.8 both spatially and in redshift is suggestive of these being the same object, but not conclusive. It is just as likely that they are two neighboring sources rather than the same source. The redshift difference amounts to $\sim$ 15,000 km/s, and corresponds to 1.7$\sigma$ assuming the uncertainty on the \cite{2006ApJ...653.1004R} redshift is sufficiently lower than ours. Moreover, interpreting SUUSS9 as a NGC 1569 analog implies that it is a Wolf-Rayet galaxy, whereas the GOODS source does not display the Wolf-Rayet characteristics in its optical spectrum \citep{2008A&A...485..657B}.  Additional data will be required to clarify the relation between these two sources, which is beyond the scope of this paper. 
While we cannot attach a formal confidence level to this statement, we do propose that the interpretation of the SUUSS 9 line  as [S\,{\sc iv}] at $z\sim 2.1$ is the most plausible postulate, followed by the less likely interpretation as the H$_2$ S(0) line at $z=0.15$.
If SUUSS 9 is indeed detected in [S\,{\sc iv}], then its [S\,{\sc iv}] line luminosity would be $8.3 \times 10^{35}$ W, whereas the dwarf galaxy NGC1569 emits $1.5 \times 10^{31}$ W. While this is a dramatic difference, the [S{\sc iv}] luminosity estimated for SUUSS9 is not extraordinary, in the sense that the quasar PG1612+261 has a comparable luminosity of $1.4 \times 10^{35}$ W.   The interpretation of the SUUSS 9 line as H$_2$ S(0) at $z=0.15$ is less demanding energetically, since it requires scaling up the emission of the intergalactic shock in Stephan's Quintet by a factor 11 only, to a line luminosity of $1.5\ .10^{33}$ W.

\begin{figure}
\plotone{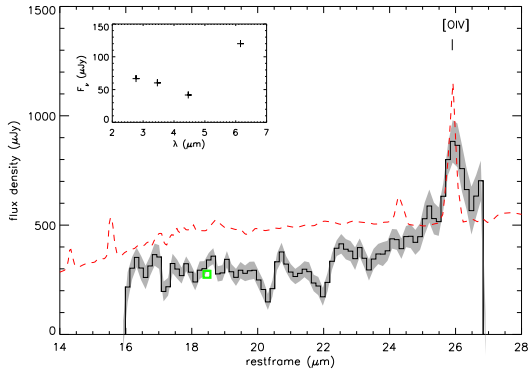} 
\caption{Spectrum of the [OIV] ``line-source'' (SUUSS 17) in restframe. 1$\sigma$-deviation is shown as the {\it gray shading}. The {\it green square} is the MIPS 24$\mu$m counterpart flux in restframe. The cross-correlation, in agreement with ancillary data, gave us a redshift of $z=0.3$ for this source. IRAC data points ({\it inset}) tend to agree with the presence of PAH emission at 6.2 $\mu$m. In {\it dashed red} we plotted the spectrum of the QSO PG1612+261 for which we obtained the best cross-correlation match. It is scaled to our source $25.9\mu m$ restframe flux density.}
\label{OIVsrc}
\end{figure}

The second line-sources (SUUSS 17) possesses a relatively well determined mid-IR continuum of about 300$\mu$Jy ($> 6 \sigma$) as can be seen in Fig \ref{OIVsrc}. This source shows one very well defined emission line at $\lambda \sim 33.7\mu m$ (observed wavelength) with a total flux of $\sim 6\times 10^{-19}\ \rm{W\ m^{-2}}$, and no other significant spectral feature over the LL1 wavelength range (20 to 35$\mu$m). The two absorption-like features at $26$ and $28.5\mu$m in the spectrum are unlikely to be real and are more likely originating from a combined effect of noise and confusion as this source sit in a denser region of our survey with 3 surrounding sources as close as 15''.
The line is well resolved by the IRS, with an estimated intrinsic full width at half max equivalent to $\sim 4000 \pm 1400\ \rm{km/s}$. NED, consulted in December 2008, reports three objects sufficiently close to be considered potential counterparts: 
GOODS J123658.45+621637.3 at $z=0.2993$, and 1.5" away,
GOODS J123658.82+621638.1 at $z=0.29863$, and 3.2" away,
and GOODS J123658.09+621639.4 at $z=1.01734$ and 6.2" away;
all redshifts are from \cite{2004AJ....127.3121W}. The second of these is assigned a Spitzer 24$\mu m$ detection at 264$\mu Jy$, making it the most credible counterpart to SUUSS 17.  

The cross-correlation analysis on this source yields several candidate solutions for different templates and does not allow us to secure a reliable redshift. However, using the $z=0.3$ redshift as prior we find the best matching template to be the spectrum of pg1612+261 (QSO, \citeauthor{2007ApJ...669..841S} \citeyear{2007ApJ...669..841S}) with an IRS determined redshift of 0.3. This redshift implies that the strong emission line in the IRS (LL1) spectrum is either [\rm{O\,\textsc{iv}}] (25.89$\mu m$) or [\rm{Fe\,\textsc{ii}}] (25.99$\mu m$). Moreover, at $z=0.3$, the IRAC 8$\mu m$ band falls on top of the 6.2$\mu m$ feature and would be expected to cause an excess compared to the shorter wavelength IRAC bands. Such an excess is evident in the inset in Fig \ref{OIVsrc}, and we tentatively interpret it as the presence of PAH emission.  The difficulty with this interpretation however is that [\rm{O\,\textsc{iv}}] is observed with  $<1000\ \rm{km/s}$ width in known QSOs (e.g. \citeauthor{2008ApJ...674L...9D} \citeyear{2008ApJ...674L...9D}), and the line width observed for SUUSS 17, typical of broad-line region emission, is not expected for high-ionization species like [\rm{O\,\textsc{iv}}].  It is also difficult to explain the large observed line width as resulting from the combination of [\rm{Fe\,\textsc{ii}}] and [\rm{O\,\textsc{iv}}] emission, for the separation between the two lines is only 0.1$\mu m$, or  900 km/s.  One might invoke the possibility of a broadened [\rm{Fe\,\textsc{ii}}] rather than the [\rm{O\,\textsc{iv}}], but one would then expect the 18.7$\mu m$ line of [\rm{S\,\textsc{iii}}] to be at least comparable, and typically several times stronger than [\rm{Fe\,\textsc{ii}}] (Dale et al 2009, in press). There is no evidence of the [\rm{S\,\textsc{iii}}] line in Fig \ref{OIVsrc}, which argues against this possibility. The 17.94$\mu m$ line of [\rm{Fe\,\textsc{ii}}] is typically weaker than the 25.99$\mu m$ line, so we would not expect it to appear. Another strong argument against SUUSS 17 displaying a broadened [\rm{Fe\,\textsc{ii}}] line is that this this line is rarely seen as the dominant emission line from QSOs (e.g. \citeauthor{2008ApJ...674L...9D} \citeyear{2008ApJ...674L...9D}), and the lack of a strong [\rm{S\,\textsc{iii}}] line is more consistent with the observed line being [\rm{O\,\textsc{iv}}] (e.g. \citeauthor{2006ApJ...647..161O} \citeyear{2006ApJ...647..161O}).

Another possibility is that the 33.7$\mu m$ emission is an Aromatic feature rather than a broadened line.  We find this improbable, since other PAH features would be expected in the same spectral range: for example if we assume this emission feature is the 11.3$\mu m$ PAH (at $z=1.98$) then emission at 7.7$\mu m$ restframe should be observed at  $\lambda = 23 \mu m$.  While the overwhelming majority of sources follow this expectation, Leipski et al (2009, submitted) show a clear counter-example in M84 (3C272.1), a FR-I radio galaxy with an IRS spectrum displaying the 11.3$\mu m$ feature but none of the features at 6 to 9$\mu m$.  However, given the strong evidence of an independent $z=0.3$ determination for a spatially coincident source with the matching continuum flux density, we conclude that this possibility of $z=1.98$ is much less likely than the $z=0.3$ identification.

With the [\rm{O\,\textsc{iv}}] at $z=0.3$ identification, the large line width suggests the presence of a strong QSO. GOODS J123658.82+621638.1 is also detected with Chandra \citep{2003AJ....126..539A}.   This is a soft X-ray emitting source and falls in the "normal galaxy" category of \cite{2004AJ....128.2048B} classification.  In a more recent analysis by \cite{2007MNRAS.377..203G}, the ratio of x-ray to infrared luminosity also places this source in the middle of the range for star-forming galaxies. Note also that  prominent  Aromatics at 6.2$\mu m$ (Fig \ref{OIVsrc} inset) also argues for a strong star formation contribution to the luminosity of this source. This apparent  contradiction with the measured line width in the spectrum may be the result of high dust attenuation affecting the x-ray luminosity, or simply due to multiple sources contributing to the IRS spectrum. The latter would not be surprising as the GOODS catalog contains two sources at $z\sim 0.3$ within 3.2" of SUUSS 17.  Here again, the additional work  needed to resolve these ambiguities is beyond the scope of this paper.

\section{Discussion}

This survey offers us a way to characterize the Mid-IR population of galaxies in a small region of the sky down to very deep detection levels. While any survey samples the Universe imperfectly, this survey avoids the bias of a narrow bandpass sampling complex SEDs. No bias is expected from the detection method used as it is only related to the integrated fluxes of the sources in the field and does not rely on any specific spectral feature detection. 

We investigate the possibility that the analysis method presented here favors sources with specific types and redshifts in our field, potentially resulting in an {\it a posteriori} bias in the results. As explained in Sect. 3.1, the cross-correlation method is sensitive to the variance of the signal in the spectral window with respect to the variance of the noise. This means that the method gives better results when applied to spectra that show significant structure such as strong emission lines, Aromatic Features or silicate absorption. Considering that the spectral window in our sample is limited to either 14-20$\mu$m or 20-35$\mu$m this could lead to inhomogeneities or biases over redshift and type of sources for which the cross-correlation method was successful. If this effect is present, the only efficient way to remove it is to add multi-wavelength data (IRAC fluxes, near-IR and optical bands ... ) to the spectra used in the cross-correlation algorithm thus lowering the probability of missing significant spectral information. 

We discuss below the luminosities of the sources detected in this survey, then their mid-IR spectral character, and place them in the context of other surveys.  We then contrast this unbiased spectral survey with single-band continuum surveys.

\subsection{IR luminosities and spectral types}
\label{SectLir}

\begin{figure}
\plotone{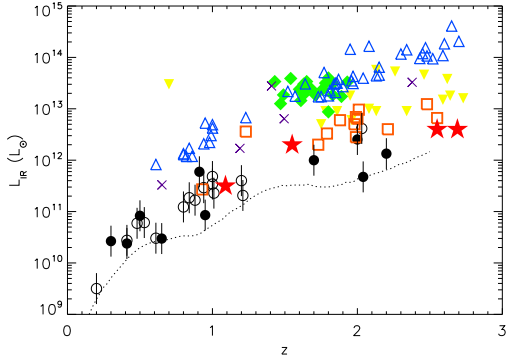}
\caption{IR bolometric luminosities computed for our sample from MIPS 24$\mu$m fluxes using a code provided by R. Chary implementing \cite{2001ApJ...556..562C} and \cite{2002ApJ...576..159D} models. Uncertainties are of 50\%. We overplot IR luminosities of several other IRS surveys of high-redshifts galaxies. \emph{Filled dots} are SUUSS sources from Field~1 and \emph{empty dots} are SUUSS sources from Field~2, \emph{yellow filled triangles} are 17 ULIRGS from the IRS GTO \citep{2005ApJ...622L.105H}, \emph{blue open triangles} are 47 ULIRGs from the XFLS \citep{2007ApJ...658..778Y}, {\it green diamonds} are bump 2/3 sources from \cite{2008ApJ...677..957F}, \emph{orange open squares} are 13 SMGs from the HDF-North SCUBA Super-map \citep{2008ApJ...675.1171P}, {\it purple crosses} are 5 SMGs from \cite{2007ApJ...655L..65M}, and \emph{red stars} are from \cite{2007ApJ...659..941T}. The {\it dotted line} is the luminosity limit of the SUUSS survey ($f_{\nu}(24\mu m) = 47\mu Jy$). }
\label{Lir}
\end{figure}

Since we only have mid-IR data, we can not compute accurate total IR-luminosities. We could use the best template for each source but some of the templates have no far-IR data and $42\%$ of the sources are not identified. 
We estimate IR bolometric luminosities from MIPS 24$\mu$m flux measurements using a code made available online by R. Chary. The code selects the most appropriated SED from \cite{2001ApJ...556..562C} and \cite{2002ApJ...576..159D} models based on the 24$\mu$m flux, fits the SEDs and computes the total 8-1000$\mu$m IR luminosities, then averages them together. Figure \ref{Lir} shows the source luminosities as a function of redshift. The luminosities range from $3.2 \times 10^9 L_\odot $ to $4.1 \times 10^{12} L_\odot$ with a mean of $5.2 \times 10^{11} L_\odot$. Even the 4 sources at $z \ge 2$ have an impressively low mean luminosity of $7.5 \times 10^{11} L_\odot$. When compared to existing IRS spectroscopic studies of high-redshift galaxies (Houck et al. 2005, Pope et al. 2006, Yan et al. 2007), our sample proves to be one of the deepest in this range of work (see Fig  \ref{Lir}). 

\begin{figure*}
\plottwo{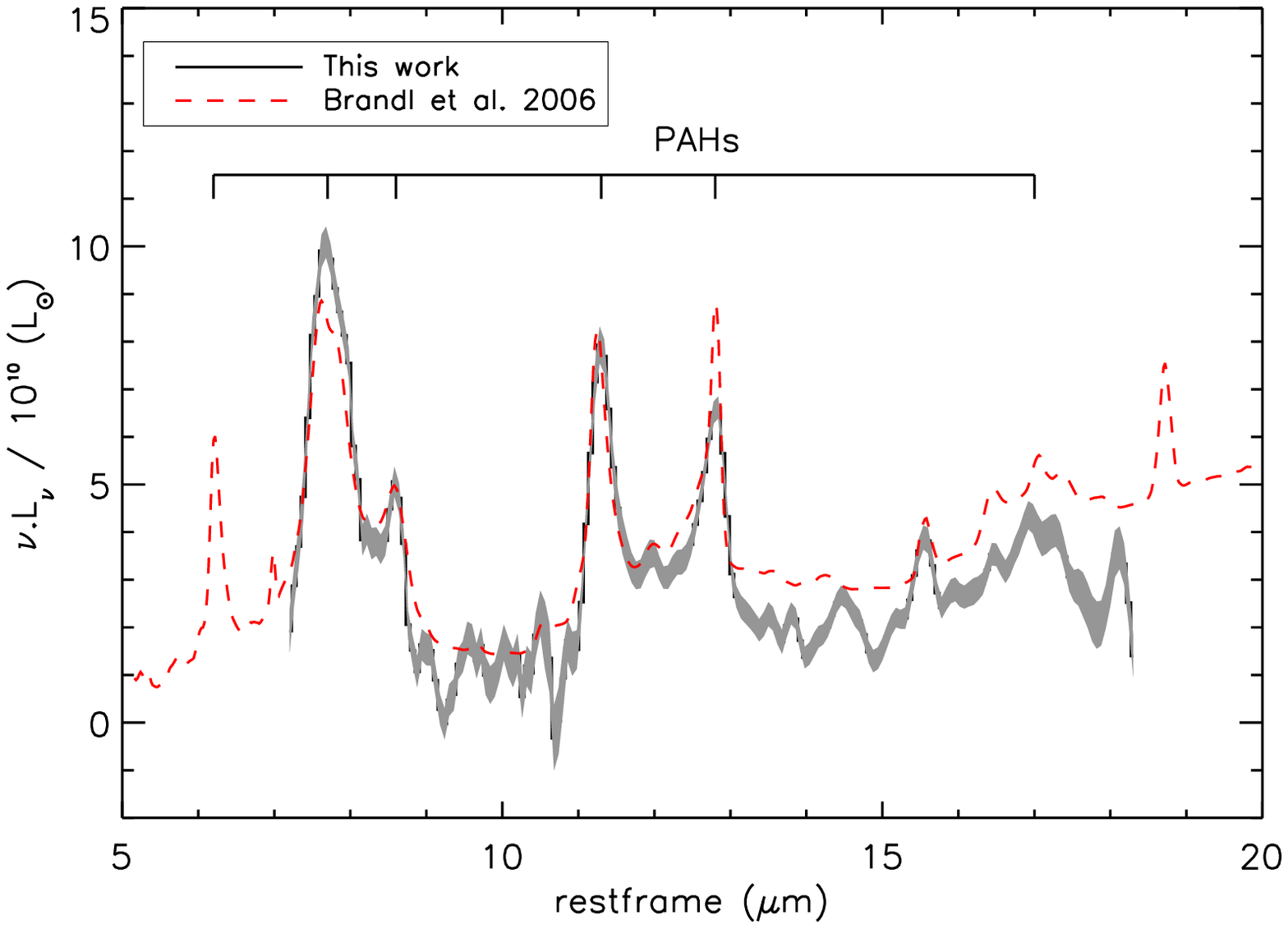}{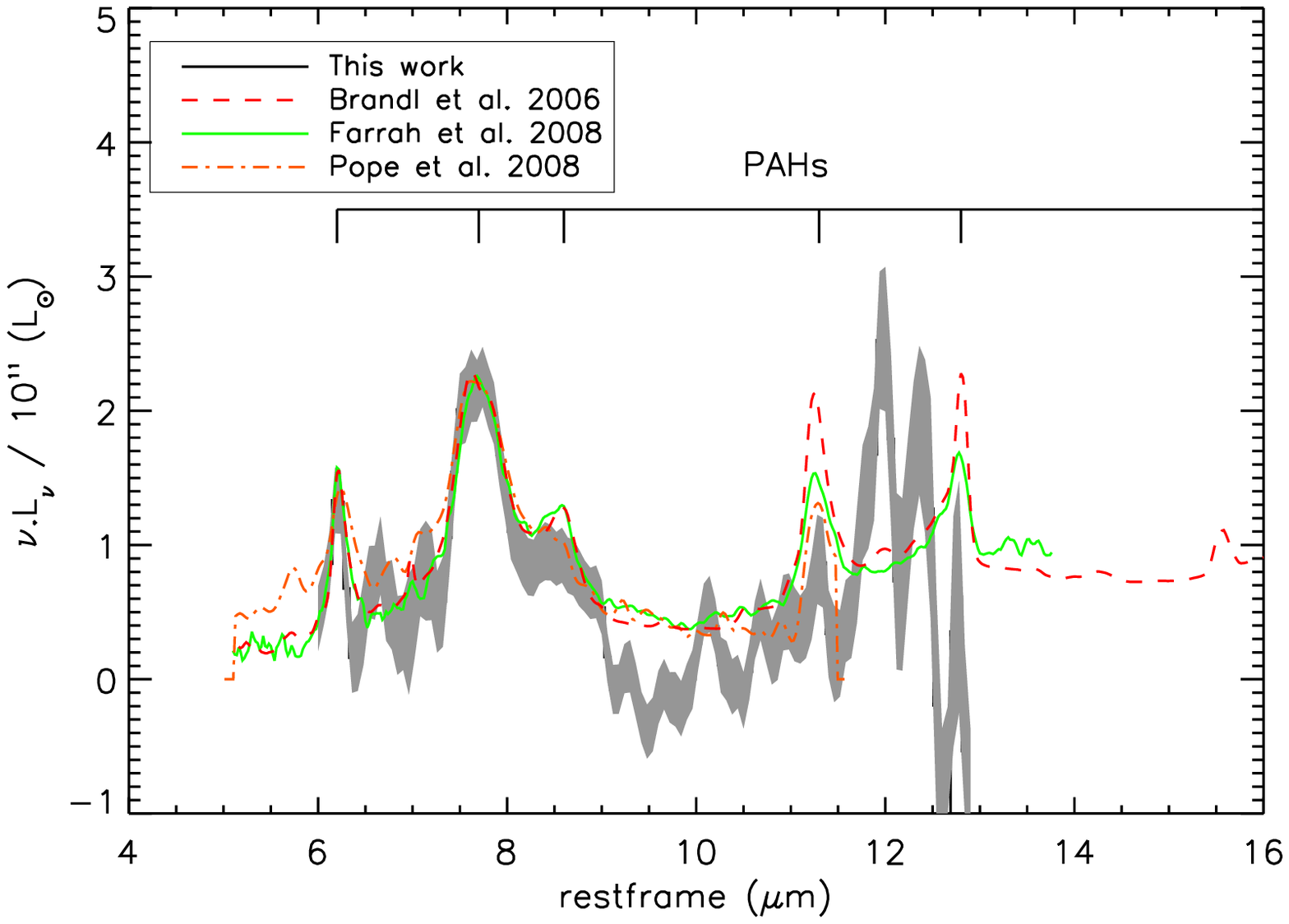}
\caption{\emph{Left}, composite spectrum at $z = 1\pm 0.2$. \emph{Right}, composite spectrum at $z\sim 2$. In {\it gray shading} we plot the 1$\sigma$ dispersion for each composite. In {\it dashed red}, we overplot the ``starburst average spectrum'' from \cite{2006ApJ...653.1129B} scaled to our composites $8.6\mu m$ and $7.7\mu m$ luminosities at $z=1$ and $z\sim 2$ respectively. We also plot in {\it green} average spectrum from \cite{2008ApJ...677..957F} and in {\it orange} $z\sim 2$ average spectrum from \cite{2008ApJ...675.1171P}. Both composites are dominated by PAH emission at $7.7\mu m$ and agree very well with the starburst template. The $11.3\mu m$ PAH detection in the $z\sim 2$ composite is not clear. This part of the composite ($\lambda > 10\mu m$) is particularly affected by the noise.}
\label{CompZ12}
\end{figure*}

We selected all sources at $0.8 < z <1.1$ (8 spectra) and at $1.7 < z < 2.2$ (4 spectra, the ``line-source'' SUUSS 9 discussed in Sect \ref{SectLineSrc} was discarded from this sample) and computed corresponding composites (Fig \ref{CompZ12}). Each spectrum was scaled to its sample mean luminosity. The composite at $z\sim 1$ is dominated by Aromatic Features. We detect the 7.7, 8.6, 11.3 emission features as well as the [\rm{Ne\,\textsc{ii}}] 12.8 $\mu$m line. We see generally agreement with the AGN-free starburst template from Brandl et al. (2006). Due to lack of spectral coverage in this $z\sim 1$ sample, we have no data at $\lambda$ greater than 17$\mu$m.

In the composite at $z\sim 2$ we can clearly see 6.2, 7.7 and 8.6 $\mu$m PAHs. In comparison to $z\sim 1$, this sample of sources is half an order of magnitude more luminous. We also can not confirm or rule out the presence of any emission at 11.3 $\mu$m, although this feature is very often detected across the entire sample of this study. This may be due to the small number of sources entering the $z\sim 2$ composite together with increased noise at this restframe wavelength. The shape and relative strengths of the 7.7 and 8.6 $\mu$m features is however conserved between $z\sim 1$ and $z\sim 2$. The 4 sources that enter this composite all show a clear stellar bump in the IRAC channels which is expected for starburst galaxies between $z = 1.7$ (bump 2) and $z = 2.2$ ($\sim$ bump 3). As seen by \cite{2008ApJ...677..957F} in their sample of 32 high-redshift ultraluminous infrared galaxies selected as bump 2 sources (mainly), composite spectra of such sources are best described by a starburst component, namely the template produced by \cite{2006ApJ...653.1129B}. Little silicate absorption is seen in the \citeauthor{2008ApJ...677..957F} sample and none is seen in our composites either. The same observation is made on the population of submillimeter-selected high redshift infrared galaxies presented in \cite{2007ApJ...655L..65M} and \cite{2008ApJ...675.1171P} where a starburst scenario (\citeauthor{2006ApJ...653.1129B} \citeyear{2006ApJ...653.1129B} template or M82) with low-AGN contribution better fits the composite spectra. 
Previous work by \cite{2005ApJ...622L.105H} or \cite{2007ApJ...664..713S} find prominent and/or non-negligible SiO absorption in their sources (55\% and 33\% of the samples respectively). Our $z\sim 2$ composite is surprisingly quite similar to the composite of $z\sim 2$ sources presented in \cite{2007ApJ...664..713S} with the exception of the slight SiO absorption. Overall, we observe mid-IR composite spectra generally similar to those derived from other samples of high-redshift ($1 < z < 2.5$) infrared galaxies with very different selection criteria and bolometric luminosities, and generally favoring a starburst character. 

\subsection{Characterizing the population (AGN vs starburst ?)}

Table \ref{SrcStatTab} summarizes the statistics of spectral types assigned for objects detected in this survey.  We assigned each spectral template used to one of three classes, namely strong aromatics, silicate absorption and mixed signatures. The counts of objects falling in each category are shown in columns 3, 5 and 4 respectively; column 7 accounts for sources which could not be matched using the spectral correlation method. The two sources discussed in Sect \ref{SectLineSrc} were gathered in a fourth class (column 6).
Among the sources with accepted matches, about half (14 out of 28) are dominated by the aromatic features and another third (8 out of 28) have a mixed character.
About 15\% (4 out of 26) have substantial silicate absorption, indicative of  high optical depth, against a dominant continuum suggesting a major AGN contribution.  Out of the total sample of 45  sources, 17 could not be matched to any templates in our library.  This might be due to their being dominated by a featureless continuum, or to their low signal-to-noise ratio.   If the former is the dominant cause, then these 17 might be
primarily AGN-dominated sources, and the statistics among matched spectra are biased towards starburst signatures.

\begin{figure*}
\plottwo{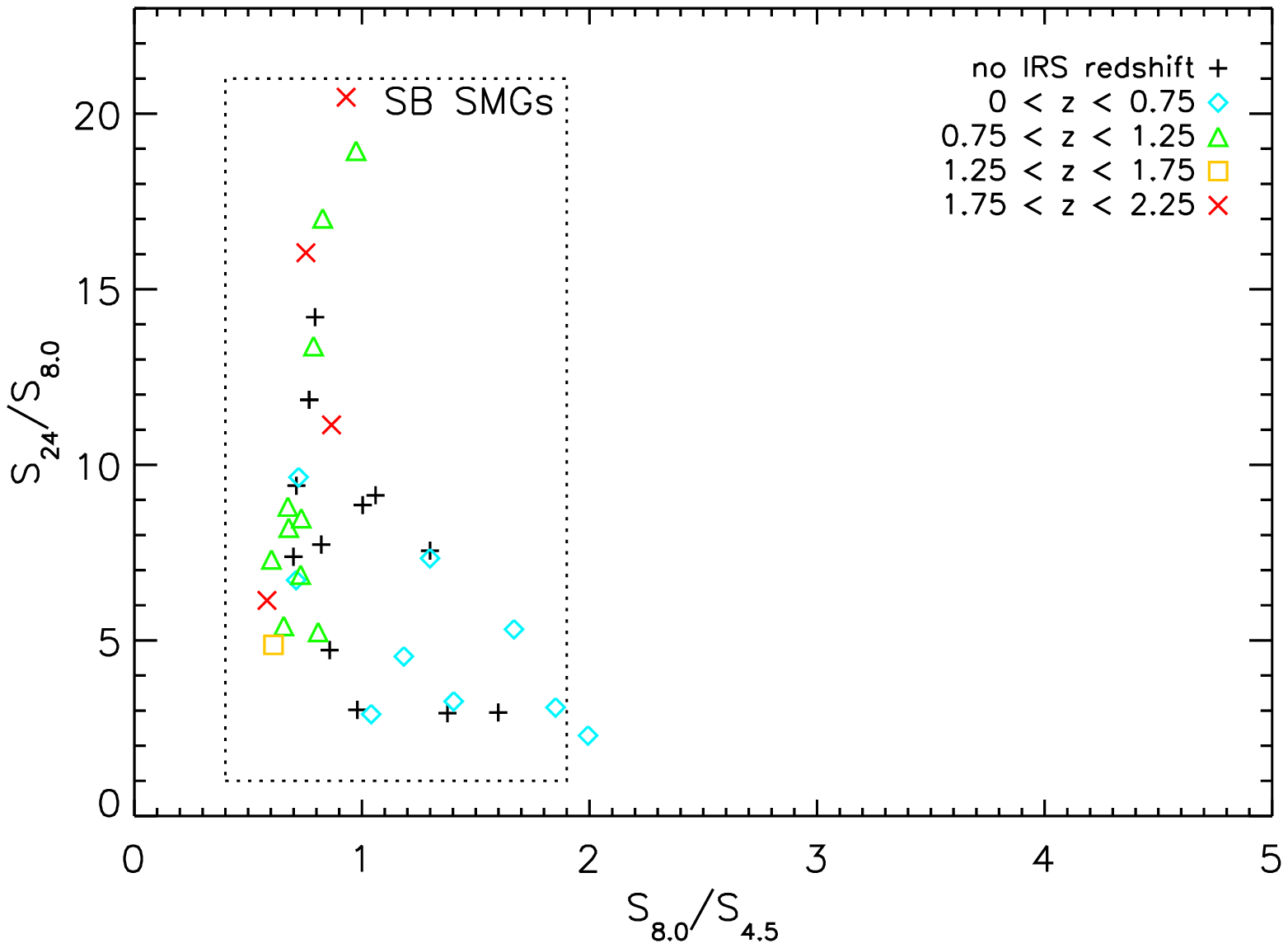}{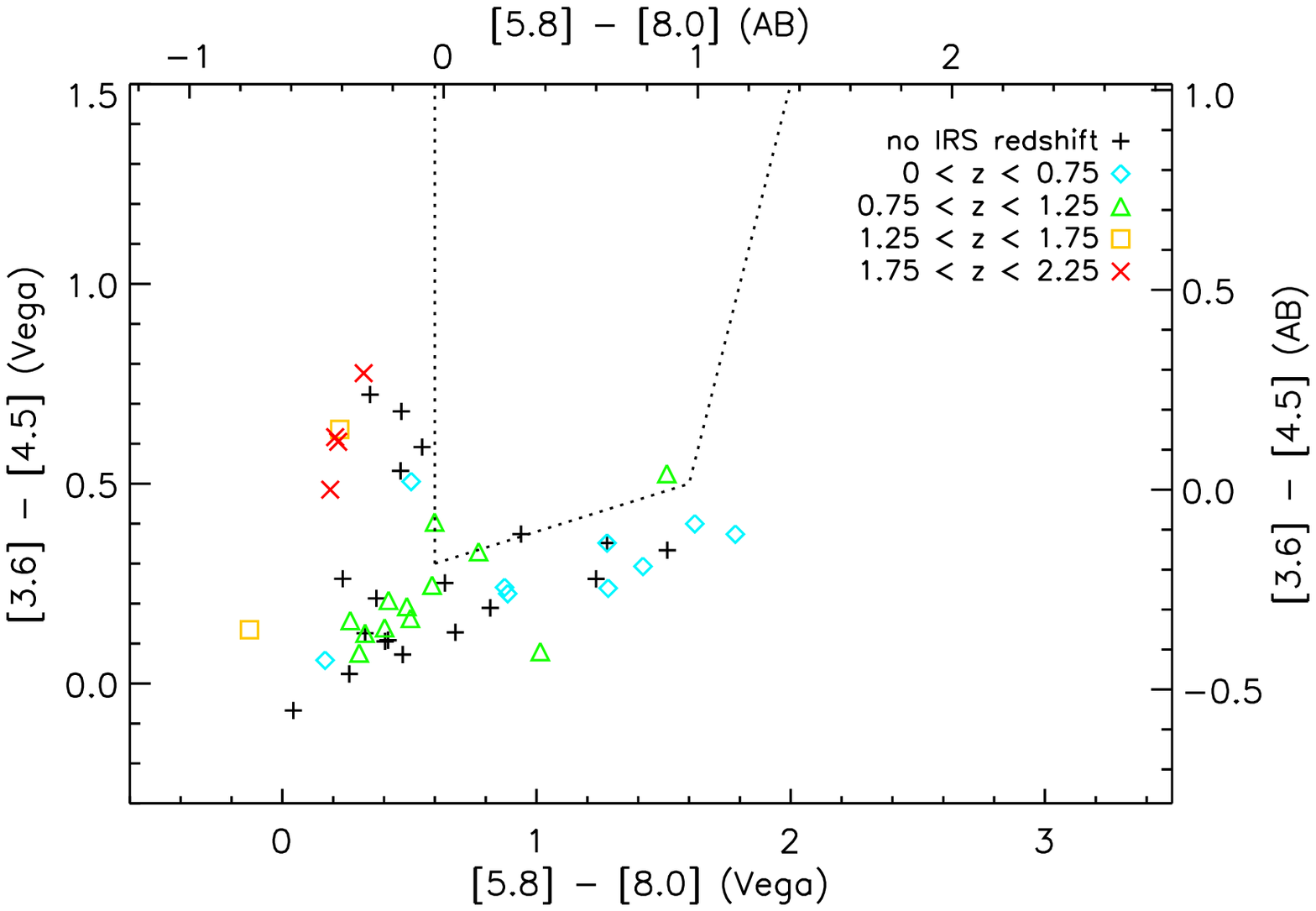}
\caption{Spitzer color-color diagram characterization of the IRS Ultradeep survey sample of galaxies. \emph{Left}, AGN diagnostic proposed by \cite{2008ApJ...675.1171P} determined from a sample of 13 SMGs from z$\sim$1 to 2.5 (see also \citealp{2004ApJS..154..124I}). \emph{Right}, AGN diagnostic using IRAC colors, explained in \cite{2005ApJ...631..163S}. The Broad-Line AGN selection criteria are represented by the {\it dotted} line. The lower and left-hand axes are scaled to Vega magnitudes whereas the top and right-hand axes refer to AB magnitudes.}
\label{AGNDiagnostic}
\end{figure*}

In order to rule out a bias in favor of starburst galaxies among matched spectra, we used additional diagnostics based on IRAC and MIPS color-color plots (Fig \ref{AGNDiagnostic}). They can be used to separate AGN-dominated sources up to $z \sim 2$ (e.g. \citeauthor{2008arXiv0806.4610D} \citeyear{2008arXiv0806.4610D}). According to these diagnostics,  no clear AGN-dominated system is detected in our total sample of 45 sources, which is consistent with the redshift and spectral-type determination among sources matching templates. The  sources in our sample therefore seem to be mostly dominated by star-formation, and this result is not a bias due to the spectral matching technique.

We have also  used X-ray data for this region from the Chandra Deep Field North 2Ms catalog \citep{2003AJ....126..539A} to look for X-ray counterparts to our sources.  We found four matches inside a 3 arcsecond matching radius. Following the classification proposed by \cite{2004AJ....128.2048B}, two of them are potential X-ray obscured AGNs (SUUSS 31 and 41), and one is a starburst galaxy  clearly identified as such with our dataset (SUUSS 20). The last X-ray counterpart is the [\rm{O\,\textsc{iv}}] line-source previously described in Sect. 3.4. Three additionnal counterparts were identified using the supplementary catalog; these are sources that were not detected with a high enough significance to be included in the main 2Ms X-ray catalog. They are fainter sources but have an optical counterpart which makes them highly likely to be real X-ray sources and potential candidates for faint AGN. One of them presents PAH emission with possible silicate absorption, one is clearly continuum dominated (with 17$\mu$m PAH emission detected), and the last one has deeper silicate absorption which give us two other possible AGN identifications. Tentatively, we estimate an AGN fraction of  4 out of 45 (9\%) of our sample, based on X-ray detections.  Given the sample size, this is consistent with the spectral typing results summarized above.

The IRS GTO observations of \cite{2005ApJ...622L.105H} unveiled a population of sources dominated by strongly obscured (SiO absorption) AGNs, some presenting PAH signatures even though faint.  The \cite{2007ApJ...658..778Y} sample of 52 ULIRGs has more diverse properties with roughly a third of the sources presenting strong PAHs, another third showing strong silicate absorption and the rest being continuum sources sometimes with weak features. These sources show large AGN contributions (80\% of the sample has some level of AGN continuum signature) with a significant fraction of sources having a starburst signature as well (\citeauthor{2007ApJ...658..778Y} \citeyear{2007ApJ...658..778Y}; \citeauthor{2007ApJ...664..713S} \citeyear{2007ApJ...664..713S}). Those two samples have very different AGN fractions and SiO absorption depths in comparison to our sample. However, those samples also derive from quite different selection criteria and flux densities.  We will address below the relation between the AGN fraction differences   and the sample selection.

Previous studies have shown submillimeter galaxies at high redshifts to be dominated by starburst emission (\citeauthor{2004ApJS..154..130E} \citeyear{2004ApJS..154..130E}; \citeauthor{2004ApJS..154..124I} \citeyear{2004ApJS..154..124I}; \citeauthor{2006MNRAS.370.1185P} \citeyear{2006MNRAS.370.1185P}). 
Thus a comparison of our sample with the \cite{2008ApJ...675.1171P} sample may be useful. As we already discussed, both samples have very similar composite spectra at $1<z<2.5$. While X-ray classification tends to show the presence of AGN in about 40\% of those sources, star formation is found to account for more than half the bolometric luminosity.  An interesting way of computing mid-IR photometric redshifts was proposed in \cite{2006MNRAS.370.1185P} using all four IRAC bands and MIPS 24$\mu$m fluxes to give an estimate of the redshift. This method assumes a fairly constant SED shape across the sample and thus is spectral-type dependent. We produced a similar formula for our sample by fitting to the redshift obtained with the cross-correlation method:
\begin{eqnarray}
z_{\rm{IR}}= a+ b \cdot \rm{log(S_{3.6})}+ c \cdot \rm{log(S_{5.8})} \nonumber \\
       \mbox{}+d \cdot \rm{log(S_{8.0})}+e \cdot \rm{log(S_{24})}
\label{eqzPope}
\end{eqnarray}
where the fitting parameters values are $1.7$, $-2.6$, $3.3$, $-1.0$ and $0.1$ for $a$, $b$, $c$, $d$ and $e$ respectively. 
We find the deviation of the redshift statistic to be of $\sigma \left(\Delta z / \left(1+z\right) \right) = 0.22$ which is rather large compared to the dispersion of $\sigma \left(\Delta z / \left(1+z\right) \right) = 0.07$ found in \cite{2006MNRAS.370.1185P} , perhaps explained by the fact that our SEDs that enter the fit have a wide variety of spectral types. However, no selection could be found that, when applied to our sample, reduced the dispersion. It is therefore very likely that there is a larger diversity of SEDs in our sample compared to the 13 SMGs studied by \cite{2008ApJ...675.1171P}, even though both have very similar average mid-IR spectra.

\subsection{Unbiased Spectral Surveys: A Different Probe}
\label{UnbiasedSurvey}

While our survey covers a small solid angle and the data are collected over two separate spectral windows, it still yields a somewhat different perspective on the mid-IR universe at $z\leq 2$ and illustrates some of the distinguishing characteristics of wide-band spectral surveys. The basic statistics, summarized in Table \ref{SrcStatTab}, point to a relative paucity of AGN signature among SUUSS sources compared to other deep spectral surveys such as \cite{2006ApJ...651..101W} or \cite{2007ApJ...658..778Y}. The same lack of AGN signatures is observed in IRAC color-color diagnostic plots of SUUSS sources, so it is very unlikely to reflect a bias introduced by the spectral correlation technique.  This difference is partly due to additional selection criteria imposed on the other surveys, partly due to the fainter flux densities and luminosities of the SUUSS sample, but probably also in  part due to the reduced selectivity of an unbiased spectral survey. 

The Yan et al and Weedman et al samples were both selected for high $24\mu m/R$ ratios, with an additional selection for high $24/8\mu m$ ratios by Yan et al.  These choices were aimed at biasing towards high redshifts and luminosities, and indeed they achieve that as can be seen in Fig \ref{Lir}.  Aromatic Features dominate the spectra of no more than a third of those samples. The selection for high $24\mu m / 8\mu m$ or $24\mu m / R$ favors high obscuration, which is mildly anti-correlated with Aromatic signatures \citep{2007ApJ...664..713S}, and might therefore reduce the incidence of Aromatic Features in those samples.

On the other hand, the SUUSS sample has a median $f_\nu(24\mu m)  \sim 100 \mu Jy$, whereas the other samples are  brighter by one order of magnitude for Yan et al, and almost 3 orders of magnitude for Weedman et al.  The lower fluxes lead to lower luminosities, and would generally favor star formation powered systems over AGN, translating into a preponderance of Aromatic-Feature rich spectra (Papovich et al 2007). Given the evidence at hand, we favor this interpretation as the dominant factor in determining the spectral characteristics of this sample. 

The IRAC colors of the SUUSS sample also point to a lack of objects residing in the AGN-dominated part of that diagnostic plot (Fig \ref{AGNDiagnostic}). The same diagnostic is used over all the GOODS-North sources with a MIPS 24$\mu$m flux in between 45 and 100$\mu$Jy (Fig \ref{AGNDiagnosticGoods}). Out of the 823 sources that fall into this category only 93 (11\%) are found in the AGN-dominated part of the diagram. The effect observed in the SUUSS sample is thus most probably due to its lower mean luminosity, which results from the sensitivity achieved but also from the choice of a particularly dark survey field.

Other factors  may affect the character of SUUSS sources.  One possibility is that spectra with more features, i.e larger $\sigma_S$, are easier to pick out in a spectrally dispersed survey than in a traditional single-band survey, since a feature would have to coincide with the survey filter to generate a signal. Thus single-band surveys would favor continuum-dominated sources at all redshifts,  bias against those whose absorption features  fall  into the band, and bias for sources whose emission features fall into the band. This may explain  the  preponderance of AGN in 24$\mu$m selected samples (e.g. Daddi et al 2007).  Fig \ref{SpecSNR}, right-hand panel illustrates the lack of selectivity in this survey, as it shows most sources peak away from 24$\mu$m, and indeed have higher fluxes where detected than at 24$\mu$m.

The 24$\mu m$  survey obviously missed the single-line source SUUSS 9, since the line fell outside the 24$\mu m$ band.
Since we found one such source in the whole survey, the incidence of such sources in general must be of the order of $2-4\%$, depending on what we assume for the efficiency of our search.  Since this source does not have a 24$\mu m$ counterpart, we can only assign an upper limit to its luminosity as defined in Fig \ref{Lir}. That upper limit is roughly $5 \times 10^{11} L_\odot$, based on the 24$\mu m$ detection limit.

The wavelength coverage in this survey was either 14 to 21$\mu$m or 20 to 35$\mu$m. This limited spectral coverage has contributed to the difficulty of identifying a spectral match for more than a third of the detections, and to the difficulty of identifying the single line detected in SUUSS 9.  Future IR spectral surveys would be more productive if they covered at least a full octave in wavelength.

\begin{figure}
\plotone{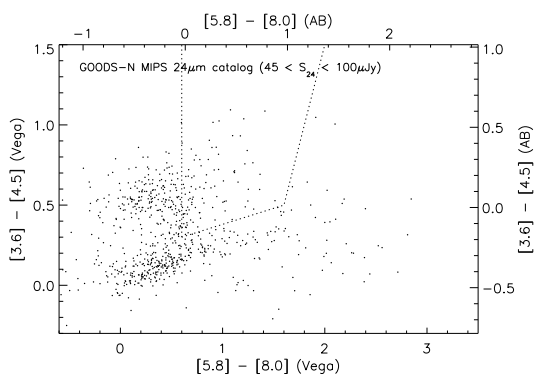}
\caption{IRAC color-color diagram for all GOODS-North sources with identified MIPS 24$\mu$m flux such as $45 < f_{\nu}(24\mu m) < 100\mu Jy$. 823 sources fall into this category and only 93 of them enter the AGN region of the diagram i.e 11\% of the sample. This low ratio indicates the importance of the detection limit on the ratio of AGN-dominated system over starburst dominated system found in flux limited surveys.}
\label{AGNDiagnosticGoods}
\end{figure}

\section{Conclusion}

We have obtained IRS spectra for 45 sources of very faint IR galaxies and determined redshifts for $\sim 60\%$ (28/45 sources) of them using a specifically designed cross-correlation method. We covered a domain of IR luminosities from about $7 \times 10^{10} L_\odot$ at $z = 0.5$ to about $10^{12} L_\odot$ at $z \sim 2.0$ poorly covered by IRS spectroscopy so far. 
At least 47\% of our sample show starburst activity as PAH signatures (up to 21/45 sources) and 31\% (14/45 sources) prove to be starburst dominated. The rest of the redshift identifications rest on silicate absorption feature (11\% or 5/45 sources). We tentatively identify only 9\% of our sample (4/45 sources) as AGN candidates. This small fraction of AGN in comparison to previous IRS spectroscopic surveys of high-redshifts galaxies is likely to originate from the fainter luminosities reached here.
We find two unusual sources with only one prominent emission line detected over the spectral range covered. One of them present a strong [OIV] line (25.9$\mu$m restframe) at $z = 0.3$ on top of a significant continuum while the other shows the [SIV] line (10.5$\mu$m restframe) at $z = 2.08$ on a very weak continuum dominated by instrumental noise. 

\acknowledgments
We are grateful to the anonymous referee for helpful suggestions which improved this paper. We would like to thank Aaron Stephen for help with X-ray data.
This work is based on observation obtained with the \emph{Spitzer Space Telescope}, which is operated by the Jet Propulsion Laboratory, California Institute of Technology, under NASA contract 1407. Support for this work was provided by NASA through an award issued by JPL/Caltech, as well as the french National Agency for Research under programs ANR-06-BLAN-0170 and ANR-05-BLAN-0289-02. This research has made use of the NASA/IPAC Extragalactic Database which is operated by JPL/Caltech, under contract with NASA.

{\it Facilities:} \facility{Spitzer (IRS)}


\begin{thebibliography}{59}
\expandafter\ifx\csname natexlab\endcsname\relax\def\natexlab#1{#1}\fi

\bibitem[{{Alexander} {et~al.}(2003){Alexander}, {Bauer}, {Brandt},
  {Schneider}, {Hornschemeier}, {Vignali}, {Barger}, {Broos}, {Cowie},
  {Garmire}, {Townsley}, {Bautz}, {Chartas}, \&
  {Sargent}}]{2003AJ....126..539A}
{Alexander}, D.~M., {Bauer}, F.~E., {Brandt}, W.~N., {et~al.} 2003, \aj, 126,
  539

\bibitem[{{Appleton} {et~al.}(2006){Appleton}, {Xu}, {Reach}, {Dopita}, {Gao},
  {Lu}, {Popescu}, {Sulentic}, {Tuffs}, \& {Yun}}]{2006ApJ...639L..51A}
{Appleton}, P.~N., {Xu}, K.~C., {Reach}, W., {et~al.} 2006, \apjl, 639, L51

\bibitem[{{Armus} {et~al.}(2006){Armus}, {Bernard-Salas}, {Spoon}, {Marshall},
  {Charmandaris}, {Higdon}, {Desai}, {Hao}, {Teplitz}, {Devost}, {Brandl},
  {Soifer}, \& {Houck}}]{2006ApJ...640..204A}
{Armus}, L., {Bernard-Salas}, J., {Spoon}, H.~W.~W., {et~al.} 2006, \apj, 640,
  204

\bibitem[{{Armus} {et~al.}(2007){Armus}, {Charmandaris}, {Bernard-Salas},
  {Spoon}, {Marshall}, {Higdon}, {Desai}, {Teplitz}, {Hao}, {Devost}, {Brandl},
  {Wu}, {Sloan}, {Soifer}, {Houck}, \& {Herter}}]{2007ApJ...656..148A}
{Armus}, L., {Charmandaris}, V., {Bernard-Salas}, J., {et~al.} 2007, \apj, 656,
  148

\bibitem[{{Armus} {et~al.}(2004){Armus}, {Charmandaris}, {Spoon}, {Houck},
  {Soifer}, {Brandl}, {Appleton}, {Teplitz}, {Higdon}, {Weedman}, {Devost},
  {Morris}, {Uchida}, {van Cleve}, {Barry}, {Sloan}, {Grillmair}, {Burgdorf},
  {Fajardo-Acosta}, {Ingalls}, {Higdon}, {Hao}, {Bernard-Salas}, {Herter},
  {Troeltzsch}, {Unruh}, \& {Winghart}}]{2004ApJS..154..178A}
{Armus}, L., {Charmandaris}, V., {Spoon}, H.~W.~W., {et~al.} 2004, \apjs, 154,
  178

\bibitem[{{Bauer} {et~al.}(2004){Bauer}, {Alexander}, {Brandt}, {Schneider},
  {Treister}, {Hornschemeier}, \& {Garmire}}]{2004AJ....128.2048B}
{Bauer}, F.~E., {Alexander}, D.~M., {Brandt}, W.~N., {et~al.} 2004, \aj, 128,
  2048

\bibitem[{{Brandl} {et~al.}(2006){Brandl}, {Bernard-Salas}, {Spoon}, {Devost},
  {Sloan}, {Guilles}, {Wu}, {Houck}, {Weedman}, {Armus}, {Appleton}, {Soifer},
  {Charmandaris}, {Hao}, {Higdon}, \& {Herter}}]{2006ApJ...653.1129B}
{Brandl}, B.~R., {Bernard-Salas}, J., {Spoon}, H.~W.~W., {et~al.} 2006, \apj,
  653, 1129

\bibitem[{{Brinchmann} {et~al.}(2008){Brinchmann}, {Kunth}, \&
  {Durret}}]{2008A&A...485..657B}
{Brinchmann}, J., {Kunth}, D., \& {Durret}, F. 2008, \aap, 485, 657

\bibitem[{{Caputi} {et~al.}(2006){Caputi}, {Dole}, {Lagache}, {McLure},
  {Puget}, {Rieke}, {Dunlop}, {Le Floc'h}, {Papovich}, \&
  {P{\'e}rez-Gonz{\'a}lez}}]{2006ApJ...637..727C}
{Caputi}, K.~I., {Dole}, H., {Lagache}, G., {et~al.} 2006, \apj, 637, 727

\bibitem[{{Caputi} {et~al.}(2007){Caputi}, {Lagache}, {Yan}, {Dole},
  {Bavouzet}, {Le Floc'h}, {Choi}, {Helou}, \& {Reddy}}]{2007ApJ...660...97C}
{Caputi}, K.~I., {Lagache}, G., {Yan}, L., {et~al.} 2007, \apj, 660, 97

\bibitem[{{Chary} \& {Elbaz}(2001)}]{2001ApJ...556..562C}
{Chary}, R. \& {Elbaz}, D. 2001, \apj, 556, 562

\bibitem[{{Chary}(2007)}]{2007ASPC..380..375C}
{Chary}, R.-R. 2007, in Astronomical Society of the Pacific Conference Series,
  Vol. 380, Deepest Astronomical Surveys, ed. J.~{Afonso}, H.~C. {Ferguson},
  B.~{Mobasher}, \& R.~{Norris}, 375--+

\bibitem[{{Dale} \& {Helou}(2002)}]{2002ApJ...576..159D}
{Dale}, D.~A. \& {Helou}, G. 2002, \apj, 576, 159

\bibitem[{{Dale} {et~al.}(2006){Dale}, {Smith}, {Armus}, {Buckalew}, {Helou},
  {Kennicutt}, {Moustakas}, {Roussel}, {Sheth}, {Bendo}, {Calzetti}, {Draine},
  {Engelbracht}, {Gordon}, {Hollenbach}, {Jarrett}, {Kewley}, {Leitherer},
  {Li}, {Malhotra}, {Murphy}, \& {Walter}}]{2006ApJ...646..161D}
{Dale}, D.~A., {Smith}, J.~D.~T., {Armus}, L., {et~al.} 2006, \apj, 646, 161

\bibitem[{{Dasyra} {et~al.}(2008){Dasyra}, {Ho}, {Armus}, {Ogle}, {Helou},
  {Peterson}, {Lutz}, {Netzer}, \& {Sturm}}]{2008ApJ...674L...9D}
{Dasyra}, K.~M., {Ho}, L.~C., {Armus}, L., {et~al.} 2008, \apjl, 674, L9

\bibitem[{{Dasyra} {et~al.}(2009){Dasyra}, {Yan}, {Helou}, {Sajina}, {Fadda},
  {Zamojski}, {Armus}, {Draine}, \& {Frayer}}]{2009arXiv0906.5271D}
{Dasyra}, K.~M., {Yan}, L., {Helou}, G., {et~al.} 2009, ArXiv e-prints

\bibitem[{{Dickinson} {et~al.}(2003){Dickinson}, {Giavalisco}, \& {The Goods
  Team}}]{2003mglh.conf..324D}
{Dickinson}, M., {Giavalisco}, M., \& {The Goods Team}. 2003, in The Mass of
  Galaxies at Low and High Redshift, ed. R.~{Bender} \& A.~{Renzini}, 324--+

\bibitem[{{Dole} {et~al.}(2001){Dole}, {Gispert}, {Lagache}, {Puget},
  {Bouchet}, {Cesarsky}, {Ciliegi}, {Clements}, {Dennefeld}, {D{\'e}sert},
  {Elbaz}, {Franceschini}, {Guiderdoni}, {Harwit}, {Lemke}, {Moorwood},
  {Oliver}, {Reach}, {Rowan-Robinson}, \& {Stickel}}]{2001A&A...372..364D}
{Dole}, H., {Gispert}, R., {Lagache}, G., {et~al.} 2001, \aap, 372, 364

\bibitem[{{Dole} {et~al.}(2004){Dole}, {Rieke}, {Lagache}, {Puget},
  {Alonso-Herrero}, {Bai}, {Blaylock}, {Egami}, {Engelbracht}, {Gordon},
  {Hines}, {Kelly}, {Le Floc'h}, {Misselt}, {Morrison}, {Muzerolle},
  {Papovich}, {P{\'e}rez-Gonz{\'a}lez}, {Rieke}, {Rigby}, {Neugebauer},
  {Stansberry}, {Su}, {Young}, {Beichman}, \& {Richards}}]{2004ApJS..154...93D}
{Dole}, H., {Rieke}, G.~H., {Lagache}, G., {et~al.} 2004, \apjs, 154, 93

\bibitem[{{Donley} {et~al.}(2008){Donley}, {Rieke}, {Perez-Gonzalez}, \&
  {Barro}}]{2008arXiv0806.4610D}
{Donley}, J.~L., {Rieke}, G.~H., {Perez-Gonzalez}, P.~G., \& {Barro}, G. 2008,
  ArXiv e-prints

\bibitem[{{Egami} {et~al.}(2004){Egami}, {Dole}, {Huang}, {P{\'e}rez-Gonzalez},
  {Le Floc'h}, {Papovich}, {Barmby}, {Ivison}, {Serjeant}, {Mortier}, {Frayer},
  {Rigopoulou}, {Lagache}, {Rieke}, {Willner}, {Alonso-Herrero}, {Bai},
  {Engelbracht}, {Fazio}, {Gordon}, {Hines}, {Misselt}, {Miyazaki}, {Morrison},
  {Rieke}, {Rigby}, \& {Wilson}}]{2004ApJS..154..130E}
{Egami}, E., {Dole}, H., {Huang}, J.-S., {et~al.} 2004, \apjs, 154, 130

\bibitem[{{Elbaz} {et~al.}(1999){Elbaz}, {Cesarsky}, {Fadda}, {Aussel},
  {D{\'e}sert}, {Franceschini}, {Flores}, {Harwit}, {Puget}, {Starck},
  {Clements}, {Danese}, {Koo}, \& {Mandolesi}}]{1999A&A...351L..37E}
{Elbaz}, D., {Cesarsky}, C.~J., {Fadda}, D., {et~al.} 1999, \aap, 351, L37

\bibitem[{{Farrah} {et~al.}(2008){Farrah}, {Lonsdale}, {Weedman}, {Spoon},
  {Rowan-Robinson}, {Polletta}, {Oliver}, {Houck}, \&
  {Smith}}]{2008ApJ...677..957F}
{Farrah}, D., {Lonsdale}, C.~J., {Weedman}, D.~W., {et~al.} 2008, \apj, 677,
  957

\bibitem[{{Georgakakis} {et~al.}(2007){Georgakakis}, {Rowan-Robinson},
  {Babbedge}, \& {Georgantopoulos}}]{2007MNRAS.377..203G}
{Georgakakis}, A., {Rowan-Robinson}, M., {Babbedge}, T.~S.~R., \&
  {Georgantopoulos}, I. 2007, \mnras, 377, 203

\bibitem[{{Helou} \& {Beichman}(1990)}]{1990LIACo..29..117H}
{Helou}, G. \& {Beichman}, C.~A. 1990, in Liege International Astrophysical
  Colloquia, Vol.~29, Liege International Astrophysical Colloquia, ed.
  B.~{Kaldeich}, 117--123

\bibitem[{{Houck} {et~al.}(2004){Houck}, {Roellig}, {van Cleve}, {Forrest},
  {Herter}, {Lawrence}, {Matthews}, {Reitsema}, {Soifer}, {Watson}, {Weedman},
  {Huisjen}, {Troeltzsch}, {Barry}, {Bernard-Salas}, {Blacken}, {Brandl},
  {Charmandaris}, {Devost}, {Gull}, {Hall}, {Henderson}, {Higdon}, {Pirger},
  {Schoenwald}, {Sloan}, {Uchida}, {Appleton}, {Armus}, {Burgdorf},
  {Fajardo-Acosta}, {Grillmair}, {Ingalls}, {Morris}, \&
  {Teplitz}}]{2004ApJS..154...18H}
{Houck}, J.~R., {Roellig}, T.~L., {van Cleve}, J., {et~al.} 2004, \apjs, 154,
  18

\bibitem[{{Houck} {et~al.}(2005){Houck}, {Soifer}, {Weedman}, {Higdon},
  {Higdon}, {Herter}, {Brown}, {Dey}, {Jannuzi}, {Le Floc'h}, {Rieke}, {Armus},
  {Charmandaris}, {Brandl}, \& {Teplitz}}]{2005ApJ...622L.105H}
{Houck}, J.~R., {Soifer}, B.~T., {Weedman}, D., {et~al.} 2005, \apjl, 622, L105

\bibitem[{{Houck} {et~al.}(2007){Houck}, {Weedman}, {Le Floc'h}, \&
  {Hao}}]{2007ApJ...671..323H}
{Houck}, J.~R., {Weedman}, D.~W., {Le Floc'h}, E., \& {Hao}, L. 2007, \apj,
  671, 323

\bibitem[{{Ivison} {et~al.}(2004){Ivison}, {Greve}, {Serjeant}, {Bertoldi},
  {Egami}, {Mortier}, {Alonso-Herrero}, {Barmby}, {Bei}, {Dole}, {Engelbracht},
  {Fazio}, {Frayer}, {Gordon}, {Hines}, {Huang}, {Le Floc'h}, {Misselt},
  {Miyazaki}, {Morrison}, {Papovich}, {P{\'e}rez-Gonz{\'a}lez}, {Rieke},
  {Rieke}, {Rigby}, {Rigopoulou}, {Smail}, {Wilson}, \&
  {Willner}}]{2004ApJS..154..124I}
{Ivison}, R.~J., {Greve}, T.~R., {Serjeant}, S., {et~al.} 2004, \apjs, 154, 124

\bibitem[{{Kim} \& {Sanders}(1998)}]{1998ApJS..119...41K}
{Kim}, D.-C. \& {Sanders}, D.~B. 1998, \apjs, 119, 41

\bibitem[{{Lagache} {et~al.}(2003){Lagache}, {Dole}, \&
  {Puget}}]{2003MNRAS.338..555L}
{Lagache}, G., {Dole}, H., \& {Puget}, J.-L. 2003, \mnras, 338, 555

\bibitem[{{Le Floc'h} {et~al.}(2005){Le Floc'h}, {Papovich}, {Dole}, {Bell},
  {Lagache}, {Rieke}, {Egami}, {P{\'e}rez-Gonz{\'a}lez}, {Alonso-Herrero},
  {Rieke}, {Blaylock}, {Engelbracht}, {Gordon}, {Hines}, {Misselt}, {Morrison},
  \& {Mould}}]{2005ApJ...632..169L}
{Le Floc'h}, E., {Papovich}, C., {Dole}, H., {et~al.} 2005, \apj, 632, 169

\bibitem[{{Le Floc'h} {et~al.}(2004){Le Floc'h}, {P{\'e}rez-Gonz{\'a}lez},
  {Rieke}, {Papovich}, {Huang}, {Barmby}, {Dole}, {Egami}, {Alonso-Herrero},
  {Wilson}, {Miyazaki}, {Rigby}, {Bei}, {Blaylock}, {Engelbracht}, {Fazio},
  {Frayer}, {Gordon}, {Hines}, {Misselt}, {Morrison}, {Muzerolle}, {Rieke},
  {Rigopoulou}, {Su}, {Willner}, \& {Young}}]{2004ApJS..154..170L}
{Le Floc'h}, E., {P{\'e}rez-Gonz{\'a}lez}, P.~G., {Rieke}, G.~H., {et~al.}
  2004, \apjs, 154, 170

\bibitem[{{Lu} {et~al.}(2003){Lu}, {Helou}, {Werner}, {Dinerstein}, {Dale},
  {Silbermann}, {Malhotra}, {Beichman}, \& {Jarrett}}]{2003ApJ...588..199L}
{Lu}, N., {Helou}, G., {Werner}, M.~W., {et~al.} 2003, \apj, 588, 199

\bibitem[{{Men{\'e}ndez-Delmestre} {et~al.}(2007){Men{\'e}ndez-Delmestre},
  {Blain}, {Alexander}, {Smail}, {Armus}, {Chapman}, {Frayer}, {Ivison}, \&
  {Teplitz}}]{2007ApJ...655L..65M}
{Men{\'e}ndez-Delmestre}, K., {Blain}, A.~W., {Alexander}, D.~M., {et~al.}
  2007, \apjl, 655, L65

\bibitem[{{Narron} {et~al.}(2007){Narron}, {Ogle}, \&
  {Laher}}]{2007ASPC..376..437N}
{Narron}, R., {Ogle}, P., \& {Laher}, R.~R. 2007, in Astronomical Society of
  the Pacific Conference Series, Vol. 376, Astronomical Data Analysis Software
  and Systems XVI, ed. R.~A. {Shaw}, F.~{Hill}, \& D.~J. {Bell}, 437--+

\bibitem[{{Ogle} {et~al.}(2007){Ogle}, {Antonucci}, {Appleton}, \&
  {Whysong}}]{2007ApJ...668..699O}
{Ogle}, P., {Antonucci}, R., {Appleton}, P.~N., \& {Whysong}, D. 2007, \apj,
  668, 699

\bibitem[{{Ogle} {et~al.}(2006){Ogle}, {Whysong}, \&
  {Antonucci}}]{2006ApJ...647..161O}
{Ogle}, P., {Whysong}, D., \& {Antonucci}, R. 2006, \apj, 647, 161

\bibitem[{{Oliver} {et~al.}(2000){Oliver}, {Rowan-Robinson}, {Alexander},
  {Almaini}, {Balcells}, {Baker}, {Barcons}, {Barden}, {Bellas-Velidis},
  {Cabrera-Guerra}, {Carballo}, {Cesarsky}, {Ciliegi}, {Clements}, {Crockett},
  {Danese}, {Dapergolas}, {Drolias}, {Eaton}, {Efstathiou}, {Egami}, {Elbaz},
  {Fadda}, {Fox}, {Franceschini}, {Genzel}, {Goldschmidt}, {Graham},
  {Gonzalez-Serrano}, {Gonzalez-Solares}, {Granato}, {Gruppioni},
  {Herbstmeier}, {H{\'e}raudeau}, {Joshi}, {Kontizas}, {Kontizas},
  {Kotilainen}, {Kunze}, {La Franca}, {Lari}, {Lawrence}, {Lemke},
  {Linden-V{\o}rnle}, {Mann}, {M{\'a}rquez}, {Masegosa}, {Mattila}, {McMahon},
  {Miley}, {Missoulis}, {Mobasher}, {Morel}, {N{\o}rgaard-Nielsen}, {Omont},
  {Papadopoulos}, {Perez-Fournon}, {Puget}, {Rigopoulou}, {Rocca-Volmerange},
  {Serjeant}, {Silva}, {Sumner}, {Surace}, {Vaisanen}, {van der Werf}, {Verma},
  {Vigroux}, {Villar-Martin}, \& {Willott}}]{2000MNRAS.316..749O}
{Oliver}, S., {Rowan-Robinson}, M., {Alexander}, D.~M., {et~al.} 2000, \mnras,
  316, 749

\bibitem[{{Papovich} {et~al.}(2007){Papovich}, {Rudnick}, {Le Floc'h}, {van
  Dokkum}, {Rieke}, {Taylor}, {Armus}, {Gawiser}, {Huang}, {Marcillac}, \&
  {Franx}}]{2007ApJ...668...45P}
{Papovich}, C., {Rudnick}, G., {Le Floc'h}, E., {et~al.} 2007, \apj, 668, 45

\bibitem[{{Pope} {et~al.}(2008){Pope}, {Chary}, {Alexander}, {Armus},
  {Dickinson}, {Elbaz}, {Frayer}, {Scott}, \& {Teplitz}}]{2008ApJ...675.1171P}
{Pope}, A., {Chary}, R.-R., {Alexander}, D.~M., {et~al.} 2008, \apj, 675, 1171

\bibitem[{{Pope} {et~al.}(2006){Pope}, {Scott}, {Dickinson}, {Chary},
  {Morrison}, {Borys}, {Sajina}, {Alexander}, {Daddi}, {Frayer}, {MacDonald},
  \& {Stern}}]{2006MNRAS.370.1185P}
{Pope}, A., {Scott}, D., {Dickinson}, M., {et~al.} 2006, \mnras, 370, 1185

\bibitem[{{Reddy} {et~al.}(2006){Reddy}, {Steidel}, {Erb}, {Shapley}, \&
  {Pettini}}]{2006ApJ...653.1004R}
{Reddy}, N.~A., {Steidel}, C.~C., {Erb}, D.~K., {Shapley}, A.~E., \& {Pettini},
  M. 2006, \apj, 653, 1004

\bibitem[{{Rieke} {et~al.}(2004){Rieke}, {Young}, {Engelbracht}, {Kelly},
  {Low}, {Haller}, {Beeman}, {Gordon}, {Stansberry}, {Misselt}, {Cadien},
  {Morrison}, {Rivlis}, {Latter}, {Noriega-Crespo}, {Padgett}, {Stapelfeldt},
  {Hines}, {Egami}, {Muzerolle}, {Alonso-Herrero}, {Blaylock}, {Dole}, {Hinz},
  {Le Floc'h}, {Papovich}, {P{\'e}rez-Gonz{\'a}lez}, {Smith}, {Su}, {Bennett},
  {Frayer}, {Henderson}, {Lu}, {Masci}, {Pesenson}, {Rebull}, {Rho}, {Keene},
  {Stolovy}, {Wachter}, {Wheaton}, {Werner}, \&
  {Richards}}]{2004ApJS..154...25R}
{Rieke}, G.~H., {Young}, E.~T., {Engelbracht}, C.~W., {et~al.} 2004, \apjs,
  154, 25

\bibitem[{{Rowan-Robinson}(2001)}]{2001ApJ...549..745R}
{Rowan-Robinson}, M. 2001, \apj, 549, 745

\bibitem[{{Rowan-Robinson} {et~al.}(2004){Rowan-Robinson}, {Lari},
  {Perez-Fournon}, {Gonzalez-Solares}, {La Franca}, {Vaccari}, {Oliver},
  {Gruppioni}, {Ciliegi}, {H{\'e}raudeau}, {Serjeant}, {Efstathiou},
  {Babbedge}, {Matute}, {Pozzi}, {Franceschini}, {Vaisanen}, {Afonso-Luis},
  {Alexander}, {Almaini}, {Baker}, {Basilakos}, {Barden}, {del Burgo},
  {Bellas-Velidis}, {Cabrera-Guerra}, {Carballo}, {Cesarsky}, {Clements},
  {Crockett}, {Danese}, {Dapergolas}, {Drolias}, {Eaton}, {Egami}, {Elbaz},
  {Fadda}, {Fox}, {Genzel}, {Goldschmidt}, {Gonzalez-Serrano}, {Graham},
  {Granato}, {Hatziminaoglou}, {Herbstmeier}, {Joshi}, {Kontizas}, {Kontizas},
  {Kotilainen}, {Kunze}, {Lawrence}, {Lemke}, {Linden-V{\o}rnle}, {Mann},
  {M{\'a}rquez}, {Masegosa}, {McMahon}, {Miley}, {Missoulis}, {Mobasher},
  {Morel}, {N{\o}rgaard-Nielsen}, {Omont}, {Papadopoulos}, {Puget},
  {Rigopoulou}, {Rocca-Volmerange}, {Sedgwick}, {Silva}, {Sumner}, {Surace},
  {Vila-Vilaro}, {van der Werf}, {Verma}, {Vigroux}, {Villar-Martin},
  {Willott}, {Carrami{\~n}ana}, \& {Mujica}}]{2004MNRAS.351.1290R}
{Rowan-Robinson}, M., {Lari}, C., {Perez-Fournon}, I., {et~al.} 2004, \mnras,
  351, 1290

\bibitem[{{Sajina} {et~al.}(2007){Sajina}, {Yan}, {Armus}, {Choi}, {Fadda},
  {Helou}, \& {Spoon}}]{2007ApJ...664..713S}
{Sajina}, A., {Yan}, L., {Armus}, L., {et~al.} 2007, \apj, 664, 713

\bibitem[{{Shi} {et~al.}(2007){Shi}, {Ogle}, {Rieke}, {Antonucci}, {Hines},
  {Smith}, {Low}, {Bouwman}, \& {Willmer}}]{2007ApJ...669..841S}
{Shi}, Y., {Ogle}, P., {Rieke}, G.~H., {et~al.} 2007, \apj, 669, 841

\bibitem[{{Smith} {et~al.}(2007{\natexlab{a}}){Smith}, {Armus}, {Dale},
  {Roussel}, {Sheth}, {Buckalew}, {Jarrett}, {Helou}, \&
  {Kennicutt}}]{2007PASP..119.1133S}
{Smith}, J.~D.~T., {Armus}, L., {Dale}, D.~A., {et~al.} 2007{\natexlab{a}},
  \pasp, 119, 1133

\bibitem[{{Smith} {et~al.}(2007{\natexlab{b}}){Smith}, {Draine}, {Dale},
  {Moustakas}, {Kennicutt}, {Helou}, {Armus}, {Roussel}, {Sheth}, {Bendo},
  {Buckalew}, {Calzetti}, {Engelbracht}, {Gordon}, {Hollenbach}, {Li},
  {Malhotra}, {Murphy}, \& {Walter}}]{2007ApJ...656..770S}
{Smith}, J.~D.~T., {Draine}, B.~T., {Dale}, D.~A., {et~al.} 2007{\natexlab{b}},
  \apj, 656, 770

\bibitem[{{Soifer} {et~al.}(2008){Soifer}, {Helou}, \&
  {Werner}}]{2008ARA&A..46..201S}
{Soifer}, B.~T., {Helou}, G., \& {Werner}, M. 2008, \araa, 46, 201

\bibitem[{{Stern} {et~al.}(2005){Stern}, {Eisenhardt}, {Gorjian}, {Kochanek},
  {Caldwell}, {Eisenstein}, {Brodwin}, {Brown}, {Cool}, {Dey}, {Green},
  {Jannuzi}, {Murray}, {Pahre}, \& {Willner}}]{2005ApJ...631..163S}
{Stern}, D., {Eisenhardt}, P., {Gorjian}, V., {et~al.} 2005, \apj, 631, 163

\bibitem[{{Teplitz} {et~al.}(2007){Teplitz}, {Desai}, {Armus}, {Chary},
  {Marshall}, {Colbert}, {Frayer}, {Pope}, {Blain}, {Spoon}, {Charmandaris}, \&
  {Scott}}]{2007ApJ...659..941T}
{Teplitz}, H.~I., {Desai}, V., {Armus}, L., {et~al.} 2007, \apj, 659, 941

\bibitem[{{Weedman} {et~al.}(2006){Weedman}, {Soifer}, {Hao}, {Higdon},
  {Higdon}, {Houck}, {Le Floc'h}, {Brown}, {Dey}, {Jannuzi}, {Rieke}, {Desai},
  {Bian}, {Thompson}, {Armus}, {Teplitz}, {Eisenhardt}, \&
  {Willner}}]{2006ApJ...651..101W}
{Weedman}, D.~W., {Soifer}, B.~T., {Hao}, L., {et~al.} 2006, \apj, 651, 101

\bibitem[{{Werner} {et~al.}(2004){Werner}, {Gallagher}, \&
  {Irace}}]{2004AdSpR..34..600W}
{Werner}, M.~W., {Gallagher}, D.~B., \& {Irace}, W.~R. 2004, Advances in Space
  Research, 34, 600

\bibitem[{{Wirth} {et~al.}(2004){Wirth}, {Willmer}, {Amico}, {Chaffee},
  {Goodrich}, {Kwok}, {Lyke}, {Mader}, {Tran}, {Barger}, {Cowie}, {Capak},
  {Coil}, {Cooper}, {Conrad}, {Davis}, {Faber}, {Hu}, {Koo}, {Le Mignant},
  {Newman}, \& {Songaila}}]{2004AJ....127.3121W}
{Wirth}, G.~D., {Willmer}, C.~N.~A., {Amico}, P., {et~al.} 2004, \aj, 127, 3121

\bibitem[{{Wu} {et~al.}(2006){Wu}, {Charmandaris}, {Hao}, {Brandl},
  {Bernard-Salas}, {Spoon}, \& {Houck}}]{2006ApJ...639..157W}
{Wu}, Y., {Charmandaris}, V., {Hao}, L., {et~al.} 2006, \apj, 639, 157

\bibitem[{{Yan} {et~al.}(2004){Yan}, {Choi}, {Fadda}, {Marleau}, {Soifer},
  {Im}, {Armus}, {Frayer}, {Storrie-Lombardi}, {Thompson}, {Teplitz}, {Helou},
  {Appleton}, {Chapman}, {Fan}, {Heinrichsen}, {Lacy}, {Shupe}, {Squires},
  {Surace}, \& {Wilson}}]{2004ApJS..154...75Y}
{Yan}, L., {Choi}, P.~I., {Fadda}, D., {et~al.} 2004, \apjs, 154, 75

\bibitem[{{Yan} {et~al.}(2007){Yan}, {Sajina}, {Fadda}, {Choi}, {Armus},
  {Helou}, {Teplitz}, {Frayer}, \& {Surace}}]{2007ApJ...658..778Y}
{Yan}, L., {Sajina}, A., {Fadda}, D., {et~al.} 2007, \apj, 658, 778

\end{thebibliography}

\pagebreak

\begin{deluxetable}{lccccccc}
\tabletypesize{ \footnotesize}
\tablecaption{Source extraction information. \label{SrcExtrTab}}
\tablehead{ 
\colhead{ID} & 
\colhead{Field} &  
\colhead{$RA_{IRS}$} &  
\colhead{$DEC_{IRS}$} & 
\colhead{SNR \tablenotemark{a}} & 
\colhead{Peak SNR \tablenotemark{b}} & 
\colhead{Peak Wavelength \tablenotemark{b}} & 
\colhead{$\Delta\lambda_{2\sigma}$ \tablenotemark{c}} \\
 & & degrees & degrees & & $\mu m$ & $\mu m$ & $\mu m$
}
\startdata
SUUSS 1      &     LL1 &  189.176 &  62.2894 &  13.3 &   5.7 &   32.22 &    4.67 \\
SUUSS 2      &     LL1 &  189.183 &  62.2848 &  45.9 &  10.2 &   31.67 &   11.43 \\
SUUSS 3      &     LL1 &  189.187 &  62.2873 &   5.0 &   7.5 &   23.00 &    2.90 \\
SUUSS 4      &     LL1 &  189.191 &  62.2835 &   7.7 &   4.7 &   31.67 &    3.54 \\
SUUSS 5      &     LL1 &  189.198 &  62.2844 &   3.1 &   3.8 &   26.58 &    1.77 \\
SUUSS 6      &     LL1 &  189.205 &  62.2947 &   2.2 &   1.5 &   31.67 &    0.16 \\
SUUSS 7      &     LL1 &  189.208 &  62.2764 &   7.9 &   5.2 &   29.11 &    3.06 \\
SUUSS 8      &     LL1 &  189.210 &  62.2827 &   5.3 &   6.2 &   31.85 &    3.38 \\
SUUSS 9\tablenotemark{d}      &     LL1 &  189.210 &  62.2869 &   1.2 &   3.2 &   33.50 &    0.16 \\
SUUSS 10    &     LL1 &  189.217 &  62.2936 &   3.6 &   5.9 &   31.12 &    1.77 \\
SUUSS 11    &     LL1 &  189.225 &  62.2907 &  11.5 &   8.7 &   22.65 &    5.64 \\
SUUSS 12    &     LL1 &  189.229 &  62.2829 &  27.5 &   8.6 &   27.84 &    7.89 \\
SUUSS 13    &     LL1 &  189.232 &  62.2822 &   3.8 &   7.1 &   24.43 &    4.19 \\
SUUSS 14    &     LL1 &  189.237 &  62.2788 &  19.3 &   6.3 &   24.07 &    7.25 \\
SUUSS 15    &     LL1 &  189.243 &  62.2638 &  22.6 &   6.2 &   33.50 &    7.73 \\
SUUSS 16    &     LL1 &  189.244 &  62.2634 &  19.8 &   5.2 &   33.50 &    7.08 \\
SUUSS 17    &     LL1 &  189.244 &  62.2771 &  67.6 &  15.4 &   33.32 &   12.56 \\
SUUSS 18    &     LL1 &  189.246 &  62.2610 &   6.8 &   7.4 &   30.57 &    5.64 \\
SUUSS 19    &     LL1 &  189.251 &  62.2687 &  21.4 &   8.1 &   31.85 &    8.21 \\
SUUSS 20    &     LL1 &  189.251 &  62.2714 &  93.6 &  29.4 &   24.43 &   12.56 \\
\\
\hline
\\
SUUSS 21    &     LL2 &  189.259 &  62.2532 &  14.3 &   6.6 &   15.28 &    3.08 \\
SUUSS 22    &     LL2 &  189.263 &  62.2519 &  10.3 &   4.3 &   20.32 &    1.08 \\
SUUSS 23    &     LL2 &  189.275 &  62.2552 &   6.2 &   3.9 &   15.92 &    1.08 \\
SUUSS 24    &     LL2 &  189.283 &  62.2582 &   8.8 &   8.9 &   18.94 &    2.00 \\
SUUSS 25    &     LL2 &  189.284 &  62.2539 &  37.6 &  10.4 &   15.83 &    4.75 \\
SUUSS 26    &     LL2 &  189.288 &  62.2524 &  31.9 &  17.4 &   17.29 &    4.67 \\
SUUSS 27    &     LL2 &  189.299 &  62.2538 &  11.0 &   5.5 &   16.56 &    2.58 \\
SUUSS 28    &     LL2 &  189.305 &  62.2548 &  22.8 &   8.8 &   16.28 &    3.33 \\
SUUSS 29    &     LL2 &  189.306 &  62.2472 &  32.3 &  16.4 &   14.91 &    3.92 \\
SUUSS 30    &     LL2 &  189.307 &  62.2531 &  33.5 &  15.4 &   17.11 &    4.42 \\
SUUSS 31    &     LL2 &  189.309 &  62.2403 &  14.5 &   7.2 &   19.95 &    2.83 \\
SUUSS 32    &     LL2 &  189.315 &  62.2377 &  14.9 &   7.9 &   19.77 &    2.58 \\
SUUSS 33    &     LL2 &  189.320 &  62.2292 &  15.9 &   7.8 &   18.03 &    3.00 \\
SUUSS 34    &     LL2 &  189.320 &  62.2307 &  12.1 &   9.6 &   18.03 &    2.83 \\
SUUSS 35    &     LL2 &  189.322 &  62.2478 &   4.8 &   6.2 &   15.00 &    1.67 \\
SUUSS 36    &     LL2 &  189.324 &  62.2328 &   6.7 &   8.4 &   18.12 &    1.75 \\
SUUSS 37    &     LL2 &  189.324 &  62.2439 &  32.5 &  12.3 &   15.83 &    4.75 \\
SUUSS 38    &     LL2 &  189.330 &  62.2261 &   6.8 &  10.2 &   18.58 &    2.25 \\
SUUSS 39    &     LL2 &  189.333 &  62.2269 &  24.5 &  14.9 &   16.65 &    2.50 \\
SUUSS 40    &     LL2 &  189.334 &  62.2300 &   5.4 &   4.8 &   18.48 &    1.17 \\
SUUSS 41    &     LL2 &  189.335 &  62.2313 &  11.3 &   5.7 &   15.18 &    2.58 \\
SUUSS 42    &     LL2 &  189.340 &  62.2291 &  24.3 &  17.5 &   15.37 &    2.75 \\
SUUSS 43    &     LL2 &  189.340 &  62.2415 &   6.9 &   5.8 &   18.12 &    1.67 \\
SUUSS 44    &     LL2 &  189.345 &  62.2318 &  34.5 &  20.3 &   15.37 &    3.67 \\
SUUSS 45    &     LL2 &  189.346 &  62.2385 &  21.7 &  11.4 &   15.09 &    3.50 \\
\enddata
\tablenotetext{a}{Signal-to-noise ratio estimation when fluxes are integrated over the whole spectral range available (14-20$\mu$m for LL2 and 20-35$\mu$m for LL1).}
\tablenotetext{b}{Maximum Signal-to-noise ratio achieved in each spectrum and the wavelength at which it peaks. The spectra were smoothed using a comprehensive hanning window over 3 channels in order to avoid contribution from any potentially remaining hot pixel value.}
\tablenotetext{c}{Contracted Spectral bandwidth over which the spectra have flux densities higher than 2$\sigma$. }
\tablenotetext{d}{SUUSS 9 is the [SIV] line source discussed in sect. 4.3 which explains the low integrated SNR compared to its peak SNR.}
\end{deluxetable}

\clearpage

\begin{deluxetable}{lccccccc}
\tabletypesize{ \footnotesize}
\tablecaption{Spitzer Multi-Wavelength Data. \label{SrcMultiwTab}}
\tablehead{ 
\colhead{name} &    
\colhead{$RA_{IRS}$} &  
\colhead{$DEC_{IRS}$} & 
\colhead{$S_{3.6}$ \tablenotemark{a}}     & 
\colhead{$S_{4.5}$ \tablenotemark{a}}      & 
\colhead{$S_{5.8}$ \tablenotemark{a}}      & 
\colhead{$S_8$ \tablenotemark{a}}      & 
\colhead{$S_{24}$ \tablenotemark{b}} \\
 & degrees & degrees & $\mu Jy$ & $\mu Jy$ & $\mu Jy$ & $\mu Jy$ & $\mu Jy$ }
\startdata
SUUSS 1  &  189.176 &  62.2894 &   33.60$\pm$   0.04 &   38.60$\pm$   0.05 &   34.40$\pm$   0.30 &   23.60$\pm$   0.32 &  115.0$\pm$   5.5 \\
SUUSS 2  &  189.183 &  62.2848 &   31.00$\pm$   0.04 &   24.70$\pm$   0.05 &   22.70$\pm$   0.30 &   41.20$\pm$   0.32 &  219.0$\pm$   5.9 \\
SUUSS 3  &  189.187 &  62.2873 &   12.00$\pm$   0.03 &   13.40$\pm$   0.05 &   14.80$\pm$   0.30 &   10.10$\pm$   0.32 &  162.0$\pm$   4.2 \\
SUUSS 4  &  189.191 &  62.2835 &   29.30$\pm$   0.04 &   22.70$\pm$   0.05 &   18.20$\pm$   0.30 &   14.90$\pm$   0.32 &   80.6$\pm$   4.5 \\
SUUSS 5  &  189.198 &  62.2844 &    5.10$\pm$   0.03 &    6.35$\pm$   0.05 &    8.32$\pm$   0.30 &    6.37$\pm$   0.32 &   56.4$\pm$   4.9 \\
SUUSS 6  &  189.205 &  62.2947 &    5.13$\pm$   0.03 &    5.66$\pm$   0.05 &    6.49$\pm$   0.33 &    6.00$\pm$   0.35 &   54.8$\pm$   5.4 \\
SUUSS 7  &  189.208 &  62.2764 &   18.30$\pm$   0.03 &   14.90$\pm$   0.05 &   11.80$\pm$   0.30 &   20.50$\pm$   0.33 &   60.0$\pm$   5.9 \\
SUUSS 8  &  189.210 &  62.2827 &   10.60$\pm$   0.03 &   10.80$\pm$   0.05 &    8.64$\pm$   0.31 &    7.68$\pm$   0.33 &   51.6$\pm$   6.8 \\
SUUSS 9  &  189.210 &  62.2869 &    \nodata &    \nodata &    \nodata &    \nodata &      \nodata \\
SUUSS 10 &  189.217 &  62.2936 &    3.56$\pm$   0.03 &    2.14$\pm$   0.05 &    0.92$\pm$   0.34 &    0.54$\pm$   0.36 &      \nodata \\
SUUSS 11 &  189.225 &  62.2907 &    7.68$\pm$   0.03 &    8.02$\pm$   0.05 &    7.20$\pm$   0.34 &    6.16$\pm$   0.37 &   73.0$\pm$   5.8 \\
SUUSS 12 &  189.229 &  62.2829 &   45.70$\pm$   0.04 &   34.80$\pm$   0.06 &   28.80$\pm$   0.33 &   34.10$\pm$   0.36 &  103.0$\pm$   5.7 \\
SUUSS 13 &  189.232 &  62.2822 &   16.70$\pm$   0.04 &   13.00$\pm$   0.05 &   11.60$\pm$   0.33 &    9.09$\pm$   0.36 &   67.1$\pm$   5.0 \\
SUUSS 14 &  189.237 &  62.2788 &    6.84$\pm$   0.03 &    7.72$\pm$   0.05 &    9.91$\pm$   0.33 &    6.69$\pm$   0.35 &   74.5$\pm$   5.7 \\
SUUSS 15 &  189.243 &  62.2638 &   12.10$\pm$   0.03 &   10.70$\pm$   0.05 &    7.69$\pm$   0.30 &   13.90$\pm$   0.32 &  102.0$\pm$   5.5 \\
SUUSS 16 &  189.244 &  62.2634 &   12.10$\pm$   0.03 &   10.70$\pm$   0.05 &    7.69$\pm$   0.30 &   13.90$\pm$   0.32 &  105.0$\pm$   6.1 \\
SUUSS 17 &  189.244 &  62.2771 &   66.70$\pm$   0.05 &   60.20$\pm$   0.06 &   41.70$\pm$   0.34 &  120.00$\pm$   0.37 &  275.0$\pm$   5.7 \\
SUUSS 18 &  189.246 &  62.2610 &   13.20$\pm$   0.03 &   13.20$\pm$   0.05 &   11.60$\pm$   0.29 &    7.69$\pm$   0.32 &   47.2$\pm$   6.3 \\
SUUSS 19 &  189.251 &  62.2687 &   15.70$\pm$   0.03 &   11.30$\pm$   0.05 &    8.33$\pm$   0.32 &    8.69$\pm$   0.34 &  103.0$\pm$   8.5 \\
SUUSS 20 &  189.251 &  62.2714 &   36.60$\pm$   0.04 &   27.20$\pm$   0.06 &   29.90$\pm$   0.33 &   26.50$\pm$   0.35 &  502.0$\pm$   7.7 \\
\\
\hline
\\
SUUSS 21 &  189.259 &  62.2532 &   15.50$\pm$   0.03 &   10.60$\pm$   0.05 &    8.76$\pm$   0.29 &    7.55$\pm$   0.31 &   71.0$\pm$   6.2 \\
SUUSS 22 &  189.263 &  62.2519 &    6.58$\pm$   0.03 &    5.94$\pm$   0.05 &    2.66$\pm$   0.29 &    3.52$\pm$   0.31 &      \nodata \\
SUUSS 23 &  189.275 &  62.2552 &   13.20$\pm$   0.03 &    9.33$\pm$   0.05 &    9.37$\pm$   0.31 &    7.66$\pm$   0.33 &   59.2$\pm$   6.3 \\
SUUSS 24 &  189.283 &  62.2582 &    8.79$\pm$   0.03 &   11.50$\pm$   0.05 &   14.30$\pm$   0.32 &   10.70$\pm$   0.36 &  219.0$\pm$   6.5 \\
SUUSS 25 &  189.284 &  62.2539 &   27.40$\pm$   0.04 &   18.80$\pm$   0.05 &   20.10$\pm$   0.32 &   14.80$\pm$   0.35 &  198.0$\pm$   5.9 \\
SUUSS 26 &  189.288 &  62.2524 &   26.40$\pm$   0.04 &   24.50$\pm$   0.06 &   18.50$\pm$   0.32 &   17.90$\pm$   0.35 &  123.0$\pm$   5.4 \\
SUUSS 27 &  189.299 &  62.2538 &   21.50$\pm$   0.04 &   18.70$\pm$   0.06 &   13.30$\pm$   0.34 &   29.90$\pm$   0.38 &   88.1$\pm$   6.2 \\
SUUSS 28 &  189.305 &  62.2548 &   16.50$\pm$   0.04 &   12.00$\pm$   0.06 &   12.30$\pm$   0.35 &    9.93$\pm$   0.41 &  169.0$\pm$   6.3 \\
SUUSS 29 &  189.306 &  62.2472 &   74.10$\pm$   0.05 &   54.80$\pm$   0.06 &   46.30$\pm$   0.34 &   33.00$\pm$   0.38 &  241.0$\pm$   7.0 \\
SUUSS 30 &  189.307 &  62.2531 &   34.70$\pm$   0.04 &   27.70$\pm$   0.06 &   26.30$\pm$   0.35 &   32.80$\pm$   0.41 &  149.0$\pm$   6.2 \\
SUUSS 31 &  189.309 &  62.2403 &   31.50$\pm$   0.04 &   25.40$\pm$   0.06 &   21.70$\pm$   0.33 &   21.80$\pm$   0.36 &  103.0$\pm$   6.2 \\
SUUSS 32 &  189.315 &  62.2377 &   16.30$\pm$   0.04 &   11.00$\pm$   0.06 &   12.20$\pm$   0.33 &    7.94$\pm$   0.37 &   76.6$\pm$   6.5 \\
SUUSS 33 &  189.320 &  62.2292 &    2.32$\pm$   0.03 &    1.68$\pm$   0.05 &    1.34$\pm$   0.31 &    0.66$\pm$   0.35 &      \nodata \\
SUUSS 34 &  189.320 &  62.2307 &   18.00$\pm$   0.04 &   15.60$\pm$   0.05 &   11.10$\pm$   0.32 &   12.60$\pm$   0.36 &   66.0$\pm$   5.3 \\
SUUSS 35 &  189.322 &  62.2478 &   10.20$\pm$   0.04 &    7.19$\pm$   0.06 &    7.06$\pm$   0.36 &    5.71$\pm$   0.42 &   81.1$\pm$   6.4 \\
SUUSS 36 &  189.324 &  62.2328 &   14.20$\pm$   0.04 &   11.90$\pm$   0.06 &    8.11$\pm$   0.33 &   16.70$\pm$   0.37 &   54.5$\pm$   5.6 \\
SUUSS 37 &  189.324 &  62.2439 &   22.50$\pm$   0.04 &   20.80$\pm$   0.06 &   15.50$\pm$   0.36 &   38.50$\pm$   0.42 &  119.0$\pm$   6.7 \\
SUUSS 38 &  189.330 &  62.2261 &    1.33$\pm$   0.03 &    1.38$\pm$   0.05 &    0.833$\pm$   0.31 &    1.87$\pm$   0.36 &      \nodata \\
SUUSS 39 &  189.333 &  62.2269 &   77.90$\pm$   0.05 &   61.30$\pm$   0.06 &   50.60$\pm$   0.33 &   63.80$\pm$   0.37 &  185.0$\pm$   4.6 \\
SUUSS 40 &  189.334 &  62.2300 &   35.50$\pm$   0.04 &   25.50$\pm$   0.06 &   23.00$\pm$   0.34 &   17.30$\pm$   0.39 &      \nodata \\
SUUSS 41 &  189.335 &  62.2313 &   35.50$\pm$   0.04 &   25.50$\pm$   0.06 &   23.00$\pm$   0.34 &   17.30$\pm$   0.39 &  142.0$\pm$   5.6 \\
SUUSS 42 &  189.340 &  62.2291 &   43.50$\pm$   0.05 &   34.90$\pm$   0.06 &   26.70$\pm$   0.34 &   25.60$\pm$   0.39 &  217.0$\pm$   6.4 \\
SUUSS 43 &  189.340 &  62.2415 &    6.52$\pm$   0.04 &    5.31$\pm$   0.06 &    4.34$\pm$   0.38 &    3.01$\pm$   0.43 &      \nodata \\
SUUSS 44 &  189.345 &  62.2318 &    2.60$\pm$   0.04 &    1.79$\pm$   0.06 &    1.17$\pm$   0.35 &    1.66$\pm$   0.42 &  301.0$\pm$   4.7 \\
SUUSS 45 &  189.346 &  62.2385 &   32.60$\pm$   0.04 &   24.90$\pm$   0.07 &   19.20$\pm$   0.38 &   16.80$\pm$   0.43 &  148.0$\pm$   5.8 \\
\enddata
\tablenotetext{a}{IRAC photometry from GOODS legacy program, \cite{2003mglh.conf..324D}}
\tablenotetext{b}{MIPS 24$\mu m$ photometry from GOODS-North, \cite{2007ASPC..380..375C}}
\end{deluxetable}

\clearpage

\begin{deluxetable}{lccccc}
\tabletypesize{\footnotesize}
\tablecaption{Templates IR properties. \label{TplCcorrTab}}
\tablehead{
\colhead{Template} & 
\colhead{z} & 
\colhead{$6.2\mu m$ EQW} & 
\colhead{$7.7\mu m$ EQW} &
\colhead{$log(L_{IR} [L_\odot])$} & 
\colhead{Ref.}  \\
\colhead{} & \colhead{}  & \colhead{$\mu m$}  & \colhead{$\mu m$} & \colhead{} & \colhead{} 
}
\startdata
Mrk231 & 0.042 & 0.011 & \nodata & 12.57 & 1 \\
Mrk273 & 0.038 & 0.171 & \nodata & 12.15 & 1 \\
Mrk463 & 0.051 & \nodata & 0.029 & 11.70 & 2 \\
Mrk1014 & 0.163 & 0.048 & 0.135 & 12.53 & 2, 7 \\
NGC6240 & 0.025 & 0.52 & 2.60 & 11.85 & 3 \\
UGC5101 & 0.039 & 0.188 & 0.419 & 12.00 & 1, 2 \\
Arp220 & 0.018 & 0.253 & \nodata & 12.16 & 1 \\
IRAS 05189 & 0.042 & 0.035 & \nodata & 12.16 & 1 \\
IRAS 08572 & 0.058 & $<$0.012 & \nodata & 12.14 & 1 \\
IRAS 12112 & 0.073 & 0.517 & 0.569 & 12.33 & 1 \\
IRAS 14348 & 0.083 & 0.254 & \nodata & 12.35 & 1 \\
IRAS 15250 & 0.055 & 0.023 & \nodata & 12.05 & 1 \\
IRAS 22491 & 0.077 & 0.594 & 0.671 & 12.19 & 1 \\
NGC1569 & $\sim$0 & \nodata & 0.402 & 8.76  & 4 \\
3C120 & 0.033 & \nodata & \nodata & \nodata & 5 \\
PG1612+261 & 0.131 & \nodata & \nodata & \nodata & 5 \\
PAH 1\tablenotemark{a} & \nodata & \nodata & \nodata & \nodata & 6 \\
PAH 2\tablenotemark{a} & \nodata & \nodata & \nodata & \nodata & 6 \\ 
PAH 3\tablenotemark{a} & \nodata & \nodata & \nodata & \nodata & 6 \\
PAH 4\tablenotemark{a} & \nodata & \nodata & \nodata & \nodata & 6 \\
PAH 5\tablenotemark{a} & \nodata & \nodata & \nodata & \nodata & 6 \\
\enddata
\tablenotetext{a}{As explained in \cite{2007ApJ...656..770S}, those 5 templates are composite spectra computed in arbitrary units of $\nu I_{\nu}$. We thus do not have corresponding redshift, luminosity or equivalent width.}
\tablerefs{(1) \cite{2007ApJ...656..148A}; (2) \cite{2004ApJS..154..178A}; (3) \cite{2006ApJ...640..204A}; (4) \cite{2006ApJ...639..157W}; (5) P. Ogle, private communication; (6) \cite{2007ApJ...656..770S}; (7) \cite{1998ApJS..119...41K}.}
\end{deluxetable}

\begin{deluxetable}{lcccccc}
\tabletypesize{\footnotesize}
\tablecaption{Cross-correlation analysis results. \label{SrcCcorrTab}}
\tablehead{
\colhead{ID} & 
\colhead{Field} & 
\colhead{$z_{IRS}$} &
\colhead{$z_{ref}$ \tablenotemark{a}} & 
\colhead{$z_{ref}$ type \tablenotemark{a}} &
\colhead{template}  & 
\colhead{spectral type \tablenotemark{b}} \\
& & & & & &
}
\startdata
SUUSS 1   &      1 &   1.70 &   1.59 &      P &  NGC6240 & mixed \\
SUUSS 2   &      1 &   0.50 &   0.50 &      S &    PAH4 & PAH \\
SUUSS 3   &      1 &   2.03 &   2.03 &      S &   Arp220 & mixed \\
SUUSS 4   &      1 &   0.95 &   1.01 &      P &   PAH4 & PAH \\
SUUSS 8   &      1 &   0.65 &    \nodata & \nodata &   PAH2 & PAH \\
SUUSS 9   &      1 &   2.08 &    \nodata & \nodata &  NGC1569 & line \\
SUUSS 14  &      1 &   2.20 &   2.67 &      P &   Mrk273 & mixed \\
SUUSS 15  &      1 &   0.41 &   0.46 &      S &    PAH3 & PAH \\
SUUSS 17  &      1 &   0.30 &   0.30 &      S &  pg1612+261 & line \\
SUUSS 18  &      1 &   2.04 &    \nodata & \nodata &    IRAS 15250 & SiO \\
SUUSS 20  &      1 &   0.91 &   0.91 &      S &    PAH4 & PAH \\
\\
\hline
\\
SUUSS 24  &      2 &   2.03 &   2.15 &      P &    PAH5 & PAH \\
SUUSS 25  &      2 &   0.84 &   0.84 &      S &    PAH4 & PAH \\
SUUSS 26  &      2 &   1.20 &   1.22 &      S &   Mrk231 & SiO \\
SUUSS 28  &      2 &   0.87 &   0.81 &      P &  UGC5101 & mixed \\
SUUSS 29  &      2 &   0.94 &   0.94 &      S &   PAH3 & PAH \\
SUUSS 30  &      2 &   0.53 &   0.52 &      S &    PAH5 & PAH \\
SUUSS 32  &      2 &   0.20 &    \nodata & \nodata &   PAH2 & PAH \\
SUUSS 33  &      2 &   1.34 &    \nodata & \nodata &  NGC6240 & mixed \\
SUUSS 34  &      2 &   1.21 &    \nodata & \nodata &   Mrk231 & SiO \\
SUUSS 36  &      2 &   0.61 &    \nodata & \nodata &    PAH4 & PAH \\
SUUSS 37  &      2 &   0.41 &   0.41 &      S &    PAH2 & PAH \\
SUUSS 38  &      2 &   1.23 &    \nodata & \nodata &  Mrk463 & SiO \\
SUUSS 39  &      2 &   0.48 &   0.48 &      S &    PAH4 & PAH \\
SUUSS 41  &      2 &   0.80 &   0.77 &      P &  Arp220 & mixed \\
SUUSS 42  &      2 &   1.00 &   1.02 &      S &   Mrk273 & mixed \\
SUUSS 44  &      2 &   1.00 &   1.02 &      S &    IRAS 22491 & mixed \\
SUUSS 45  &      2 &   1.01 &   0.96 &      P &   PAH1 & PAH \\
\\
\hline
\\
SUUSS 12\tablenotemark{c}   &      1 &  \nodata &   0.50 &   S &  \nodata & \nodata \\
SUUSS 16\tablenotemark{c}   &      1 &  \nodata &   0.46 &   S &  \nodata & \nodata \\
SUUSS 19\tablenotemark{c}   &      1 &  \nodata &   1.01 &   S &  \nodata & \nodata \\
SUUSS 27\tablenotemark{c}   &      2 &  \nodata &   0.30 &   S &  \nodata & \nodata \\
SUUSS 31\tablenotemark{c}   &      2 &  \nodata &   0.47 &   S &  \nodata & \nodata \\
\enddata
\tablenotetext{a}{Spectroscopic (S) redshifts are from the {\it Treasurey Keck Redshift Survey} (Wirth et al., 2004) and photometric (P) redshifts are from Caputi et al. (2006).}
\tablenotetext{b}{Sort the main spectral signature identified in the MIR spectra: {\it PAH} for strong aromatic features, {\it SiO} for silicate absorption, {\it mixed} for intermediate spectra, and {\it line} for the 2 line sources discussed in Sect \ref{SectLineSrc}.}
\tablenotetext{c}{Sources with no conclusive identification from the cross-correlation method but for which we had ancilliary redshifts (same reference).}
\end{deluxetable}

\begin{deluxetable}{lcccccc}
\tabletypesize{\footnotesize}
\tablecaption{SUUSS sample statistics. \label{SrcStatTab}}
\tablehead{
\colhead{order} & 
\colhead{\# detections} & 
\colhead{strong} &
\colhead{mixed}  & 
\colhead{silicate} & 
\colhead{strong line} &
\colhead{featureless /} \\
& & aromatics & signatures & absorption &  & Continuum
}
\startdata
 20-35$\mu$m &         20 &      5 &      3 &      1 &           2 &           9 \\
 14-20$\mu$m &         25 &      9 &      5 &      3 &           0 &           8 \\
\enddata
\end{deluxetable}




\addtocounter{figure}{-7}
\begin{figure*}
\plotone{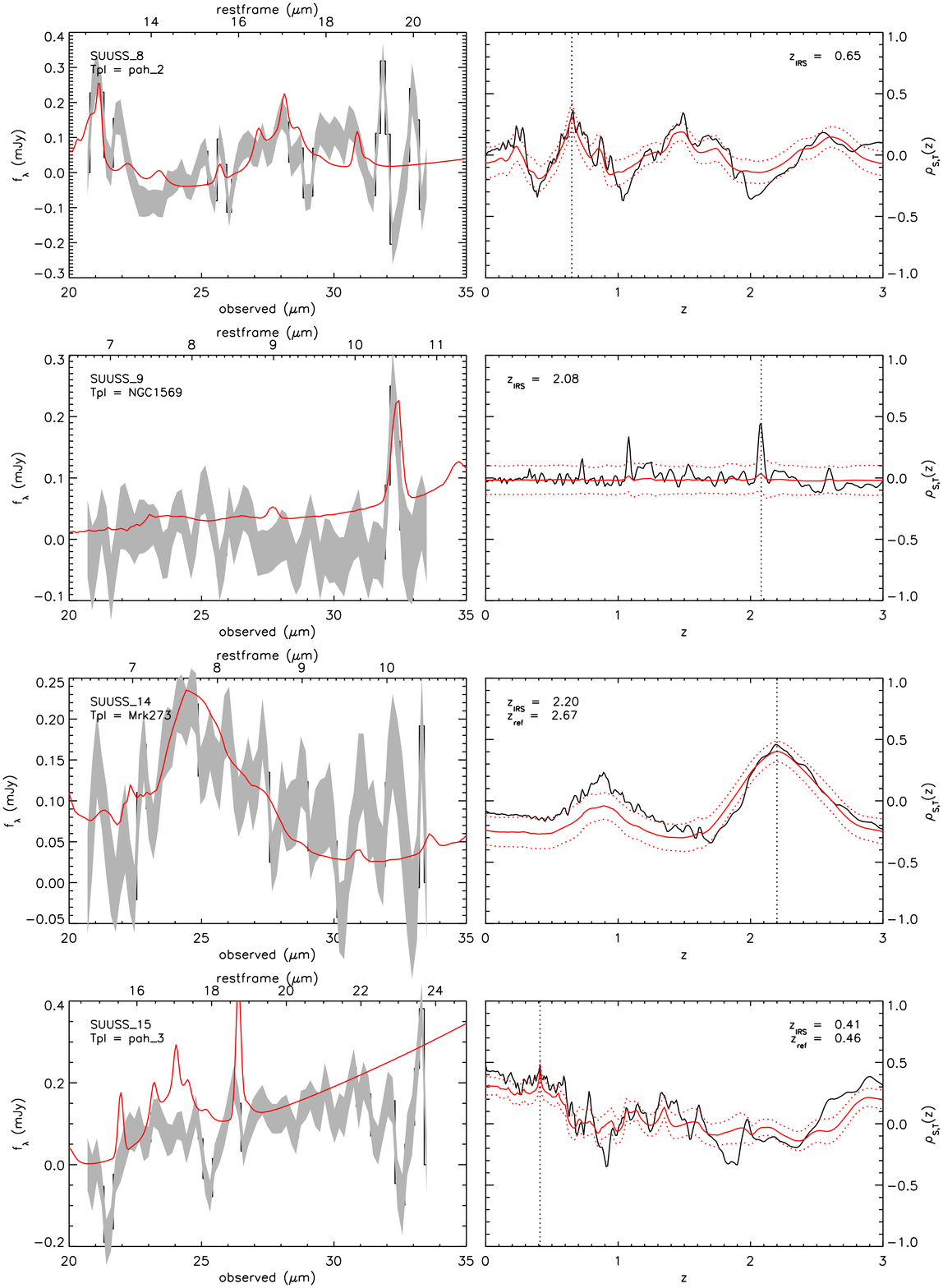}
\caption{\emph{continued}}
\end{figure*}

\addtocounter{figure}{-1}
\begin{figure*}
\plotone{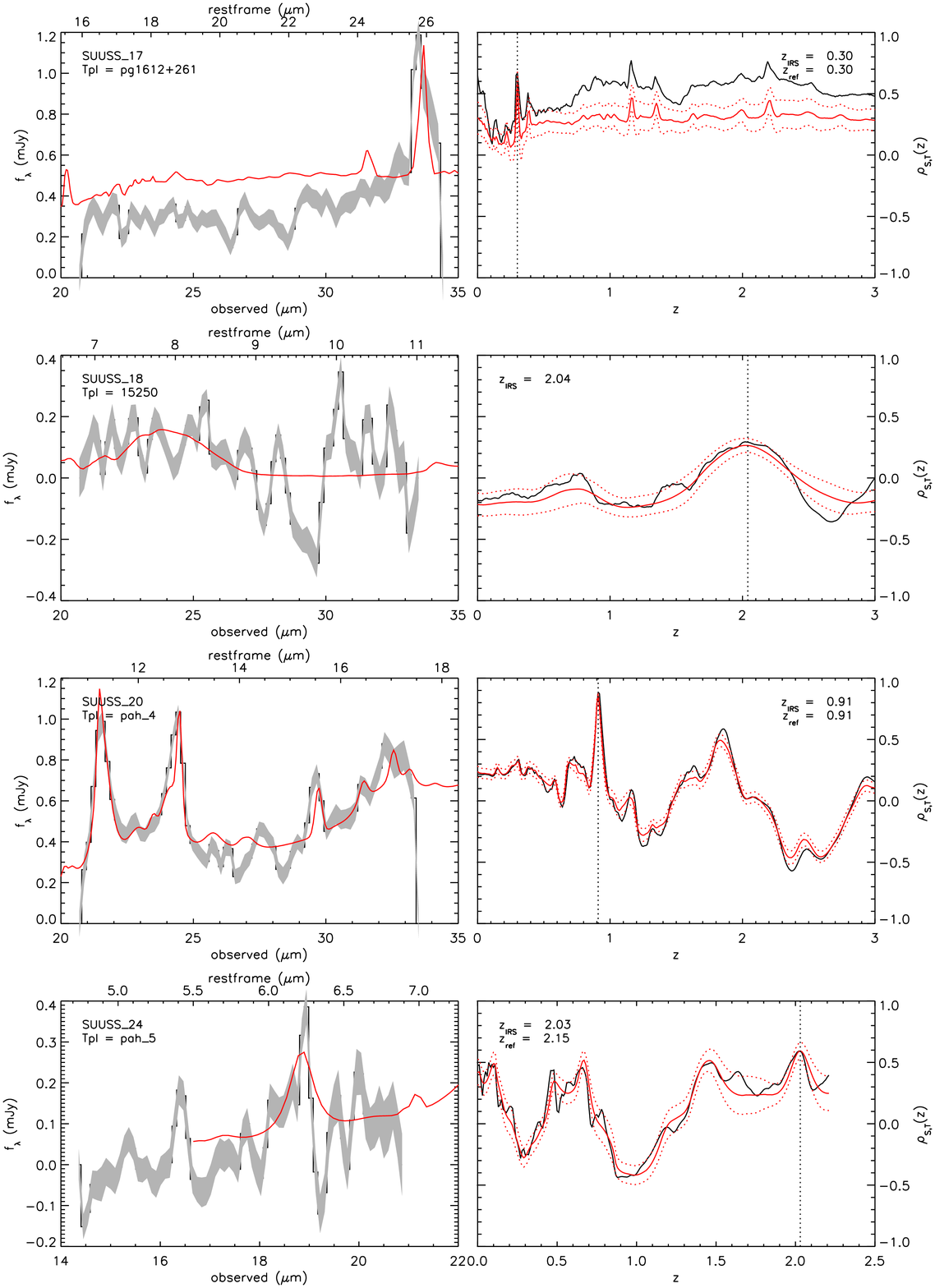}
\caption{\emph{continued}}
\end{figure*}

\addtocounter{figure}{-1}
\begin{figure*}
\plotone{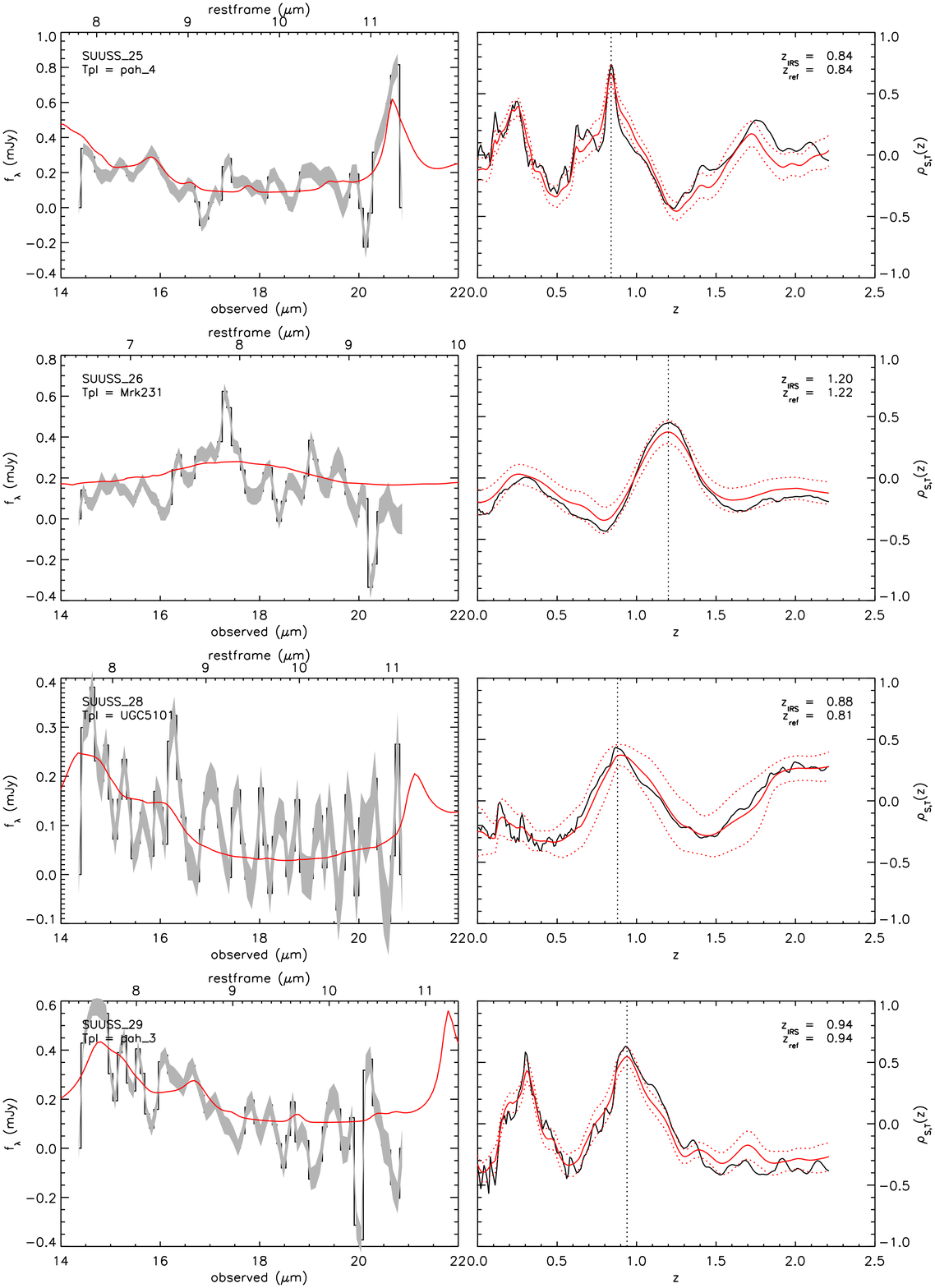}
\caption{\emph{continued}}
\end{figure*}

\addtocounter{figure}{-1}
\begin{figure*}
\plotone{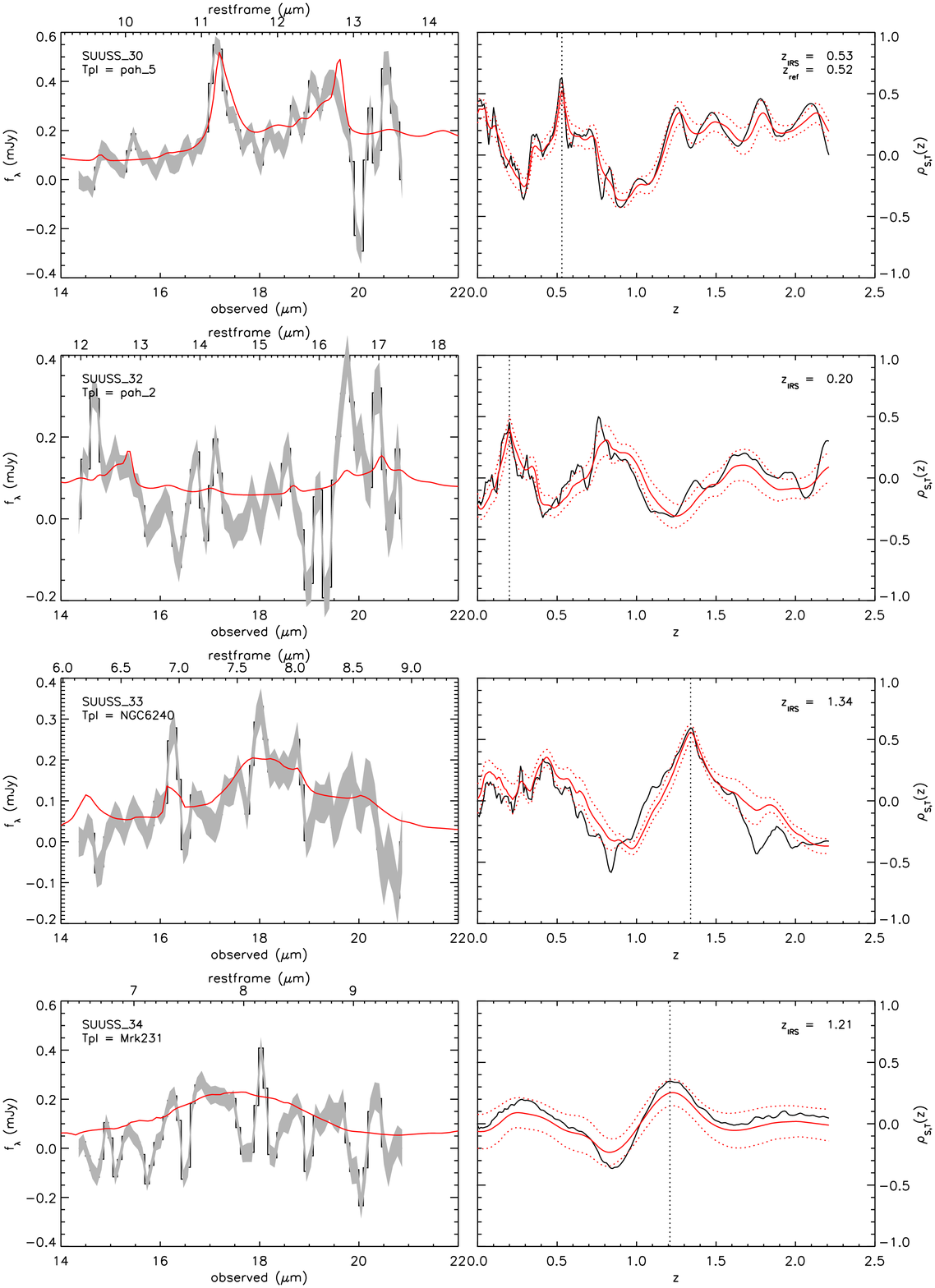}
\caption{\emph{continued}}
\end{figure*}

\addtocounter{figure}{-1}
\begin{figure*}
\plotone{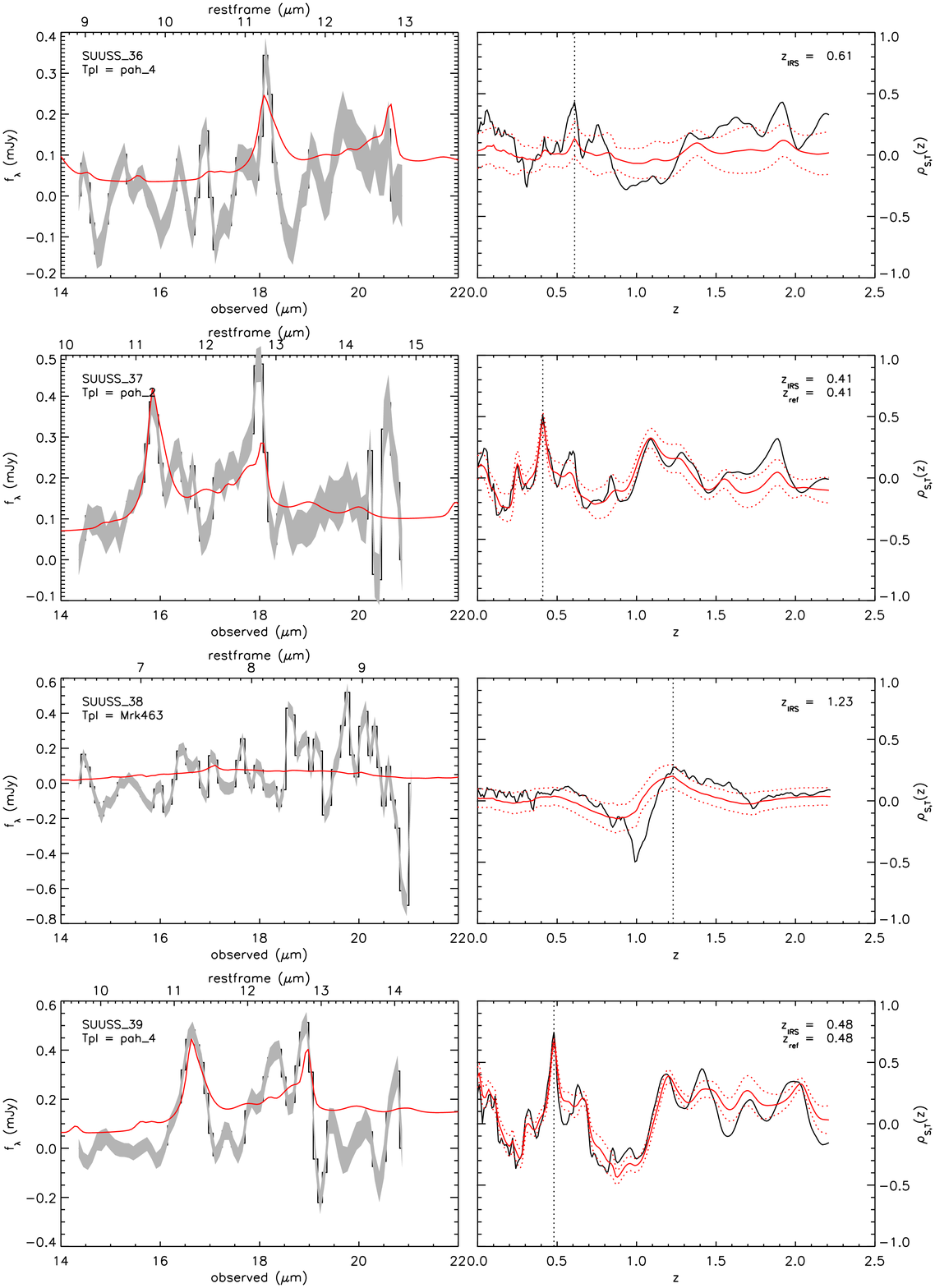}
\caption{\emph{continued}}
\end{figure*}

\addtocounter{figure}{-1}
\begin{figure*}
\plotone{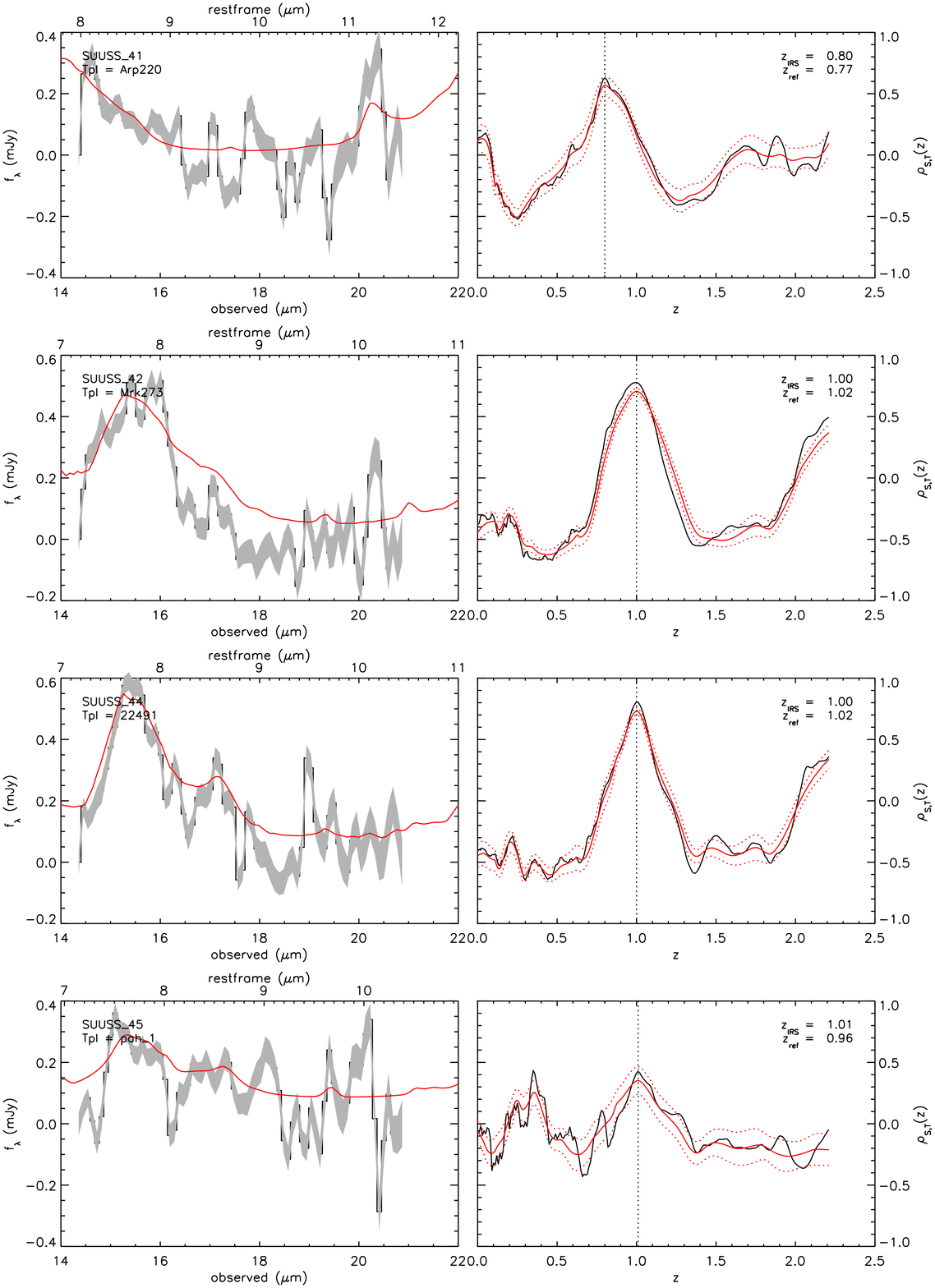}
\caption{\emph{continued}}
\end{figure*}


\end{document}